# Deep Inelastic Scattering of Polarized Electrons by Polarized $^3$He and the Study of the Neutron Spin Structure *

The E142 Collaboration

P.L. Anthony[7,12], R.G. Arnold[1], H.R. Band[16], H. Borel[5], P.E. Bosted[1], V. Breton[2], G.D. Cates[11], T.E. Chupp[8], F.S. Dietrich[7], J. Dunne[1], R. Erbacher[13], J. Fellbaum[1], H. Fonvieille[2], R. Gearhart[12], R. Holmes[14], E.W. Hughes[4,12], J.R. Johnson[16], D. Kawall[13], C. Keppel[1], S.E. Kuhn[13,10], R.M. Lombard-Nelsen[5], J. Marroncle[5], T. Maruyama[16,12], W. Meyer[12], Z.-E. Meziani[13,15], H. Middleton[11], J. Morgenstern[5], N.R. Newbury[11], G.G. Petratos[6,12], R. Pitthan[12], R. Prepost[16], Y. Roblin[2], S.E. Rock[1], S.H. Rokni[12], G. Shapiro[3], T. Smith[8], P.A. Souder[14], M. Spengos[1], F. Staley[5], L.M. Stuart[12], Z.M. Szalata[1], Y. Terrien[5], A.K. Thompson[9], J.L. White[1,12], M. Woods[12], J. Xu[14], C.C. Young[12], G. Zapalac[16]

[1] American University, Washington, DC 20016

[2] LPC IN2P3/CNRS, Univ. Blaise Pascal, F-63170 Aubiere Cedex, France

[3] University of California and LBL, Berkeley, CA 94720

[4] California Institute of Technology, Pasadena, CA 91125

[5] Centre d'Etudes de Saclay, DAPNIA/SPhN, F-91191 Gif-sur-Yvette, France

[6] Kent State University, Kent, OH 44242

[7] Lawrence Livermore National Laboratory, Livermore, CA 94550

[8] University of Michigan, Ann Arbor, MI 48109

[9] National Institute of Standards and Technology, Gaithersburg, MD 20899

[10] Old Dominion University, Norfolk, VA 23529

[11] Princeton University, Princeton, NJ 08544

[12] Stanford Linear Accelerator Center, Stanford, CA 94309

[13] Stanford University, Stanford, CA 94305

[14] Syracuse University, Syracuse, NY 13210

*Work supported by Department of Energy contract DE–AC03–76SF00515.

[15] Temple University, Philadelphia, PA 19122
[16] University of Wisconsin, Madison, WI 53706

**Abstract**

The neutron longitudinal and transverse asymmetries $A_1^n$ and $A_2^n$ have been extracted from deep inelastic scattering of polarized electrons by a polarized $^3$He target at incident energies of 19.42, 22.66 and 25.51 GeV. The measurement allows for the determination of the neutron spin structure functions $g_1^n(x,Q^2)$ and $g_2^n(x,Q^2)$ over the range $0.03 < x < 0.6$ at an average $Q^2$ of 2 (GeV/$c$)$^2$. The data are used for the evaluation of the Ellis-Jaffe and Bjorken sum rules. The neutron spin structure function $g_1^n(x,Q^2)$ is small and negative within the range of our measurement, yielding an integral $\int_{0.03}^{0.6} g_1^n(x)dx = -0.028 \pm 0.006\ (stat) \pm 0.006\ (syst)$. Assuming Regge behavior at low $x$, we extract $\Gamma_1^n = \int_0^1 g_1^n(x)dx = -0.031 \pm 0.006\ (stat) \pm 0.009\ (syst)$. Combined with previous proton integral results from SLAC experiment E143, we find $\Gamma_1^p - \Gamma_1^n = 0.160 \pm 0.015$ in agreement with the Bjorken sum rule prediction $\Gamma_1^p - \Gamma_1^n = 0.176 \pm 0.008$ at a $Q^2$ value of 3 (GeV/$c$)$^2$ evaluated using $\alpha_s = 0.32 \pm 0.05$.

# 1 Introduction

During the past twenty five years, experiments measuring spin-averaged deep inelastic scattering of electrons, muons and neutrinos have provided a wealth of knowledge about the nature of QCD and the structure of the nucleon in terms of quarks and gluons. Among the highlights are the determination of scaling violations[1, 2, 3] from structure functions as predicted by QCD, leading to a value of the strong coupling constant $\alpha_s$[4], and the test of the Gross-Lewellyn-Smith sum rule[5], in which QCD radiative corrections are verified within experimental errors. More recently, polarized deep inelastic scattering experiments, which probe the spin orientation of the nucleon's constituents, are providing a new window on QCD and the structure of the nucleon.

Pioneering experiments[6, 7, 8] with polarized electrons and protons, performed at SLAC in the late 1970's and early 1980's in a limited $x$ range, revealed large spin dependence in deep inelastic e-p scattering. These large effects were predicted by Bjorken[9] and by simple SU(6) quark models. More recently, results from the CERN EMC experiment[10] over a wider $x$ range have sparked considerable interest in the field because the data suggest surprisingly that quarks contribute relatively little to the spin of the proton and that the strange sea quark polarization is significant.

A central motivation for these experiments is a pair of sum rules. The first, due to Bjorken[9], is a QCD prediction that invokes isospin symmetry to relate the spin-dependent structure functions to the neutron beta-decay axial coupling constant $g_A$. The experimental test of this sum rule requires data on both the proton and the neutron. Advances in technology for producing highly polarized beams and targets make possible increasingly precise measurements. A second sum rule, due to Ellis and Jaffe[11], which has more theoretical uncertainty, applies to the proton and neutron separately. Assuming SU(3) symmetry, data from either the neutron or proton can be used to determine the contributions of each quark flavor to the spin of the nucleon. It is the apparent disagreement of the EMC proton data with the prediction of the Ellis-Jaffe sum rule that led to the striking conclusions mentioned above. This paper reports on a precision determination of the neutron spin structure function $g_1^n$ using a polarized $^3$He target.

In spin-dependent deep inelastic scattering[12] one measures the quantity

$$A_1(x, Q^2) = \frac{\sigma_{\frac{1}{2}} - \sigma_{\frac{3}{2}}}{\sigma_{\frac{1}{2}} + \sigma_{\frac{3}{2}}} \tag{1}$$

where $\sigma_{\frac{3}{2}(\frac{1}{2})}$ is the absorption cross section for virtual photons with total $J_z = \frac{3}{2}(\frac{1}{2})$ for the final state.

In the case of a target of Dirac particles, $A_1$ is unity. In QCD, the nucleon may be described in terms of a set of quark momentum distributions $q_i(x, Q^2)$, where $x$ is the fraction of nucleon momentum carried by the struck quark and $Q^2$ is the squared four-momentum transfer to the nucleon. The index $i$ includes $u$, $d$, and $s$ quarks and antiquarks. Thus for the nucleon, $A_1$ measures how the individual quarks, weighted by the square of their charges, are aligned relative to the nucleon as a whole.

The $q_i(x, Q^2)$ have been evaluated from the measured values of the spin averaged structure function $F_1(x, Q^2)$ for various leptons and nucleon targets. However, the determination of $A_1(x, Q^2)$ requires that quark momentum distributions be decomposed in terms of spin. Thus $q_i^{\uparrow}(x, Q^2)$ ($q_i^{\downarrow}(x, Q^2)$) give the probability that a quark of type $i$ has a fraction $x$ of the nucleon's momentum with its spin parallel (antiparallel) to that of the nucleon. Then

$$\Delta q_i(x, Q^2) \equiv q_i^{\uparrow}(x, Q^2) - q_i^{\downarrow}(x, Q^2) + \overline{q}_i^{\uparrow}(x, Q^2) - \overline{q}_i^{\downarrow}(x, Q^2) \tag{2}$$

and

$$A_1(x, Q^2) = \frac{\sum e_i^2 \Delta q_i(x, Q^2)}{\sum e_i^2 (q_i(x, Q^2) + \overline{q}_i(x, Q^2))} \simeq \frac{g_1(x, Q^2)}{F_1(x, Q^2)}. \tag{3}$$

where $e_i$ is the charge of the $i^{th}$ quark.

The latter equation defines the spin-dependent structure function $g_1(x, Q^2)$. A more precise definition of $g_1$ and the relevant kinematics is given in Section **4** below.

In the nonrelativistic quark model, the spins of the quarks relative to the spin of the nucleon are given by the SU(6) wavefunctions, resulting in the predictions $A_1^p = \frac{5}{9}$ and $A_1^n = 0$. This simple picture holds approximately at $x \approx 0.3$. At low $x$, $A_1^p$ decreases due to the dominance of sea quarks which one might naively expect to have small polarization. In this region, Regge theory suggests that $A_1 \sim x^\alpha$ with $1 < \alpha < 1.5$. At large $x$, $A_1 \to 1$ according to perturbative QCD arguments[13, 14]. Nucleon models have been constructed incorporating these general features and yield reasonable fits to the data[15, 16, 17] for the proton.

The Bjorken sum rule, however, applies to the integrals of $g_1$ :

$$\Gamma_1^{p(n)}(Q^2) = \int_0^1 g_1^{p(n)}(x, Q^2)dx. \tag{4}$$

The goal of the experiments is to measure $g_1^{(p)n}$ over as wide a kinematic range as possible to extract a value for $\Gamma_1^{(p)n}$.

The paper is organized as follows. Section **2** defines the theoretical framework used in polarized deep inelastic scattering and tests of the Bjorken and Ellis-Jaffe sum rules. Section **3** discusses the experimental method and data collection. Section **4** describes the analysis leading to the raw asymmetries used to extract the physics asymmetries $A_1$ and $A_2$ and spin structure functions $g_1$ and $g_2$ of $^3$He. Section **5** reports on dilution factor studies and radiative corrections, while section **6** reports on the $^3$He results and the nuclear corrections used to extract the virtual photon-neutron asymmetries $A_1^n$ and $A_2^n$ and the spin structure functions $g_1^n$ and $g_2^n$. Section **7** describes the physics implications of the results and conclusions are presented in Section **8**.

# 2 Theoretical framework

## 2.1 Bjorken Sum Rule

The Bjorken sum rule prediction, which relates high energy electromagnetic scattering to the low energy beta decay of the neutron, was derived in the high $Q^2$ limit by Bjorken[9] using current algebra, and later shown to be a rigorous QCD prediction[3] with calculable radiative corrections for finite $Q^2$[18, 19, 20]. This sum rule may be derived in QCD by using the Operator Product Expansion (OPE)[21]. The OPE relates integrals of quark momentum distributions $\Delta q_i(x, Q^2)$ to matrix elements of single operators such as

$$G_{An}^{q_i} s_\mu \equiv \frac{1}{2}\langle n, s|\overline{q}_i\gamma_\mu\gamma_5 q_i|n, s\rangle. \tag{5}$$

where $|n, s\rangle$ represents a *neutron* $(n)$ with spin $s$. The $G_{An}^{q_i}$ are constants independent of $Q^2$, although they do depend on the choice of renormalization scale $\mu$. There are different $G_A$'s corresponding to each combination of quarks and baryons in the lowest octet. They are related by isospin and SU(3) symmetry so that in the limit that SU(3) is exact only three independent quantities remain,

$$\begin{aligned} G_{Ap}^u &= \Delta u \\ G_{Ap}^d &= \Delta d \\ G_{Ap}^s &= \Delta s. \end{aligned} \tag{6}$$

These are the matrix elements of the *proton*. With the latter notation, one must be very careful to distinguish $\Delta q$ from $\Delta q(x, Q^2)$.

Useful linear combinations of the matrix elements are:

$$\begin{aligned} a_3 = g_A &= G_{Ap}^u - G_{Ap}^d; \\ a_8 &= G_{Ap}^u + G_{Ap}^d - 2G_{Ap}^s; \\ a_0 = \Delta\Sigma &= G_{Ap}^u + G_{Ap}^d + G_{Ap}^s. \end{aligned} \tag{7}$$

Similar combinations for the quark momentum distributions are:

$$\begin{aligned} \Delta q_3(x, Q^2) &= \Delta u(x, Q^2) - \Delta d(x, Q^2); \\ \Delta q_8(x, Q^2) &= \Delta u(x, Q^2) + \Delta d(x, Q^2) - 2\Delta s(x, Q^2); \\ \Delta q_0(x, Q^2) &= \Delta u(x, Q^2) + \Delta d(x, Q^2) + \Delta s(x, Q^2). \end{aligned} \tag{8}$$

Here $\Delta q_{3(8)}(x, Q^2)$ is the nonsinglet combination and $\Delta q_0(x, Q^2)$ is the singlet combination. An immediate advantage of this notation is that $a_3$ and $a_8$ are constants independent of the renormalization scale $\mu^2$. $\Delta\Sigma(\mu^2)$ on the other hand depends on the scale and special care must be used when interpreting this quantity. Using the above notations, the main result of OPE gives

$$\int_0^1 \Delta q_{3(8)}(x,Q^2)dx =$$
$$a_{3(8)}\Big(1+\Sigma_{n=1}^{\infty} C_n(\tfrac{\alpha_s(Q^2)}{\pi})^n\Big)+\Sigma_{m=1}^{\infty}\tfrac{K_m}{Q^{2m}}. \quad (9)$$

The perturbative QCD series in $\alpha_s(Q^2)$ describes high energy or short distance effects and has been recently evaluated exactly up to third order in QCD. The power series in $1/Q^2$, in Eq. 9 contains the "higher twist" terms. These terms describe long-distance, non-perturbative behavior that involves, among other effects, the details of the wavefunctions of the quarks in the nucleon. The calculation of some of these terms has been the subject of recent literature [22]. It is expected that the contributions of these terms are small for the $Q^2$ range discussed in this paper.

The spin dependent structure function $g_1(x,Q^2)$ measures the difference in number of partons with helicity parallel versus antiparallel to the helicity of the nucleon weighted by the square of the parton charge. Explicitly, we have

$$g_1^{p(n)}(x,Q^2) = \frac{1}{2}\sum e_i^2 \Delta q_i(x,Q^2) =$$
$$\tfrac{1}{2}[\tfrac{4}{9}(\tfrac{1}{9})\Delta u(x,Q^2)+\tfrac{1}{9}(\tfrac{4}{9})\Delta d(x,Q^2)+\tfrac{1}{9}\Delta s(x,Q^2)]. \quad (10)$$

The Bjorken sum rule follows from Eqs. 9 and 10[23]:

$$\Gamma^{Bj}(Q^2) \equiv \Gamma_1^p(Q^2)-\Gamma_1^n(Q^2) =$$
$$\tfrac{1}{6}g_A[1-\tfrac{\alpha_s(Q^2)}{\pi}-\binom{3.58}{3.25}(\tfrac{\alpha_s(Q^2)}{\pi})^2-\binom{20.2}{13.8}(\tfrac{\alpha_s(Q^2)}{\pi})^3+\cdots] \quad (11)$$

where the upper (lower) numbers are for three (four) quark flavors, and higher twist terms have been neglected. The number of active quark flavors is determined by the number of quarks with $m_q < Q$, taking $m_c = 1.5$ GeV and $m_b = 4.5$ GeV. For our case we use three flavors, since the effects of charm are expected to turn on slowly.

Measuring $\Gamma^{Bj}$ at different $Q^2$ provides a sensitive test of QCD and its radiative corrections. It is one of two QCD sum rules (the Gross-Llewellyn-Smith being the other) where the right hand side of the sum rule is accurately known. With sufficiently precise data, one can extract $\alpha_S$ based on the Bjorken sum rule and compare with the determination of $\alpha_S$ from other processes.

## 2.2 The Nucleon Sum Rule

The sum rule for a single nucleon is [25]:

$$\begin{aligned}
\Gamma_1^{p(n)}(Q^2) &= \\
&\tfrac{1}{12}\Big[\Big(\pm g_A + \tfrac{1}{3}a_8\Big)\Big(1 - \tfrac{\alpha_s}{\pi} - \binom{3.58}{3.25}\big(\tfrac{\alpha_s}{\pi}\big)^2 - \binom{20.2}{13.8}\big(\tfrac{\alpha_s}{\pi}\big)^3 \cdots\Big) \\
&\quad + \tfrac{4}{3}\Delta\Sigma(\mu^2 = Q^2)\Big(1 - \tfrac{\alpha_s}{\pi} - \binom{1.10}{0.07}\big(\tfrac{\alpha_s}{\pi}\big)^2 - \cdots\Big)\Big], \qquad (12)
\end{aligned}$$

where the upper (lower) coefficients are for three (four) flavors. To leading order, this expression requires a new nucleon matrix element, $(a_8 + 4\Delta\Sigma)$, which can only be estimated from nucleon models. As pointed out by M. Gourdin [24], $a_8 = 3F - D$ as determined from baryon beta decay if flavor SU(3) symmetry is assumed. Then $\Delta\Sigma$ may be determined from the sum rule. $\Delta\Sigma$ is an important input for nucleon models. Much of the excitement in the field arises from the unexpected EMC[10] result that $\Delta\Sigma \approx 0$. In the nonrelativistic quark model, $\Delta\Sigma = 1$. However the motion of the quarks should give a suppression similar to the suppression of $g_A$ from 5/3 to the experimental value of 1.2, yielding $\Delta\Sigma \approx 0.7$. In addition, gluons and orbital angular momentum may make substantial contributions to the spin of the proton. The present world average of $\Delta\Sigma \sim 0.3$ was not anticipated by most authors prior to the measurements.

In addition, $\Delta\Sigma$ is also needed for predicting elastic scattering cross sections. Two examples of physical processes involving $\Delta\Sigma$ are neutrino scattering and the scattering of possible supersymmetric particles.

Eq. 12 involves singlet operators, which in leading order of QCD includes the Adler-Bell-Jackiw (ABJ) anomaly[26, 27] because the relevant anomalous dimension is nonzero. One result is the scale dependence of $\Delta\Sigma(\mu^2)$. Any physical process dependent on $\Delta\Sigma$ must also involve other $\mu^2$-dependent factors such that the result is independent of $\mu^2$. For example, part of the neutrino-proton elastic scattering cross section arises from the current[28]

$$J^0_\nu = z^0(\mu^2)\Delta\Sigma(\mu^2)s_\nu. \tag{13}$$

Here $z^0 = \frac{1}{2}(1+\Delta(\mu^2))$, the weak charge of the proton. This quantity is just 1/2 in the SU(2)×U(1) electroweak theory and is scale-dependent. The scale dependencies of $z^0$ and $\Delta\Sigma$ cancel so that $J^0_\nu$ is scale-independent and thus a measurable physical quantity. Moreover, choosing $\mu^2 = M_Z^2$ helps minimize $\Delta(\mu^2)$. Strikingly, it is $\Delta\Sigma(\mu^2)$ with $\mu^2 > 2$ (GeV/c)$^2$ that is needed to compute neutrino scattering at $Q^2 \approx 0$.

For our results, we have chosen to use the scale $\mu^2 = Q^2$ to avoid passing over quark thresholds. If experiments performed at different $Q^2$ are compared, the scale of one or both of the experiments should be changed according to the formula

$$\begin{aligned}\Delta\Sigma(\mu_1^2) &= \Delta\Sigma(\mu_2^2)\times \\ &[1 + \binom{.667}{.960}\left(\frac{\alpha_s(\mu_1^2)-\alpha_s(\mu_2^2)}{\pi}\right) + \binom{1.21}{1.97}\left(\frac{\alpha_s^2(\mu_1^2)-\alpha_s^2(\mu_2^2)}{\pi^2}\right)].\end{aligned} \tag{14}$$

Ideally, one would choose a universal scale: $\mu^2 = M_P^2$ or $\mu^2 = M_Z^2$, however, the first option suffers from the fact that $\alpha_s(M_P)$ is not well known and the second option requires running the scale across several mass thresholds which yields even more complex expressions. Another option commonly used is to define a quantity $\Delta\Sigma_{inv}$ and use the formula [25]

$$\begin{aligned}&\Gamma_1^{p(n)}(Q^2) = \\ &\frac{1}{12}[\left(\pm g_A + \frac{1}{3}a_8\right)(1 - \frac{\alpha_s}{\pi} - \binom{3.58}{3.25}(\frac{\alpha_s}{\pi})^2 - \binom{20.2}{13.8}(\frac{\alpha_s}{\pi})^3\cdots) \\ &\quad + \frac{4}{3}\Delta\Sigma_{inv}(1 - \binom{0.33}{0.04}\frac{\alpha_s}{\pi} - \binom{0.55}{-0.54}(\frac{\alpha_s}{\pi})^2 - \cdots)].\end{aligned} \tag{15}$$

We will also quote $\Delta\Sigma_{inv}$ for our data.

The individual quark contributions can be extracted using either Eq. 12 or Eq. 15 along with $a_8 = 3F - D$:

$$\begin{aligned}\Delta u &= [2\Delta\Sigma + a_8 + 3g_A]/6; \\ \Delta d &= [2\Delta\Sigma + a_8 - 3g_A]/6; \\ \Delta s &= [\Delta\Sigma - a_8]/3.\end{aligned} \tag{16}$$

### 2.3 The Ellis-Jaffe Sum Rule

Prior to the establishment of QCD, Ellis and Jaffe [11] made a numerical prediction for the nucleon sum rule by arguing that $\Delta s = G_A^s = 0$. This gives the additional relation

$$a_8 = \Delta\Sigma \tag{17}$$

which, when combined with the value for $a_8$ extracted from hyperon decay, provides values for all of the needed matrix elements. As pointed out by Jaffe[29], this relation is rather curious in the context of QCD because $a_8$ is scale independent and $\Delta\Sigma(\mu^2)$ is scale dependent. Hence the prediction of the Ellis Jaffe sum rule depends upon the scale chosen. Experiment E142 has used $\mu^2 = 2$ (GeV/c)$^2$. A more common choice is to set $a_8 = \Delta\Sigma_{inv}$. The difference is about 0.005 in $\Gamma_1^n$ and should be accounted for when comparing different experiments.

A second issue with the Ellis-Jaffe sum rule is that Eq. 17 implies a large contribution to $\Gamma_1^{p(n)}$ in Eq. 12 coming from $a_8$ so that the error $\delta a_8$ becomes important. The determination of $G_A^s = \Delta s$ from the nucleon sum rule also suffers from this problem.

Given the importance of $a_8$, it is worth going into some detail about its origin. By assuming flavor SU(3) for the nucleon octet, we have

$$\langle n, s|\overline{\psi}\gamma_\mu\gamma_5 V_+\psi|\Delta\Sigma^-, s\rangle \approx \langle n, s|\overline{\psi}\gamma_\mu\gamma_5 V_3\psi|n, s\rangle, \tag{18}$$

or in terms of quark operators

$$\begin{aligned} g_A(\Sigma^- \to n)2s_\mu &\equiv \langle n, s|\overline{u}\gamma_\mu\gamma_5 s|\Sigma^-, s\rangle \approx \\ &\langle n, s|\overline{u}\gamma_\mu\gamma_5 u - \overline{s}\gamma_\mu\gamma_5 s|n, s\rangle \equiv G_{An}^u - G_{An}^s \approx \\ &\langle p, s|\overline{d}\gamma_\mu\gamma_5 d - \overline{s}\gamma_\mu\gamma_5 s|p, s\rangle = (G_{Ap}^d - G_{Ap}^s)2s_\mu. \end{aligned} \tag{19}$$

Thus the axial matrix elements for $\Sigma$ decay can be related to proton matrix elements. Similar results hold for the other hyperon decays. To average over many hyperon decay measurements, the following relation is used:

$$F = \frac{1}{2}(\langle G^u_{Ap}\rangle - \langle G^s_{Ap}\rangle) \text{ and } D = \frac{1}{2}(\langle G^u_{Ap}\rangle - 2\langle G^d_{Ap}\rangle + \langle G^s_{Ap}\rangle). \tag{20}$$

Jaffe and Manohar [30] assign a generous error $3F - D = 0.60 \pm 0.12$ while Ratcliffe[31] quotes a range of values form 0.53 to 0.83 based on various assumptions about $SU(3)_f$ breaking and which decays to use. For the purpose of this paper, we will use $a_8 = 3F - D = 0.58 \pm 0.12$, which is the updated central value of Close and Roberts [32] but has the generous error of Jaffe and Manohar[30]. The net result is that the prediction of the Ellis-Jaffe sum rule for $Q^2$=2 (GeV/$c$)$^2$ is $\Gamma_1^n = -0.011 \pm 0.016$. The 0.12 uncertainty on $3F - D$ translates to a 0.06 uncertainty on $\Delta s$, which is not negligible compared to typical world averages of $\Delta s \sim -0.1$.

An alternative definition, equivalent in the limit of exact SU(3), that is often used in the literature is $F + D \equiv g_A(n \to p) = 1.2573 \pm 0.0028$. Then hyperon data are used to obtain $F/D$. The world average, based on the analysis of Close and Roberts[32], is $F/D = 0.575 \pm 0.016$. This result yields $\Gamma_1^n = -0.011 \pm 0.005$, with a substantially smaller uncertainty, while the uncertainty on $\Delta s$ changes from 0.06 to 0.04.

The above two alternatives illustrate how the prediction of the Ellis-Jaffe sum rule is sensitive to the various assumptions chosen.

## 3 The E142 experiment

The experiment discussed in this paper was performed to measure for the first time the virtual photon-nucleon spin asymmetries, $A_1^n$ and $A_2^n$, in deep inelastic scattering of polarized electrons by polarized $^3$He. From these asymmetries, the neutron spin structure functions $g_1^n$ and $g_2^n$ are extracted. The experiment relied on the production and delivery of a high energy polarized electron beam at the Stanford Linear Accelerator Center (SLAC). The polarized incident electrons were delivered to End Station A where they scattered off a polarized $^3$He target and were detected in magnetic spectrometers. The experiment, named SLAC E142, collected data over a period of six weeks in November and December of 1992. An overview of the primary technical achievements of the experiment are presented in Table 1. Details of the polarized electron beam polarimetry, polarized target and magnetic spectrometers are discussed in subsequent sections.

Previous results on the spin structure function $g_1^n$ from this experiment have been published[33]. This paper reports on a more thorough analysis of the results leading to a change in the previously published results for $g_1^n$. Among the new information presented in the present paper are results on the neutron transverse spin structure function $g_2^n$, and the results on the $Q^2$ dependence of the neutron longitudinal spin structure function $g_1^n$.

### 3.1 The Polarized Electron Beam

The SLAC polarized electron source[34], using an AlGaAs photocathode at a temperature of 0°C, produced the polarized electron beam for this experiment [35, 36]. Polarized electrons were produced by illuminating the photocathode with circularly polarized light at a wavelength near the band-gap edge of the photocathode material. AlGaAs with 13% Al, rather than GaAs, was chosen as the photocathode since the larger band gap of the AlGaAs cathode was a better match to the available flashlamp pumped dye laser operating at a wavelength of 715 nm. The electron helicity was changed randomly pulse by pulse by controlling the circular polarization of the excitation light. Using this cathode, the polarized source produced an electron beam polarization of about 36%.

Electrons from the source were accelerated to energies ranging from 19 to 26 GeV and directed onto the polarized $^3$He target. The SLAC accelerator operated with pulses of approximately 1 $\mu$sec duration at a rate of 120 Hz. The beam current was quite high, operating at typically $2\times10^{11}$ electrons per pulse. The spectrometers collected typically 2 events per pulse from the polarized $^3$He target, yielding approximately 300 million events for the experiment.

The experiment collected data at three discrete energies of 19.42, 22.66, 25.51 GeV with an energy acceptance of typically 0.5%. Since the primary beam undergoes a 24.5° bend before reaching the experimental target, the electron spin precesses more than the momentum by an amount:

$$\Delta\theta = \pi \cdot \frac{24.5^\circ}{180^\circ} \cdot \frac{g_e - 2}{2} \cdot \frac{E}{m} \tag{21}$$

where $E$ is the beam energy, $m$ is the electron mass, $g_e$ is the electron gyromagnetic ratio and $\Delta\theta$ is the angle between the electron spin and momentum at the target. When $\theta$ is an integer multiple of $\pi$, the electron spin is longitudinal at the target. The energies 19.42 and 22.66 GeV satisfy this condition exactly, while the energy 25.51 corresponds to 93% of the maximum available polarization. This latter energy was the maximum energy that the beam line magnets could support at the time.

The beam spot size was typically 2 to 4 mm at the target. Studies of the spot position and radius near the target as measured by a wire array found no dependence on beam helicity at the level of better than $\pm$ 0.01 mm. From models of the variation of the target window thicknesses, it was determined that this implied that false asymmetries due to a possible helicity-dependent motion of the electron beam position would be significantly less than $10^{-4}$.

The electron beam polarization was determined using single-arm Møller polarimetry. The high peak beam currents precluded the detection of double arm coincidences. The cross section for spin dependent elastic electron-electron scattering is given by [37]:

$$d\sigma/d\Omega = (d\sigma_0/d\Omega)\ (1 + \sum_{i,j} P_B^i A_{ij} P_T^j) \tag{22}$$

where $P_B^i$ are the components of the beam polarization and $P_T^j$ are the components of the target polarization. The z axis is along the beam direction and the y axis is chosen normal to the scattering plane. The cross section is given by the unpolarized cross section $d\sigma_0/d\Omega$, and the asymmetry terms $A_{ij}$. If $P_T$ is independently known, the above expression may be used to determine the beam polarization $P_B$.

To lowest order, the fully relativistic unpolarized laboratory cross section is given by:

$$(d\sigma_0/d\Omega)_L = \left[\frac{\alpha(1+\cos\theta_{CM})(3+\cos^2\theta_{CM})}{2m\sin^2\theta_{CM}}\right]^2 . \qquad (23)$$

For the measurement of longitudinal polarization with a longitudinally polarized target foil, the only relevant asymmetry term is $A_{zz}$ given by:

$$A_{zz} = -\frac{(7+\cos^2\theta_{CM})\sin^2\theta_{CM}}{(3+\cos^2\theta_{CM})^2} . \qquad (24)$$

Here $\theta_{CM}$ is the center-of-mass scattering angle, m is the electron mass, and $\alpha$ is the fine structure constant. The asymmetry maximum is at $\theta_{CM} = 90°$ where the unpolarized laboratory cross section is 0.179 b/sr and $A_{zz} = -7/9$.

Most if not all Møller polarimeters utilize thin ferromagnetic foils as the polarized electron target. The distinction between the free target electrons of the previous formulae and the bound atomic electrons of the physical target was ignored until recently when Levchuk [38] pointed out that the analyzing power of Møller polarimeters may have significant corrections due to the electron orbital motion of the target foil electrons. Atomic electrons have momentum distributions which are different for different atomic shells. Electrons in the outer shells have small momenta but those from the inner shells have momenta up to 100 keV. Although small compared to a beam energy of 22.66 GeV, these momenta are not small compared to the electron rest mass and can alter the center of mass energy and thus the scattering angle in the lab frame by up to 10 %. The relative angular smearing correction for polarized and unpolarized electrons is shown in Fig. 1. The effect causes different line shapes for scatters from different shells. Since the polarized target electrons are only in the 3d (M) shell, the fraction of signal from the polarized target electrons and thus the expected Møller asymmetry varies over the Møller scattering elastic peak. Inclusion of this effect has been shown to modify the analyzing power of Møller polarimeters by up to 15% [38, 39, 40] depending on the exact geometry of the polarimeter. Inclusion of this effect modifies the analyzing power of the E142 Møller polarimeter by 5%.

The E-142 Møller polarimeter shown in Fig. 2 consisted of a scattering target chamber containing several magnetized foils, a collimator to define the scattering angle and angular acceptance, a magnet to measure the momentum of the scattered electrons, a segmented detector array to detect the scattered electrons, and a data acquisition system.

The magnetized target foils were made of Vacoflux [41], an alloy of 49% Co, 49% Fe, and 2% Va by weight. The foils were 3 cm wide by 35 cm long and were mounted at a 20.7° angle with respect to the beam. Three foils of approximate thickness 20, 30, and 50 $\mu$m were installed. Nearly all the Møller data were taken with the two thinner foils. The foils were magnetized by Helmholtz coils providing a 100 G field along the beam direction. The polarity of the coils was typically reversed between Møller data runs to alternate the sign of the foil polarization and to minimize systematic errors.

The polarization $P_T$ of the target electrons was determined from the relation:

$$P_T = \frac{M}{n_e \mu_B} \times (\frac{g' - 1}{g'}) \times (\frac{g_e}{g_e - 1}) \ , \tag{25}$$

where M is the bulk magnetization in the foil, $n_e$ is the electron density, $g_e = 2.002319$ is the free electron gyromagnetic ratio, and $\mu_B = 9.273 \times 10^{-21}$ G-cm$^3$ is the Bohr magneton. The factor involving the magnetomechanical ratio ($g'$) includes the correction for the orbital contribution to the magnetization. Interpolating between the measured $g'$ values of Fe and Co, the $g'$ of Vacoflux was calculated to be $1.889 \pm 0.005$. Substituting into the above equation yields:

$$P_T = \frac{M}{n_e \mu_B} \times (0.94011 \pm 0.00280). \tag{26}$$

The magnetization M was determined by direct flux measurements using a precise integrating voltmeter (IVM) connected to a pickup coil placed around the foils. From Faraday's law, as the external H field is swept between –H and +H, the integral of the induced voltage over time can be related to the B and H fields and hence the magnetization M through:

$$4\pi M = B - H = \frac{\int_{in} V dt - \int_{out} V dt}{2 \times N_T \times A_{foil}} \ , \qquad (27)$$

where *in* and *out* refer to flux measurements with the foil in and out, and $A_{foil}$ is the cross-section area of the foil. Recognizing that the foil density can be determined from the measured mass of the foil, length and area $A_{foil}$, the foil polarization $P_T$ can be determined from Eq. 26 and Eq. 27.

A 21 radiation length thick tungsten mask located 7.11 meter from the Møller target restricted the scattering angles of the particles entering the polarimeter detector to $5.0 \leq \theta \leq 10.5$ mrad. The azimuthal acceptance of the rectangular mask opening depended on the scattering angle and varied from $\pm 0.14$ to $\pm 0.068$ rad with respect to the horizontal plane. The scattered particles next passed through a 0.25 mm thick Mylar vacuum window and entered a large aperture spectrometer magnet. The 1.83 meter long magnet was typically run at a $\int Bdl = 14.5$ kG-m for the 22.66 GeV data. The spectrometer setting selects Møller scattered electrons at 10 GeV/c corresponding to a center of mass scattering angle of 97°. The field integral was adjusted during the experiment to position the Møller peak in the detector, to compensate for different beam energies and to maximize signal and reduce background.

The main beam passed through a 33 mm round hole in the mask and continued down the E-142 beam line. The beam exiting the central hole in the mask contained large numbers of low energy bremsstrahlung electrons produced by the target foil. Large magnetic fields would bend these particles out of the beamline generating unacceptable backgrounds in the detector. To reduce the field along the beamline a 7.6 cm by 30.5 cm soft iron septum with a 5 cm by 5 cm hole for the beam was inserted in the magnet gap. The septum reduced the B field seen by the beam by about a factor of 100 reducing $\int Bdl$ to approximately 150 G-m.

After exiting the magnet the Møller scattered electrons traveled through a He bag to the detector located 22.4 meters from the target. The detector consisted of 37 gas proportional tubes embedded in lead. Each 4 mm diameter brass tube contained a 40 micron wire strung through the center. The tubes were placed in two parallel rows 7.9 mm apart. The first row was behind 36.8 mm of lead. The second row was 6.9 mm behind the front row and offset by 3.9 mm giving an effective segmentation of 3.9 mm in the horizontal (scattering) plane. The lead absorbed soft photon backgrounds and amplified the Møller signal. Since the momenta and scattering angle of the Møller scatters are correlated, the scatters fall in a tilted stripe at the detector. The detector was oriented so that the tubes were parallel to the Møller stripe. The active detector length of 4 cm corresponded to a momentum acceptance of 2.9%. The entire detector package was mounted on a vertical mover allowing different momenta to be selected.

The signal in each tube was integrated over the 1 $\mu$sec long beam pulse by a charge integrating preamplifier. The data, together with the sign of the beam polarization, were recorded by a peak sensing analog to digital converter (ADC) system. The beam polarization was randomly reversed between pulses to reduce systematic errors. The number of Møller electrons detected per pulse varied with current and target thickness, but was typically 70 per pulse. A typical Møller run lasted 150 seconds and contained a million Møller electron scatters.

### 3.2 Beam Polarization Analysis

For each Møller run an average pulse height for each detector channel was calculated for both right (R) and left (L) handed incident beam. The pulse to pulse variance of the ADC values was used to estimate the error in the average pulse height. These averages and errors were recorded with relevant beam currents, detector and target positions, and magnet settings. Typical measured distributions for R+L and R–L are shown in Fig. 3. The R+L distribution (Fig. 3a) shows an elastic scattering peak with a radiative tail on top of an unpolarized background. The signal to background ratio varied with shielding conditions and beam parameters from $\approx 2$ at the beginning of the experiment to $\approx 7$ after shielding improvements made during the experiment. The R–L distribution (Fig. 3b) is to a good approximation pure Møller electron scattering and shows only a radiative tail with no background.

The beam polarization was determined by fitting the observed elastic scattering and asymmetry distributions to line shapes based on Møller scattering plus a background component. Although several techniques were used as cross checks, the full data sets were analyzed with a technique which derived the Møller component of the line shapes from the measured R–L distributions and used this shape together with a quadratic background component to fit the observed scattering distributions. In this technique, all of the observed R–L line shape is attributed to Møller scattered electrons and the background is assumed to be unpolarized.

The analysis technique used the observed R–L line-shape and the angular smearing functions shown in Fig. 1 to generate a predicted R+L line-shape for Møller scatters. For zero target momenta the R–L line-shape and the R+L line-shapes are identical except for backgrounds. The trial R+L line shape was generated from the observed R–L line-shape by first correcting for the angular smearing due to the polarized target electrons and then convoluting the result with the smearing correction for all (polarized and unpolarized) target electrons. Additional corrections were made for the variation of the cross section and change of azimuthal acceptance with scattering angle and for the variation in the value of the Møller scattering asymmetry over the angular acceptance of the detector. The observed R+L distribution was then fit by the predicted line-shape plus a quadratic background. The solid line is Fig. 3a shows the resultant fitted line shape for the typical runs.

The measured longitudinal beam polarization $P_B$ is shown in Fig. 4 for Møller polarimeter data runs covering the last 5 weeks of the experiment. Only runs with the Møller peak well centered and with statistical errors less than 5% are displayed. The lower polarization of the 25.5 GeV data is evident showing the effect of the non-optimal beam energy. Correcting for the beam energy, an average beam polarization was calculated for each of the target foils averaging over the different beam ener-gies. The average beam polarization determined from the 20 $\mu$m foil data was $0.360 \pm 0.002$ and from the 30 $\mu$m foil data was $0.354 \pm 0.001$ where the errors are statistical only. The foil averages differ by 1.5% , within the 1.7% systematic error on the foil polarization. The beam polarization did not exhibit any time dependence over the duration of the experiment.

In addition to the systematic error in the foil polarization there is a contribution to the overall systematic error from the uncertainty in the modeling of the scattering kinematics, lineshapes, asymmetries, detector linearity, and preamp-ADC linearity. The various contributions to the systematic error are summarized in Table 2. Adding the systematic uncertainties in quadrature yields an overall systematic error of 3.1% relative. The resulting longitudinal beam polarization averaged over the target foils is then $P_B = (0.357 \pm 0.001 \pm 0.011) \times \cos\left(\pi \frac{E(GeV)}{3.237}\right)$.

### 3.3 The Polarized $^3$He Target

The experiment used a polarized $^3$He target to extract the neutron spin structure function. The polarized $^3$He target relies on the technique of spin-exchange optical pumping[42, 43, 44]. Spin-exchange optical pumping refers to a two step process in which, (1) rubidium (Rb) atoms are polarized by optical pumping, and (2) the electronic polarization of the Rb atoms is transferred to the nuclei of the $^3$He atoms by spin-exchange collisions. The optical pumping is accomplished by driving transitions from the Rb $5^2S_{1/2}$ ground state to the $5^2P_{1/2}$ first excited state using circularly polarized light from lasers. The wavelength of this transition, often referred to as the Rb $D_1$ line, is 795 nm. Within a timescale of milliseconds, one of the two substates of the ground state is selectively depopulated, resulting in very high atomic polarization[45]. The spin exchange takes place when the polarized Rb atoms undergo binary collisions with the $^3$He atoms. The $^3$He electrons, being paired in the $^1S_0$ ground state, do not participate in the collision from a spin point of view. The spin-$\frac{1}{2}$ $^3$He nucleus, however, interacts with the Rb valence electron through hyperfine interactions, which can result in a mutual spin flip. As long as the Rb vapor is continu-ally being polarized, this results in a gradual transfer of angular momentum to the $^3$He nuclei.

The time evolution of the $^3$He polarization, assuming the $^3$He polarization $P_{\mathrm{He}} = 0$ at $t = 0$, is given by

$$P_{\mathrm{He}}(t) = \langle P_{Rb} \rangle \left( \frac{\gamma_{SE}}{\gamma_{SE} + \Gamma_R} \right) \left( 1 - e^{-(\gamma_{SE} + \Gamma_R)t} \right) \tag{28}$$

where $\gamma_{SE}$ is the spin-exchange rate per $^3$He atom between the Rb and $^3$He, $\Gamma_R$ is the relaxation rate of the $^3$He nuclear polarization through all channels other than spin exchange with Rb, and $\langle P_{Rb} \rangle$ is the average polarization of a Rb atom.[46] The spin exchange rate $\gamma_{\mathrm{SE}}$ is defined by

$$\gamma_{\mathrm{SE}} \equiv \langle \sigma_{\mathrm{SE}} v \rangle [\mathrm{Rb}]_{\mathrm{A}} \tag{29}$$

where $\langle \sigma_{\mathrm{SE}} v \rangle = 1.2 \times 10^{-19}$ cm$^3$/sec is the velocity-averaged spin-exchange cross section for Rb – $^3$He collisions[46, 47] and $[\mathrm{Rb}]_{\mathrm{A}}$ is the average Rb number density seen by a $^3$He atom. We operated at Rb number densities such that the spin-exchange time constant $\gamma_{\mathrm{SE}}^{-1}$ was typically 10 to 30 hrs and the time constant for build-up of $^3$He nuclear polarization, $(\gamma_{\mathrm{SE}} + \Gamma_R)^{-1}$ ranged from about 9 to 20 hours. A typical spin-up polarization curve is shown in Fig. 5.

In order to achieve the highest $^3$He polarization, we attempted to maximize $\gamma_{\mathrm{SE}}$ and minimize $\Gamma_R$. From Eq. 29, maximizing $\gamma_{\mathrm{SE}}$ implies increasing the alkali-metal number density, which in turn requires more laser power [46, 48]. For a fixed volume of polarized Rb, the number of photons needed per second must compensate for the number of Rb spins destroyed per second. In total, we used five lasers, which collectively provided about 16–22 W and achieved a value of $\gamma_{\mathrm{SE}} \approx 1/12$ hours.

There are several processes which contribute to the $^3$He relaxation rate $\Gamma_R$. An important example is relaxation that occurs during $^3$He–$^3$He collisions due to the dipole interaction between the two $^3$He nuclei[49]. Dipole induced relaxation provides a lower bound to $\Gamma_R$, and has been calculated to be

$$\Gamma_{\mathrm{dipolar}} = \frac{1}{744 \text{ hours}} [^3\mathrm{He}], \tag{30}$$

at 23°C where $[^3\mathrm{He}]$ is the number density of $^3$He in amagats (an amagat is a unit of density corresponding to 1 atm at 0°C)[49].

The relaxation rate varies with temperature, implying a maximum relaxation time constant of ~100 hours for the $^3$He densities and temperatures found in our target. Another important contribution to $\Gamma_R$ is relaxation that occurs during wall collisions, a relaxation rate we will designate $\Gamma_{\mathrm{wall}}$. Both $\Gamma_{\mathrm{dipolar}}$ and $\Gamma_{\mathrm{wall}}$ are intrinsic to a given target cell, making it useful to define the quantity

$$\Gamma_{\mathrm{cell}} = \Gamma_{\mathrm{dipolar}} + \Gamma_{\mathrm{wall}} \tag{31}$$

that accounts for all relaxation mechanisms that are associated with a specific cell. For the three cells actually used in our experiment, $\Gamma_{\mathrm{cell}}^{-1}$ varied between 53 and 65 hours.

In addition to $\Gamma_{\mathrm{cell}}$ there are interactions not inherent to the target cell which further increase the nuclear relaxation rate. Inhomogeneities in the magnetic field that provides an alignment axis for the $^3$He nuclear polarization induce relaxation according to

$$\Gamma_{\nabla B} = D\left(\frac{|\nabla B_x|^2 + |\nabla B_y|^2}{B_0^2}\right), \tag{32}$$

where $D$ is the diffusion constant of the $^3$He in the target, $B_0$ is the magnitude of the alignment field, assumed to lie along the z–axis, and $B_{\mathrm{x}}$ and $B_{\mathrm{y}}$ are the components of the magnetic field transverse to $B_{\mathrm{z}}$[50]. This effect was very small and we calculated $\Gamma_{\nabla B}^{-1} \sim 400$ hours in our target.

During the experiment, nuclear relaxation was also induced by the presence of ionizing radiation from the electron beam, a phenomenon which is well understood both theoretically[51] and experimentally[52]. When a $^3$He atom is ionized, the hyperfine interaction couples the nuclear spin to the unpaired electron spin which can in general be depolarizing. Furthermore, electrons from other $^3$He atoms can be transferred to the original ion, creating the potential for depolarizing other atoms. The depolarization rate $\Gamma_{\mathrm{beam}}$ therefore depends on the ionization rate of the $^3$He and the average number of $^3$He nuclear depolarizations per $^3$He ion created. The relaxation time $\Gamma_{\mathrm{beam}}^{-1}$ for our experiment inferred from the $\sim 10\%$ relative drop in $^3$He polarization at our maximum beam current of 3.3 $\mu$A is 100–200 hours, consistent with the predicted time constant of 170 hours.

When all the relaxation mechanisms are included, the total $^3$He nuclear relaxation rate is given by

$$\Gamma_R = \Gamma_{\rm cell} + \Gamma_{\rm beam} + \Gamma_{\nabla B}. \tag{33}$$

From the previous discussion we see that $\Gamma_R^{-1}$ was in the range of about 40 hours. With $\gamma_{\rm SE}^{-1} \approx 25$ hours, equation (28) predicts that the maximum polarization, given by $\gamma_{\rm SE}/(\gamma_{\rm SE}+\Gamma_R)$, is about 0.62.

A schematic of the target system is shown in Fig. 6. The central feature of the polarized $^3$He target is the glass cell containing $\sim$8.4 atm of $^3$He (as measured at 20°C), several milligrams of Rb metal, and $\sim$ 65 Torr of $N_2$. The $N_2$ aids in the optical pumping by causing radiationless quenching of the Rb atoms when they are in the excited state[45]. The target cells were based on a double chamber design,[53] comprising an upper "pumping chamber" in which the optical pumping and spin exchange took place, and a lower "target chamber" through which the electron beam passed. The Rb was contained almost entirely in the upper pumping chamber, which was the only chamber that was heated (to achieve the desired Rb number density) when the target was in operation. The upper and lower chambers had roughly the same volume (70 cm$^3$ and 90 cm$^3$, respectively) for a total volume of about 160 cm$^3$, and were connected by a 60 mm long, 9 mm inside diameter "transfer tube". The target chamber had a length of about 30 cm, a diameter that was roughly 2 cm, and thinned rounded convex end windows. The average window thickness for the cells used in the experiment was 112 $\mu$m per window.

The pumping chamber was enclosed by an oven, with heating supplied by flowing hot air, the temperature of which determined the Rb number density. The oven, and all other items which were near the target cell, were made of non-magnetic materials so as not to interfere with the NMR polarimetry. The oven was made of a high temperature plastic called Nylatron GS, and was operated at temperatures of about 160 to 165 °C, which corresponds to a Rb number density of $1.7-2.2\times10^{14}$ atoms/cm$^3$[54]. It was found that higher temperatures resulted in leaks forming in the oven. The colder target chamber, at $\sim$ 65 °C, had a Rb density that was about three orders of magnitude lower. The quantity $[\mathrm{Rb}]_A$ that was referred to earlier is the volume weighted *average* of the Rb number density over both chambers. For the temperatures at which we operated, the pressure in the cell was 11 atmospheres, and the density in the target chamber was about 8.9 Amagats.

The optical pumping was accomplished using five titanium-sapphire lasers pumped by five argon ion lasers. The beams were passed through $\lambda/4$ plates to achieve circular polarization, and were arranged to get a reasonable filling of the pumping chamber's cross section. The laser system was housed in a protective "laser hut" in a high radiation area near the target. Access for laser tuning during the experiment was limited, but was generally necessary only once every few days.

A set of 1.4 m diameter Helmholtz coils, coaxial with the electron beam, produced a 20 to 40 G alignment field for the $^3$He nuclear polarization. The field strength was chosen to be large enough to (1) suppress the effects of ambient magnetic field inhomogeneities and (2) to facilitate a nuclear magnetic resonance (NMR) measurement of the Boltzmann equilibrium polarization of protons in water for polarimetry purposes. The $^3$He nuclear polarization was measured using the NMR technique of adiabatic fast passage (AFP)[55]. The AFP system used, in addition to the main field coils, a set of 46 cm diameter Helmholtz rf drive coils and an orthogonal set of smaller pick-up coils, both of which are pictured in Fig. 6. A second set of Helmholtz coils transverse to the electron beam axis was used to rotate the target polarization and for operation with a polarization transverse to the beam.

The glass target cell, the oven, the rf coils, the pickup coils, and other assorted target components were located inside a vacuum chamber in order to reduce the background event rates from non-target materials. Small cooling jets of $^4$He were directed at the thin entrance and exit windows of the target chamber as a precaution against the thin glass breaking due to excessive heating from the electron beam.

The production of target cells with long intrinsic relaxation times $\Gamma_{\mathrm{cell}}^{-1}$ proved to be a challenging task. The cells were made out of aluminosilicate glass (Corning 1720) since such glass is known to have favorable spin-relaxation properties[56, 57]. The use of aluminosilicate glass, however, was not sufficient for obtaining long $\Gamma_{\mathrm{cell}}^{-1}$'s. We found it necessary, for instance, to "re-size" tubing before incorporating the tubing into the final cell construction. In this process, tubing of some initial diameter is brought to a molten state and blown to a new diameter while being turned on a glass lathe. It is our belief that this process results in a more pristine surface, presumably with fewer contaminants and defects.

For filling with $^3$He gas, the cells were attached to a high vacuum system ($\sim 10^{-7}$ to $10^{-8}$ Torr) and given long bake-outs under vacuum for 3 to 6 days at 475 °C. The Rb was distilled into the cell with a hand-held torch from a side arm of the vacuum system. During the distillation, the cells remained open to the vacuum pumps so that any material outgassed due to the heat of the torch was pumped away. Next, a small amount of nitrogen (99.9995% pure) was frozen into the cell. Finally, the initially 99.995% chemically pure $^3$He was introduced into the cell through a trap at liquid $^4$He temperature. This cryogenic trap further purifies the $^3$He by condensing out any contaminants. The cells were cooled with liquid $^4$He during filling in order to achieve a high density of $^3$He while maintaining a pressure of less than one atmosphere. The cryogenic filling technique ensures that when the tube through which the cell is filled is heated, the glass will collapse on itself, thereby sealing the cell.

Out of ten cells produced with the techniques described above, $\Gamma_{\mathrm{cell}}^{-1}$ was carefully characterized in five. In these cases $\Gamma_{\mathrm{cell}}^{-1}$ was always in excess of 30 hours, and for the cells used in the experiment, $\Gamma_{\mathrm{cell}}^{-1}$ was in the range of 50 to 65 hours at room temperature. These numbers, compared to the 95 hour upper limit on $\Gamma_{\mathrm{dipolar}}^{-1}$ at 20°C, imply that most of the relaxation was caused by the unavoidable $^3$He-$^3$He dipole interaction, although some improvement in $\Gamma_{\mathrm{cell}}^{-1}$ is still possible.

Polarimetry was accomplished by comparing the AFP signals of the $^3$He with the AFP signals from water samples. The AFP scans involved applying rf at 92 kHz using the drive coils while simultaneously sweeping the main magnetic field through the resonance condition. When passing through the resonance, the nuclear spins reverse their direction, creating a measureable NMR signal in the pickup coils. Two resonance curves were obtained during each AFP measurement — one as the field was swept up, and the other as the field was swept down. Examples of resonance curves for both $^3$He and water are shown in Fig. 7. In the case of the water signal, the average of 25 scans is shown. Target polarization losses during a measurement were typically less than 0.1% relative.

When studying water, some care needs to be taken to interpret the AFP signals properly. The average proton polarization that occurs during the two AFP peaks can be written

$$P_p = \xi \tanh\left(\frac{h\nu}{2k_B T}\right), \tag{34}$$

where $h$ is Planck's constant, $\nu = 92$ kHz is the frequency of the applied rf, $k_B$ is Boltzmann's constant, and $T$ is the temperature of the sample. Basically, $P_p$ is the thermal equilibrium Boltzmann polarization that is expected at the field corresponding to the point at which resonance occurs. There is a caveat, however, in that the proton spins relax toward the *effective* magnetic field experienced in the rotating frame of the rf, which on resonance is given by the magnitude of the applied rf (in our case about 76 mG, much less than the applied static field). This effect is accounted for by the parameter $\xi$, which we have calculated to be $0.966 \pm 0.014$. If the longitudinal relaxation time $T_1$ for the protons were infinite, $\xi$ would be equal to one. As it is, however, the measured proton signal corresponds to a slightly lower polarization than one would naively expect.

Through a careful comparison with water signals, we determined a calibration of $1.61\pm$ 0.11% polarization per 10 mV of signal. As Fig. 7 shows, the $^3$He signals were extremely clean. The uncertainty in the polarimetery was thus dominated by the uncertainty in the calibration constant. The largest contribution was from the determination of the magnitude of the water signals, which were about 1.9$\mu$V. The error here was dominated by a systematic shift in the base line of the NMR signal before and after passing through resonance, an effect that is clearly visible in Fig. 7. The interpretation of the water signal, that is, the calculation of $\xi$, was another important contribution. Electronic gains, the comparison of the exact shapes of the different $^3$He and water cells, and knowledge of the exact density of the $^3$He were also important contributions. Finally, the lock-in amplifier that was used in the NMR set-up had a time constant that gave a small distortion to the AFP resonance shape, for which a small correction had to be applied. The various errors are summarized in Table 3.

During the experimental run, the $^3$He polarization was measured roughly every four hours. The results of these measurements are shown in Fig. 8. The average $^3$He polarization over the entire experiment was about 33%. During the first three weeks of the experiment, there were a few precipitous drops in the polarization due to a variety of problems, as indicated in Fig. 8. Later, however, the target polarization was very stable, running for three weeks with only slow drifts. Toward the end of the experiment, the slight drop in polarization that is evident in Fig. 8 is due to an increase in the beam current from an average of 2.1 $\mu$A to 3.4 $\mu$A.

### 3.4 The Electron Spectrometers

Electrons scattered from the $^3$He target were detected in two single-arm spectrometers. The spectrometers were centered at 4.5° and 7.0° with respect to the beam line in order to maximize the kinematic coverage for an electron beam energy of 22.7 GeV and an event selection criteria of $Q^2 > 1$ (GeV/$c$)$^2$. The momentum acceptance ranged from 7 to 20 GeV/c for both arms. A schematic of the two systems is shown in Fig. 9. Both arms used magnetic elements from the existing SLAC 8 and 20 GeV/c spectrometers.

The design of the spectrometer was driven by several requirements. The cross sections to be measured were known to be small, typically of the order of $10^{-32}$ cm$^2$/(sr·GeV). The raw counting ratio asymmetry of the two different spin orientations was also predicted to be small, of the order of $10^{-3}$ to $10^{-4}$. In order to minimize beam running time, the spin structure function measurements required spectrometers with the largest possible solid angle over a momentum acceptance range extending from 7 to 20 GeV/c. Such a momentum acceptance gives a rather wide coverage over $x$ with $Q^2 > 1$ $(\mathrm{GeV}/c)^2$ (see Fig. 10).

In addition, these small scattering angle spectrometers were designed to suppress an expected large photon background coming from the target due to bremsstrahlung, radiative Møller scattering and the decay of photoproduced $\pi^{\circ}$ mesons. Background rate calculations indicated the need for at least a "two-bounce system" (the configuration of the spectrometer should allow a photon to reach the detectors only after scattering at least twice on the magnet poles or vacuum walls) in order to keep this background at a tolerable level.

The energy resolution of the spectrometers was defined solely by the required $x$ resolution. The cross section asymmetries were not expected to exhibit any sizable dependence on momentum transfer. The energy resolution ranged from $\pm 5\%$ at $E'$=7 GeV to $\pm 4\%$ at $E'$= 18 GeV for each spectrometer. The resulting resolution in $\Delta x/x$ ranged from $\pm 8\%$ at low $x$ up to $\pm 15\%$ at the highest $x$ covered by each spectrometer ($x \leq 0.4$ in the 4.5° spectrometer, and $x \leq 0.6$ in the 7° spectrometer).

The spectrometer design[58] used two dipoles bending in opposite directions, providing a large solid angle acceptance which remains constant over a very large momentum interval. The solid angle of the "reverse-bend" dipole doublet configuration, when integrated over the 7 to 20 GeV/$c$ momentum interval, is twice that of previous "conventional" designs with the two dipoles bending in the same direction. The maximum solid angle of the two spectrometers is shown as a function of momentum in Fig.11. The reverse bend also fulfills the "two-bounce" requirement by optimizing the deflecting angles and the separation of the two dipoles. In the 7.0° spectrometer the distance between the two dipoles was 2 m and the two vertical deflection angles were 7° (down) for the front dipole and 12° (up) for the rear dipole for 12 GeV particles. This combination makes the spectrometer a "two bounce" system for photons and at the same time provides sufficient dispersion for determining the scattered particle momenta. In the 4.5° arm the deflection angles of the dipoles are the same as for the 7.0° arm but their separation is 4 m resulting also in a "two-bounce" system.

Another advantage of the reverse bend configuration is that the detector package is located at approximately the primary beam height. This convenient elevation makes the concrete structure required for shielding the detectors from room background considerably less massive compared to the conventional design with both dipoles bending up. The reduced mechanical complexity translates to significant economic benefits as both the set-up time and apparatus costs are minimized.

In the $7^\circ$ arm, the bend plane position of the scattered particles at the detectors depends weakly on their momenta as shown in Fig. 12. The particle momenta are correlated with the divergence of their trajectory at the exit of the spectrometer. This results in a loss in momentum resolution, not critical to the experiment, but spreads out the pion background, which is highly peaked at 7 GeV/c, onto a large detector area, allowing measurements at a fairly large pion rate.

The purpose of the quadrupole in the $4.5^\circ$ spectrometer was to increase the angular magnification in the non-bend plane and spread the scattered particles onto a larger detector area in this direction as can be seen in Fig. 13. In the bend plane the quadrupole focusing improves the momentum resolution of the system as both the position and divergence of the scattered particles at the exit of the spectrometer are correlated with momentum. The introduction of the quadrupole reduces the highly peaked solid angle in the range of 7 to 10 GeV/c and relaxes the instantaneous counting rates in the detectors allowing accumulation of data in parallel and at about the same rate with the $7^\circ$ spectrometer.

Each spectrometer was instrumented with a pair of gas threshold Čerenkov detectors, a segmented lead-glass calorimeter of 24 radiation lengths in a fly's eye arrangement, six planes of segmented scintillation counters grouped into two hodoscopes (front and rear) and two planes of lucite trigger counters. The electrons were distinguished from the large pion background using the pair of Čerenkov counters in coincidence. The scattered electron energies were measured by two methods. The first used the track information from the scintillator hodoscopes and the known optical properties of the magnetic spectrometer. The second one relied on energy deposition in the lead-glass calorimeter.

The two Čerenkov counters of each spectrometer [59] employed $N_2$ radiator gas with an effective length of 2 and 4 meters, respectively. Two spherical mirrors in each of the two-meter tanks and three mirrors in each of the four-meter tanks collected Čerenkov radiation over the active area of light emission. Each set of mirrors focused the Čerenkov light on one 5" R1584 Hamamatsu photomultiplier per tank. The glass mirrors were manufactured at CERN by slumping a 3 mm thick, 836 mm diameter disk of float glass into a stainless steel mold. The glass was cut to the appropriate dimensions, cleaned and then coated with 80 nm of Al followed by a protective coating of 30 nm of $MgF_2$ which is transparent down to 115 nm[60]. The measurement of the reflectivities for all ten mirrors used in the detectors yielded an average of 80% at 160 nm and 89% at 200 nm. To enhance the electron detection efficiency, each of the photomultiplier UV glass surfaces was coated with 2400nm of P-terphenyl wavelength shifter followed by a protective coating of 25 nm of $MgF_2$[61]. The fluorescence maximum of p-terphenyl of about 370 nm[62] matched well the region of high quantum efficiency of the photomultipliers. Moreover, the short 1-2 ns decay time of this emission enabled us to retain accurate timing information from the Čerenkovs. The 2 m Čerenkov counters operated at a threshold for pions of 9 GeV/c and the 4 m Čerenkov counters at a threshold of 13 GeV/c. The measured number of photoelectrons per incident electron was $\sim$ 7.5, resulting in a detection efficiency of over 99.5%.

The two scintillator hodoscopes[63] provided data for an evaluation of possible systematic errors in the lead-glass and Čerenkov counter data. They were used to identify backgrounds and to measure the pion asymmetry in order to subtract contaminations in the electron sample. The fine hodoscope segmentation ($\sim$185 scintillator elements per spectrometer) was chosen to tolerate the large expected photon and neutron backgrounds and to reconstruct with sufficient resolution the production coordinates of the scattered particles. Both horizontal and vertical planes consist of scintillator elements of 3 cm width with a "2/3" over-lap resulting in a bin width of 1 cm.

The separation of the two hodoscopes was $\sim 6.5$ m in the 4.5° arm and $\sim 4.5$ m in the 7.0° arm. The angular tracking resolution of the hodoscopes was $\pm 0.7$ mrad for the 4.5° spectrometer and $\pm 0.9$ mrad for the 7.0° spectrometer; the position tracking resolution was $\pm 0.3$ cm for both spectrometers. The angular resolutions in the non-bend plane were $\sim \pm 0.5$ mr for both spectrometers, whereas for the bend plane, it was $\sim \pm 0.5$ mr for the 4.5° arm and $\sim \pm 0.3$ mr for the 7° arm.

The momentum resolution depends on the absolute value of momentum and varied from $\pm 0.5\%$ to $\pm 2.5\%$ for the 4.5° spectrometer and from $\pm 0.6\%$ to $\pm 3.5\%$ for the 7.0° spectrometer as can be seen in Fig. 14. The figure also displays the energy resolution of the shower counter.

The initial (at the target) production coordinates $x_\circ$, $\theta_\circ$, $y_\circ$ and $\phi_\circ$, and the momentum of the particles transported through the spectrometers were reconstructed by means of reverse-order TRANSPORT[64] matrix elements using the final (at the rear hodoscope location) coordinates $x_f$, $\theta_f$, $y_f$ and $\phi_f$ of the detected particles. The very large momentum bites of the spectrometers required using a fourth-order reverse TRANSPORT expansion in $y_f$ and $\phi_f$ for reconstructing the particle momenta.

The shower counter calorimeter for each spectrometer was assembled from a selected subset of 200 (20 rows of 10 blocks) lead glass bars from a previous experiment [65]. Each bar consisted of Schott type F2 (refractive index of 1.58) lead glass with dimension $6.2 \times 6.2 \times 75$ cm$^3$ providing for 24 radiation lengths along the direction of the detected electrons. The blocks were arranged in a fly's eye configuration, stacked upon each other with a segmentation that allowed for an accumulation of data at a maximum $\pi/e$ ratio of about 20. With the two Čerenkov counters in the trigger, the contamination of the shower signals by pions was small (on the order of a few percent).

The shower counter resolution for electrons was measured in a test beam at CERN to be [66]

$$\sigma/E' \approx \pm(2.5 + 6.5/\sqrt{E'})\%. \tag{35}$$

The counters were calibrated with a sample of scattered elec-trons of 5 GeV energy in a special elastic electron-proton scattering run using a gaseous hydrogen target. Extrapolation of the calibration algorithm to higher energies was performed using the scintillator hodoscopes and the known optical properties of the spectrometers. The detailed study of the performance of the detector is described elsewhere [66].

The spectrometer set-up and detector packages proved to be robust. All Čerenkov counters ran with an acceptable average photoelectron yield, typically greater than six photoelectrons. The simple hodoscope tracking system was able to reconstruct tracks with an efficiency of greater than 80%. Subsequent sections describe in some detail the analysis and spectrometer performance.

## 3.5 The Electron Trigger

The main electron trigger [67, 68] for each spectrometer consisted of a triple coincidence between the two Čerenkovs and the sum of the shower counter signals. This trigger was 96% efficient for electron events and had a contamination of 13% of non-electrons events. Secondary triggers, prescaled to reduce the rates, consisted of various combinations of the detector elements designed to measure efficiencies and pion backgrounds.

Up to four triggers were allowed per spectrometer per beam spill. There was a 30 ns deadtime after each trigger. Each trigger gated a separate set of 205 ADC's (Lecroy 2280 system) for the shower counter. Each shower counter signal was sent to four ADC's corresponding to the four possible triggers, making a total of over 800 ADC channels per spectrometer. The hodoscope signals along with the selected elements of the trigger went to multihit TDC's (LRS 2277 system) which had a 20 ns deadtime. To reduce the load on the data aquisition, the signals to the hodoscope TDC's had a 100 ns gate provided by the trigger system. Thus each electron candidate had a 100 ns window of activity in the detector. The deadtime correction to the asymmetry was determined from a Monte Carlo model of the trigger and instantaneous beam intensity [59, 67, 68]. The average correction was about 10% in the 4.5° spectrometer and less than

4% in the 7° spectrometer.

# 4 Analysis

Data analysis was directed at extracting the electron scattering asymmetries and structure functions with a high rate spectrometer. The highest rates occurred in the 4.5° degree spectrometer arm, ranging from 1.5 to 2.5 electron triggers per pulse on average. The rate was typically less than 1 electron trigger per pulse in the 7° arm. Since the SLAC linac delivers pulses at a rate of 120 Hz, the experiment recorded a large sample of deep inelastic scattering electron events. In total, approximately 350 million electron events were collected of which 300 million were used to determine the asymmetries and spin structure functions.

The analysis focused on identifying electrons and determining their momentum in a high rate environment. Charged pions were the main source of background. The electron trigger consisted of a triple coincidence in the signals coming from the two Čerenkov counters and summed shower counter per spectrometer arm. This method served to reject the majority of charged pions, which typically enter the spectrometer out of time compared to the Čerenkov signal.

The remaining pion background originated from either very high momentum pions (typically greater than 13 GeV) or from pions that enter the spectrometer in time with an electron. Backgrounds from pion contamination within the trigger were studied by comparing the energy and momentum determination of the event from the measurements in the shower counter and hodoscope, respectively. Overall, the contamination from high energy pion events was less than 1%, since the cross-section for this process is low; whereas, backgrounds from events with an electron and pion in coincidence was also relatively small, since the electron energy cluster would typically deposit more energy than the pions, and only the highest energy cluster per trigger was kept for the analysis. Backgrounds from neutral particles were minimal, since the spectrometer was designed to accept only charged particles.

Electron events were selected if they triggered both Čerenkov counters with an ADC signal greater than one and a half photoelectrons, and deposited a minimum energy (typically greater than 5.5 GeV) as a cluster of 3 x 3 blocks in the segmented shower counter. An algorithm based on artificial intelligence techniques (cellular automaton) was developed to cluster hit blocks belonging to the same shower [69]. Events with a shower contained entirely in one block were rejected. This single-block event cut served to reject a large fraction of pion events as determined both from a GEANT simulation[70] and from studying results from the energy versus momentum comparison. A trigger from the Čerenkov counters opened a 100 nsec gate (during the 1 $\mu$sec spill) and the pulse heights from the lead glass blocks and from the phototubes of the Čerenkovs were recorded using zero-suppressing LeCroy 2282 ADCs. Only clusters with the maximum energy in the 100 nsec gate were accepted in the analysis. The ability to reject pions is studied by comparing the energy and momentum determinations of an event. Typically, pion events registered a higher momentum than energy, since the pion shower energy is usually not entirely visible in the electromagnetic calorimeter.

For extracting the asymmetries, once an electron event was selected, the shower counter was used both to identify its energy and to determine the scattering angle $\theta$ from the centroid position of the shower in the calorimeter. The position determination of the shower centroid was accurate to approximately $\pm$ 10 mm, significantly better than needed for sufficient $x$ resolution. The hodoscope tracking system was developed to calibrate the shower counter, perform systematic studies of the backgrounds and to monitor the shower counter and Čerenkov efficiencies throughout the experiment. Typically, hodoscope tracking efficiencies varied from 80 to 95% depending on the trigger rate. Noise sources from random hits due to photon background with the associated electronic deadtime were the main cause of hodoscope inefficiency. For this reason, the shower counter was the primary detector used for the analysis.

Calibration of the shower counter energy was performed using two methods. First, knowledge of the magnetic field and spatial positions of the spectrometer magnets allowed for the determination of the particle's momenta via tracking. Tracking with pristine events was used to calibrate the shower counter. The primary uncertainty in this method comes from the finite width of the hodoscope fingers and the knowledge of the position of these fingers relative to the spectrometer magnets. The uncertainty in the momentum measurement is energy dependent and estimated to be below 2.5% over the range of energies detected (Fig. 14). The second method employed a special test run performed ahead of the experiment and used a 5 GeV electron beam scattering off a hydrogen target ($H_2$) to observe the elastic peak. The location of the peak in both the hodoscope and the shower counter checked the absolute energy scale to $\pm 3\%$ as determined from magnetic measurements.

In order to investigate possible inefficiencies in the detector package, some special low rate runs were performed to measure the total cross section. Although the spectrometer was not designed for cross section measurements, such a check was useful for systematic studies. Extraction of the total cross section requires detailed knowledge of the spectrometer acceptance, central momentum deadtime and target thickness. Fig. 15 present a comparison of the results on the total cross section for the 4.5° and 7° spectrometer, respectively. Systematic errors on the measurement are large (typically 15%). The data are especially sensitive to the knowledge of the radiative corrections which can change the shape significantly. Within systematic uncertainties, there is reasonable agreement between the data and the Monte Carlo determination, which uses cross section measurements from previous SLAC experiments[71].

After event selection, contamination of the electron event samples was relatively small. Fig. 16 presents distributions of events comparing the energy measured in the shower counters, $E'$, compared to the momentum, $p$, measured by tracking using the hodoscopes. Low-energy tails ($E'/p < 0.8$) come largely from pion contamination, whereas high-energy tails ($E'/p > 1.2$) come from overlapping events with typically either two electron interactions or an electron and pion interactions. A neural network analysis was developed to study pion rejection using only the calorimeter information [70], but was not used in the final asymmetry analysis. The pion contamination of the electron event sample at low energy (E' $\approx$ 7 GeV )was found to be approximately 3% and decreasing at higher energies to less than 1%.

Corrections to the electron asymmetries from pion contamination are performed assuming zero asymmetry coming from the pions with an uncertainty of $\pm 0.15$. A special study using out of time pion events within the trigger gate revealed no evidence for a significant pion asymmetry as shown in Fig. 17. The final effect due to pion contamination is small.

An additional source of contamination to the deep inelastic scattering electron event sample arises from hadron decays producing secondary electrons. For example, if a neutral pion is produced in the final state of an interaction, it will decay into two photons which themselves can scatter and produce an electron which enters the spectrometer and simulates a true deep inelastic scattering event. Contamination due to this process is measured by reversing the polarity of the spectrometer magnets and collecting dedicated data on the positron rates. The measurement serves as a valid subtraction as long as the hadronic decay process is charge symmetric. We found that approximately 5% of the low $x$ events were contaminated from such a process. The effect decreases rapidly as $x$ increases. The behavior is similar to the pion contamination, with a larger contamination at low E$'$ and dying off quickly at higher E$'$. The asymmetry in the positron rates is measured to be zero with large uncertainties ($\approx \pm$ 30%). The largest systematic uncertainty in the lowest $x$ bin comes from this effect (see Tables 11, 12).

For the asymmetry analysis, electron events which passed the event selection cuts were divided into bins of scattered energy $E'$, scattering angle $\theta$, and relative target and beam helicities, $N^{\uparrow\downarrow(\uparrow\uparrow)}$ $(E',\theta)$. From these counts (normalized by incident charge $N_e$), raw asymmetries are formed:

$$A_{\parallel}^{raw} = \frac{(N/N_e)^{\uparrow\downarrow} - (N/N_e)^{\uparrow\uparrow}}{(N/N_e)^{\uparrow\downarrow} + (N/N_e)^{\uparrow\uparrow}} \qquad (36)$$

and for data collected with transverse target polarization,

$$A_{\perp}^{raw} = \frac{(N/N_e)^{\uparrow\rightarrow} - (N/N_e)^{\downarrow\rightarrow}}{(N/N_e)^{\uparrow\rightarrow} + (N/N_e)^{\downarrow\rightarrow}}. \quad (37)$$

On these measured raw asymmetries, corrections for the beam polarization $P_b$, target polarization $P_t$, dilution factor $f_{He}$, electronic dead time $\Delta_{dt}$, radiative corrections $\Delta_{RC}$ and kinematics are performed to extract the $^3$He parallel and transverse asymmetries :

$$\begin{aligned} A_{\|} &= \frac{(A_{\|}^{raw} + \Delta_{dt})}{P_b P_t f_{He}} + \Delta_{RC}^{\|}, \\ A_{\perp} &= \frac{(A_{\perp}^{raw} + \Delta_{dt})}{P_b P_t f_{He}} + \Delta_{RC}^{\perp}. \end{aligned} \quad (38)$$

From these asymmetries, the virtual photon-$^3$He asymmetries $A_1^{^3He}(x, Q^2)$ and $A_2^{^3He}(x, Q^2)$ are found as a function of $x$ and $Q^2$:

$$\begin{aligned} A_1 &= \frac{1}{(1+\eta\zeta)}[A_{\|}/D - (\eta/d)A_{\perp}] = \frac{A_{\|}}{D} - \eta A_2, \\ A_2 &= \frac{1}{(1+\eta\zeta)}[A_{\perp}/d + (\zeta/D)A_{\|}]. \end{aligned} \quad (39)$$

Furthermore, the spin structure functions $g_1^{^3He}(x, Q^2)$ and $g_2^{^3He}(x, Q^2)$ are extracted using the asymmetries given above,

$$\begin{aligned} g_1 &= \frac{F_2}{2x(1+R)}[A_1 + \gamma A_2], \\ g_2 &= \tfrac{F_2}{2x(1+R)}[A_2 \tfrac{1}{\gamma} - A_1] \\ &= F_2 \tfrac{1+\gamma^2}{2xD'(1+R)} \cdot \tfrac{y}{2\sin\theta}[\tfrac{E+E'\cos\theta}{E'} A_{\perp} - \sin\theta A_{\|}]. \end{aligned} \quad (40)$$

Here $R(x, Q^2) = \sigma_L/\sigma_T$ is the ratio of the longitudinal to transverse cross sections and $F_2(x, Q^2)$ is the unpolarized deep inelastic structure function. Both are functions of x and $Q^2$. The other kinematic variables are related to the incoming and outgoing scattered electron energy ($E$ and $E'$, respectively) via

$$\nu = E - E',$$
$$y = \nu/E, \tag{41}$$
$$D = \frac{E - E'\epsilon}{E(1+\epsilon R)},$$
$$D' = \frac{(1-\epsilon)(2-y)}{y[1+\epsilon R]}, \tag{42}$$

and

$$d = D\sqrt{\frac{2\epsilon}{1+\epsilon}}. \tag{43}$$

The factors $\eta$, $\zeta$ and $\gamma$ are found via

$$\eta = \frac{\epsilon\sqrt{Q^2}}{E - E'\epsilon}, \tag{44}$$
$$\zeta = \eta(\frac{1+\epsilon}{2\epsilon}), \tag{45}$$
$$\gamma = \frac{\sqrt{Q^2}}{\nu}, \tag{46}$$

where $\epsilon$ characterizes the virtual photon polarization,

$$\epsilon = \frac{1}{1 + 2(1 + \nu^2/Q^2)\tan^2(\theta/2)}. \tag{47}$$

The above relations (Eqs. 39-47) are valid for scattering off any spin 1/2 object and therefore are used for a free nucleon as well as $^3$He.

## 5 Corrections and Systematic Uncertainties

Although statistical errors typically dominate any particular $x$ and $Q^2$ bin, for evaluating sum rules the systematic uncertainties play an important role. The most important of these are the beam polarization $P_b$, the target polarization $P_t$, and the dilution factor $f_{He}$. The relationship between the raw asymmetry and the extracted physics asymmetries are given in Eq. 38. Any uncertainty in $P_t$, $P_b$ or $f$ will affect all the asymmetries together over the entire kinematic range. Similarly, this will translate into a comparable uncertainty over the integrals of $g_1^n$ and therefore the sum rule. The systematic uncertainties associated with $P_b$ and $P_t$ have been discussed in the beam and target sections (3.2 and 3.3) respectively. Corrections due to hadronic contaminations have been discussed in the previous section. Here we discuss the dilution correction, the radiative corrections and their associated systematic uncertainties.

## 5.1 Dilution Factor

The largest systematic uncertainty in the experiment (besides the low $x$ extrapolation) came from the determination of the dilution factor $f_{He}$. This factor corresponds to the fraction of events originating from $^3$He scattering versus scattering from the rest of the target and is measurable.

The polarized $^3$He target consisted of a mixture of $^3$He gas, $N_2$ gas and glass windows. The target cell contained approximately 9 atmospheres of $^3$He, $\sim$ 65 Torr of $N_2$ and glass windows with a thickness of approximately 110 microns each. The relative proportion of electron events originating from the $^3$He versus the $N_2$ versus the glass was in the ratio of approximately 10 to 1 to 20. The $^3$He nucleus itself consists of primarily a polarized neutron plus two unpolarized protons. Further dilution is accounted for to extract the polarized neutron result from $^3$He as described in section 6.2.

In order to determine the dilution factor $f_{He}$ for the three targets used in the experiment, we relied on two independent techniques. First, we measured the amount of material in the target and calculate $f_{He}$ using known cross sections. Table 4 presents a breakdown of the material in the three cells used in the experiment. The dilution factor is dependent on $x$ and $Q^2$ and takes the form,

$$f_{He}(x,Q^2) = \frac{n_{He}\sigma_{He}(x,Q^2)}{n_{He}\sigma_{He}(x,Q^2) + n_{N_2}\sigma_{N_2}(x,Q^2) + n_g\sigma_g(x,Q^2)} \tag{48}$$

where $n_i$ is the total number of nucleons found in species $i$ (He, $N_2$ or $g$ for glass) and $\sigma_i$ is the average experimental cross section

per nucleon expressed as

$$\sigma_i(x,Q^2) = \frac{P_i^{ND}(x)}{A_i}[Z_i\sigma_p(x,Q^2) + (A_i - Z_i)\sigma_n(x,Q^2)] \quad (49)$$

where $P_i^{ND}$ accounts for the nuclear dependence correction (EMC effect) using the parameterization given by Gomez et al. [72] and $A_i$ is the atomic mass number of species $i$. The target cell glass (Corning 1720) has the composition 57% $SiO_2$, 20.5% $Al_2$, 12% MgO, 5.5% CaO, 4% $B_2O_3$ and 1% $Na_2$, yielding within 1% the same number of protons and neutrons. Assuming that the ratio $R = \sigma_L/\sigma_T$ is the same for the proton and the neutron[71], we can express the dilution factor as:

$$f_{He}(x,Q^2) = \left[1 + \frac{3(1+R^{np})}{2(2+R^{np})}\left(\frac{P_g^{ND}}{P_{He}^{ND}}\frac{RC_g}{RC_{He}}\frac{n_g}{n_{He}} + \frac{P_{N_2}^{ND}}{P_{He}^{ND}}\frac{RC_{N_2}}{RC_{He}}\frac{n_{N_2}}{n_{He}}\right)\right]^{-1} \quad (50)$$

where $RC(x,Q^2)$ is a radiative correction factor which relates the Born cross section to the experimental cross section for different species. The quantity $R^{np} = F_2^n/F_2^p$ is the ratio of unpolarized structure functions for neutrons and protons. We determined the neutron structure function using $F_2^n = 2F_2^D - F_2^p$ where the proton and deuteron structure functions per nucleon are taken from the NMC fits [75] to the SLAC [74], BCDMS [73] and NMC data. Because no uncertainties are included in the NMC fit, we used the relative point-to-point uncertainties to the SLAC data [71] which is at similar kinematics as the present experiment. We also included a normalization uncertainty from the SLAC data of 2.1% for $F_2^p$ and 1.7% for $F_2^D$. Furthermore, we included a 2% uncertainty for the proton and 0.6% for the deuteron arising from the maximum deviation of the SLAC and NMC fits in the range $0.08 < x < 0.6$. For $x < 0.08$ where there is very little SLAC data, a 5% error is placed on the NMC structure functions defining the maximum deviation between the NMC data and the NMC fit. For $R = \sigma_L/\sigma_T$ we used the central values and errors given by a SLAC global analysis [71].

For this section and throughout this paper the spin-independent $^3$He structure function in the deep inelastic region is evaluated as follows:

$$F_2^{^3He}(x,Q^2) = P^{ND}(x)[2F_2^D(x,Q^2) + F_2^p(x,Q^2)] \quad (51)$$

where $P^{ND}(x)$ is an estimate of the nuclear dependence effect in $^3$He from reference [72] and differs from unity by less than 2% in our kinematic range. We have assigned an additional 1% uncertainty to $F_2^{^3He}$ due to the nuclear dependence effect. The overall systematic uncertainty from this method is dominated by the knowledge of the thickness of the glass windows. The windows of the target cells were measured using a precision tooling gauge with an accuracy of 7%. Uncertainties in the measurement due to variations of the glass thicknesses are included in estimating this uncertainty. Other contributions to the uncertainty from the nuclear dependence effect and $F_2$ are negligible.

A limitation of the method described above is that events originating from beam halo interactions with the 30 cm long side walls of the 1 cm radius target cell are not taken into account. The electron beam was centered on the target cell. Primary electrons from the beam passing 1 cm from the center could interact with the target glass walls producing additional scattered electrons.

During the experiment, several dedicated runs were performed in which the beam was steered away from the target center. Fig. 18 presents the average event rate per pulse in the spectrometer as a function of the central beam position. An increase in the event rate is evident as the beam is moved more than 3 mm from its nominal position. The variation in event rate is well-described by a quadratic function and is attributed to the beam passing through an increasingly thick part of the target endcaps. During the data taking the target was positioned at the event rate minimum. From such studies, we concluded that as long as the beam position was stable at the center of the target, the beam halo effect on the dilution factor should be small, except for the possibility of flat tails. If the beam halo has long, flat tails, then if the beam is moved off center the event rate may not change, but a constant background of unpolarized events may be present from interactions with the glass side walls.

A second independent method for determining the dilution factor was performed. Periodically throughout the experiment, data were collected in which a reference cell was placed in the electron beam. The cell consisted of the same glass type and had the same dimensions as the polarized target cells. The cell was filled with $^3$He gas at various controlled pressures.

At zero pressure the events are due to the glass endcaps, but as the cell is filled to different pressures the event rate increases. The event rate $r(x, Q^2)$ normalized to incident charge is expressed as a function of pressure as follows:

$$r(x, Q^2) = C[n_g \sigma_g(x, Q^2) + 3\frac{\mathcal{N} L P_r}{RT} \sigma_{He}(x, Q^2)], \tag{52}$$

where $C$ is a proportionality constant, $R$ the $^3$He gas constant, $T$ the temperature of the gas, $\mathcal{N}$ is Avogadro's number, $L$ the length and $P_r$ the pressure of the reference cell, respectively. When the pressure in the reference cell $P_r$ equals that of the target cell $P_{TC}$, the number density $n_{He} = 3\mathcal{N} L P_r / RT$ matches that of Eq. 48. The rates were corrected for rate-dependent electron detection efficiencies and changes in the external unpolarized radiative corrections as a function of helium pressure.

Fig. 19 presents an example of a sequence of the reference cell runs where the slope $\alpha$ can be interpreted as $n_{He}\sigma_{He}(x, Q^2)/P_{TC}$ and intercept $\beta$ as $C n_g \sigma_g(x, Q^2)$. The dilution factor is extracted as:

$$f_{^3He}(x, Q^2) = \frac{\alpha P_{TC}}{(\alpha P_{TC} + \beta)}. \tag{53}$$

Since the $^3$He pressure was directly measured and the reference cell glass window thicknesses known to better than 7%, this measurement could be compared directly to the first method described above (Fig. 20). The direct measurement of the dilution factor from these special runs would naturally take into account any possible beam halo effects. From these studies, we conclude that there was no observation of any large beam halo effects on the dilution factor. The final systematic uncertainty on the dilution factor is taken to be 8%, where the dominant uncertainty comes from the knowledge of the window thicknesses (7%), needed for the first method described above.

## 5.2 Radiative Corrections

Due to the real and virtual radiation of electrons during the scattering process, the longitudinal and transverse measured asym-metries ($A_{||}, A_{\perp}$) need further corrections known as the electromagnetic radiative corrections. The latter are performed to extract the structure functions $g_{1,2}(x, Q^2)$ and photon-nucleon asymmetries $A_{1,2}(x, Q^2)$ as defined in the Born approximation where the scattering process is described by the exchange of a single virtual photon. These corrections are cast into two categories: internal and external. The internal effects are those occurring at the nucleus responsible for the deep inelastic scattering under investigation and therefore need to be performed even for an infinitely thin target. The external effects are those which modify the energy of the incident and scattered electron via bremsstrahlung and ionization losses from interactions with other atoms before and after the deep inelastic process has occurred. The external corrections depend on target thickness. While the formalism for the spin-independent deep inelastic scattering was developed by Mo and Tsai[76, 77], that of the spin-dependent formalism was developed by Kuchto, Shumeiko and Akushevich[78, 79] and implemented in their code POLRAD. The internally radiated helicity-dependent deep inelastic scattering cross section can be decomposed into its components following reference [79]:

$$\begin{aligned} \sigma^r_{\pm}(x,y) &= \sigma^B_{\pm}(x,y)[1 + \frac{\alpha}{\pi}(\delta^{IR}_R + \delta_{vert} + \delta^l_{vac} + \delta^h_{vac})] \\ &+ \sigma^F_{\pm,in}(x,y) + \sigma^{qel}_{\pm}(x,y) + \sigma^{el}_{\pm}(x,y) \end{aligned} \tag{54}$$

where $\sigma^r_{\pm}$ is the internally radiated helicity-dependent differential cross section $(d^2\sigma/dxdy)_{\pm}$ and $(\pm)$ refers to the helicity of the electron relative to that of the target, and $\sigma^B_{\pm}$ is the helicity-dependent Born cross section of interest. The quantities $\delta_{vert}$, $\delta^{IR}_R$, $\delta^l_{vac}$ and $\delta^h_{vac}$ are the electron vertex contribution, the soft photon emission, the electron vacuum polarization and hadronic vacuum polarization contributions, respectively. The quantity $\sigma^F_{\pm,in}$ is the infrared divergence-free part of the inelastic radiative tail, and $\sigma^{qel}_{\pm}$ and $\sigma^{el}_{\pm}$ are the quasielastic and elastic radiative tail contributions, respectively.

The internal corrections were calculated using the program POLRAD version 14 which uses the new iterative method[79]. In this method, the best fit to the experimental asymmetry $A_1^{^3He}(x)$ is used to build the polarized structure functions $g_1^{^3He}(x)$. The cross sections for specific states of polarizations are then constructed and used to evaluate all the contributions of Eq. 54. Here all quantities refer to $^3He$. From this result a new $A_1$ is produced and used as an input to the next iteration step by constructing a new model for $g_1^{^3He}$

$$g_1^{(k)} = \frac{F_2(x, Q^2)}{2x[1 + R(x, Q^2)]}[A_1^{measured} + \Delta A_1(g_1^{(k-1)})], \qquad (55)$$

where $k$ is the iteration index.

The process is then repeated until convergence is reached, which occurs within three to four iterations. POLRAD was first checked against a program we developed based on the work by Kuchto and Shumeiko[78] and also against the Tsai[77] formalism for the unpolarized case. Similar results to POLRAD were found with both checks as expected.

The nuclear coherent elastic tail is evaluated using different best fits to the elastic form factors of $^3$He and found to be small. This leaves only three physical regions of significant contribution to the total internal radiative correction to be considered: the quasielastic region which starts a few MeV after the elastic peak (since no nuclear excited states are bound in $^3$He); the resonance region which partially overlaps the quasielastic tail; and finally the deep inelastic region which we have assumed to start at $W^2$ = 4 (GeV/$c$)$^2$. In Eq. 54 resonance and inelastic contributions are both included in $\sigma_{in}^F$. The internal and external radiative corrections require the knowledge of the spin independent structure functions $F_1^{He}(x, Q^2)$ and $F_2^{He}(Q^2, \nu)$ and spin dependent structure functions $g_1^{He}(x, Q^2)$ and $g_2^{He}(x, Q^2)$ over the canonical triangle region[76, 77]. The lowest $x$ bin in this measurement ($x = 0.035$) determines the largest kinematic range of $Q^2$ and $x$ over which the structure functions have to be known. It extends in the range $0.31 \leq Q^2 \leq 18.4$ (GeV/$c$)$^2$ and $0.03 \leq x \leq 1$. The variables of integration which define the canonical triangle are given by $M_x$ and $t$, in the range $M_n + m_\pi \leq M_x \leq W$ and $t_{min} \leq t \leq t_{max}$; $t \equiv Q^2$. Here $M_x$ is the invariant mass of all possible contributing scattering and W the invariant mass of the scattering of interest.

The $^3He$ spin-independent structure functions used in the quasielastic region were those of de Forest and Walecka [80]. These structure functions allow for a convenient parameterization in the evaluation of the unpolarized radiative tail. In the resonance region we chose the spin independent structure functions obtained by fitting the data in the resonance region given in reference [81], while for the deep inelastic region we used the same models for the proton and deuteron structure functions as described in subsection 5.1 to build the spin-independent structure functions of $^3$He [see Eq. 51].

The spin dependent structure functions used in the resonance region were obtained from the AO program[82] which is based on an analysis of electromagnetic transition amplitudes in that region. In the deep inelastic region, as described previously, a fit to the extracted $A_1^{^3He}$ from this experiment was used to build the first spin-dependent structure functions input to the iterative method.

The external corrections were performed by extending the procedure developed by Mo and Tsai[76, 77] for the unpolarized scattering cross sections to that of the helicity dependent scattering cross sections. It used an iterative unfolding procedure on the internal and external corrections together until convergence is reached. The procedure requires the knowledge of the internally-radiated Born helicity-dependent cross sections. The measured cross section $\sigma^m_\pm$ is expressed as the convolution of the internally radiated Born cross section with the radiation effect due to the finite thickness of the target:

$$\sigma^m_\pm(E_s, E_p) = \int_{E_s^{min}}^{E_s} dE_s' \int_{E_p}^{E_p^{max}} dE_p' I(E_s, E_s', t_{in}) \sigma^r_\pm(E_s', E_p') I(E_p', E_p, t_{out}) \quad (56)$$

where $E_s, E_p$ are the incident and detected electron energies, and $I(E_{in}, E_{out}, t)$ is the probability that an electron of energy $E_{in}$ will have an energy $E_{out}$ due to bremsstrahlung emission after having passed through $t$ radiation lengths (r.l.) of material. For the polarized case, this probability function is spin-independent for a thin target because it is dominated by forward (charge) scattering off target atoms. All kinematics parameters and the function $I$ are well-described and discussed in [77]. The entrance glass window plus half of the $^3$He thickness accounts for $t_{in} = 0.00125$ r.l. For $t_{out}$, the electrons exit the target through four discrete sections in which the amount of material the electrons traverses after scattering is different (see Table 5). These four contributions to $t_{out}$ from each region are summed. Note that $t_{out}$ is much larger than $t_{in}$, especially in the region where the scattered electrons passed through the NMR pick-up coils.

The procedure is applied for each helicity case separately (similar to the unpolarized case) and then the asymmetry $A_1^{^3He}$ is formed from the result of each helicity case:

$$A_{1Born}^{^3He} = \frac{1}{D}\left(\frac{\sigma_{\downarrow\uparrow}^{B} - \sigma_{\uparrow\uparrow}^{B}}{\sigma_{\downarrow\uparrow}^{B} + \sigma_{\uparrow\uparrow}^{B}}\right)$$
$$\equiv \frac{1}{D}\left(\frac{\sigma_{\downarrow,\uparrow}^{m} - \sigma_{\uparrow,\uparrow}^{m}}{\sigma_{\downarrow,\uparrow}^{m} + \sigma_{\uparrow,\uparrow}^{m}}\right) + \Delta_{RC}^{ext} + \Delta_{RC}^{int}. \quad (57)$$

The statistical uncertainty of the radiative corrections at each measured kinematics point follows from the statistical uncertainty of the measured rate at that point and the assumption of exact knowledge of the radiative background. Using Eq. 54 we can generalize this expression to the full radiative corrections where internal and external radiative effects are convoluted. In order to evaluate the total statistical error for each corrected kinematics point during the correction procedure, a table of fractions $f_i$ of absolute "background" contributions to the total cross sections due to the radiative elastic, quasielastic and deep inelastic tails was stored. These fractions include photon tails from the internal corrections as well as those of external contributions. However, they do not include vertex and vacuum polarization terms which are considered as non-physical background contributions.

The systematic errors in the radiative corrections are estimated by changing the input model for the asymmetries and unpolarized cross sections in the unmeasured kinematics defined by the "canonical" triangle in POLRAD. The unmeasured region, for example, includes the resonance region at low $Q^2$. In order to estimate the sensitivity of the corrections to these models, we tested our results assuming a flat asymmetry input for our raw $A_1^{^3He}$ data and compare the results to a quadratic fit input to the same data with weak constraints at the low and high $x$ regions. From the variation of the results, we estimate a 25% relative uncertainty in the internal and external radiative corrections for the 7° spectrometer data and 25% for the internal corrections to the 4.5° spectrometer data. The external corrections to the 4.5° were particularly sensitive to the cross section shape at high $x$. This correction is assigned a 35% relative uncertainty.

In Table 6 we present the radiative corrections to the data as well as the fractions necessary to evaluate the changes in statistical uncertainty on the results, while Fig. 21 shows the effect of the internal and external electromagnetic radiative corrections on the measured asymmetry $A_1^{^3He}$ needed to obtain the Born asymmetry. One sees that the corrections are quite small (typically shifting the data by 1/3 the size of the statistical error).

# 6 Experimental Results

## 6.1 Measured $^3$He Results

Results on $A_1^{^3He}$, $A_2^{^3He}$, $g_1^{^3He}$ and $g_2^{^3He}$ versus $x$ are given in Table 7 and 8. The asymmetry $A_2^{^3He}$ is determined using Eq. 39. Within our limited statistical precison, the values of $A_2^{^3He}$ are consistent with zero. Experiment E143 [83] measured $A_2^d$ and $A_2^p$ and found that the difference $A_2^n$ is consistent with zero with small uncertainties. For the rest of the analysis, we extract $A_1^{^3He}$ and $g_1^{^3He}$ using Eq. 39 and Eq. 40 with $A_2^{^3He}$ set to be identically zero with a systematic uncertainty equivalent to our measured statistical uncertainties shown in Table 8 or the positivity constraint $A_2^n < \sqrt{R}$, whichever is smaller. The uncertainty provided by the measurement of $A_\perp$ was smaller than the $\sqrt{R}$ limit except for the results in the two x bins, x=0.35 and x=0.47. Within our precision there is no obvious $Q^2$ dependence of $A_1^{^3He}$ at fixed x. The $Q^2$ averaged $A_1^{^3He}$ are given in Table 7. From the asymmetry results of $A_1^{^3He}$, the spin structure function $g_1^{^3He}$ is obtained assuming $A_2^{^3He}$=0, namely

$$g_1^{^3He}(x,Q^2) = \frac{F_2^{^3He}(x,Q^2)}{2x(1+R(x,Q^2))} A_1^{^3He}(x,Q^2). \tag{58}$$

## 6.2 Extracting the Neutron Result from $^3$He

A polarized $^3$He target can be used to extract information on polarized neutrons. The main reason is that in the naive approximation the $^3$He nucleus is considered to be a system of three nucleons in a spatially symmetric S-state. The Pauli principle constrains the overall wavefunction to be antisymmetric, and therefore the spin-isospin wavefunction must then be antisymmetric. Exchanging the two protons must yield a symmetric wavefunction, implying that the two protons are paired antisymmetrically in a spin singlet state. In this picture, the two proton spins line up anti-parallel to one another, resulting in a cancellation of spin-dependent effects coming from the protons. Naturally, the $^3$He nucleus is not exactly a system of nucleons in a spatially pure S-state, and corrections due to the other states must be implemented in order to extract the result for a pure neutron. Fairly extensive work on the $^3$He wavefunctions has been performed and these wavefunctions are used to estimate magnetic moments and to extract the degree of polarization of the neutron in $^3$He[84]. Furthermore, the determination of the neutron spin structure function from a measurement on $^3$He relies on an understanding of the reaction mechanism for the virtual photon absorption combined with the use of a realistic $^3$He wavefunction. Detailed investigations of the $^3$He inelastic spin response functions versus that of a free neutron have been carried out by three groups [85, 86, 87]. They examined the effect of the Fermi motion of nucleons and their binding in $^3$He along with the study of the electromagnetic vertex using the most realistic $^3$He wavefunction. Consistent findings have been reached among these groups, and we summarize here those relevant to our experiment.

In the deep inelastic region a neutron spin structure response and asymmetry can be extracted from that of $^3$He using a procedure in which S, S$'$ and D states of the $^3$He wavefunction are included, but no Fermi motion or binding effects are introduced:

$$g^n_{1,2} = \frac{1}{\rho_n}(g^{^3He}_{1,2} - 2\rho_p g^p_{1,2}) \tag{59}$$

$$A^n_{1,2} = \frac{F_2^{^3He}}{F_2^n}\frac{1}{\rho_n}(A_{1,2}^{^3He} - 2\frac{F_2^p}{F_2^{^3He}}\rho_p A^p_{1,2}), \tag{60}$$

where $g^n_{1,2}$, $g^p_{1,2}$ and $g_{1,2}^{^3He}$ are the spin structure functions of an effective free neutron, a free proton and $^3$He, respectively. Similarly $A^n_{1,2}$, $A^p_{1,2}$ and $A_{1,2}^{^3He}$ are the photon-target asymmetries for an effective free neutron, a free proton and $^3$He, respectively. The studies yield $\rho_n = (87 \pm 2)\%$ and $\rho_p = (-2.7 \pm 0.4)\%$ for the polarizations of the neutron and proton in $^3$He due to the S, S' and D states of the wavefunction [84, 85]. The calculations using the "exact" $^3$He wave function including the full treatment of Fermi motion and binding effects show negligible differences with the above approximation in the deep inelastic region. A precise proton measurement is important to minimize the error on the correction. We point out that our measurements have a lower limit in missing mass W$^2$ of 4 GeV$^2$, already beyond the quasielastic and resonance region which were found to be more sensitive to nuclear effects [87].

In this analysis we have used Eqs. 59 and 60 to extract the neutron asymmetry $A^n_1$ and spin dependent structure function $g^n_1$, where the proton asymmetry $A^p_1$ and spin-dependent structure function $g^p_1$ used are those measured in experiment E143 [88]. The uncertainties in the measured proton results are taken into account in extracting the results on $g^n_1$. The relative impact on the overall error bars is small for all $x$ bins.

No further corrections due to possible final-state effects have been incorporated. Nevertheless, placing limits on possible contaminations from final state nuclear effects has been one of the significant motivations for measuring the neutron spin structure function with different nuclear targets (i.e. polarized deuterium and $^3$He).

Results on $A^n_1$, $g^n_1$, $A^n_2$ and $g^n_2$ are presented in Tables 9 and 10. In Table 9 the values of $g^n_1$ are also given at constant $Q^2_0 = 2.0$ (GeV/$c$)$^2$. Table 11 and Table 12 present the detailed contribution of every correction parameter to the overall systematic uncertainty on $A^n_1$ and $g^n_1$ at each $x$ point. In order to extract the values of $g^n_1$ at one unique value of $Q^2$, we assumed that the asymmetry $A^n_1$ is $Q^2$-independent and used $A_2 = 0$, consistent with the study of the $Q^2$ dependence of our data. From the two spectrometers and the three beam energies used in this experiment, we extracted $A_1^n$ at six different values of $Q^2$. Over this modest range of $Q^2$ and within the statistical errors, we find that $A_1^n$ is consistent with being independent of $Q^2$ as seen in Fig. 22 and enumerated in Table 13. This trend is confirmed by the recent precision E143 results [90] on the proton and neutron in the equivalent $Q^2$ range. Fig. 23 and Fig. 24 show the compiled results on $A_1^n(x)$ and $xg_1^n(x)$. Fig. 25 presents the results for $A_2^n(x)$.

# 7 Neutron First Moment and Physics Implications

Integrating $g_1^n(x, Q_0^2)$ over the measured range of $x$ at a fixed value of $Q_0^2 = 2$ (GeV/$c$)$^2$, one obtains

$$\begin{aligned} \int_{0.03}^{0.6} g_1^n(x, Q_0^2) dx &= \sum_{i=1}^{8} g_1^n(x_i, Q_0^2) \Delta x_i \\ &= -0.0284 \pm 0.0061(stat) \pm 0.0059(sys) \quad (61) \end{aligned}$$

where $\Delta x_i$ are the bin widths and the $g_1^n(x_i, Q_0^2)$ are evaluated using Eq. 59. The first error is statistical and the second is systematic. The total systematic error is evaluated by adding all contributions in quadrature, assuming they come from uncorrelated sources. Table 14 lists the dominant contributions to the total systematic error of the integral in Eq. 61.

Since the data are only measured over part of the interval $0 < x < 1$, we must extrapolate in order to evaluate the integral $\Gamma_1^n(Q^2 = 2 \text{ (GeV}/c)^2) = \int_0^1 g_1^n(x, Q_0^2) dx$ over the full $x$ range. There are two regions to consider, large $x$ ( $0.6 < x \leq 1$ ) and small $x$ ($0 \leq x < 0.03$). For large $x$, within a three constituent quark description of the nucleon, the assumption of single flavor dominance at high $x$ and $Q^2$ leads to the prediction that $A_1 \to 1$ as $x \to 1$. This phenomenological result has also been derived from arguments based on perturbative QCD and a nonperturbative wavefunction describing the nucleon [13, 14].

In order to evaluate this integral we assumed that for $x > 0.6$, $A_1^n = 0.5 \pm 0.5$ and used the $F_2^n$ results from SLAC[71] rather than NMC [75] since it is based on data closer in kinematic range to the present experiment. The error assigned to $A_1$ was chosen to cover all possible behaviors of $A_1^n$ in this region including that of $g_1^n$ suggested by the quark counting rules[91] $g_1^n(x) \equiv (1-x)^3$. We find for $Q^2$=2 GeV$^2$,

$$\int_{0.6}^{1} g_1^n(x)dx = +0.003 \pm 0.003. \qquad (62)$$

For small $x$ one must rely heavily upon theory, especially noting that if $A_1 \neq 0$ as $x \to 0$, $\int g_1 dx \to \infty$. We assume the Regge theory prediction for the behavior of the nucleon $g_1$, namely $g_1(x \to 0) \propto x^{-\alpha_1}$ where the Regge intercept $\alpha_1$ can vary in the range $-0.5 < \alpha_1 < 0$[92, 93], although there are no strong theoretical grounds for these limits. In this region dominated by the sea and gluon contributions to the nucleon structure, it is thought that no difference should exist between the proton and neutron behavior. Therefore, we used the same value of $\alpha_1$ as the previous proton spin structure function experiments [88, 96], $\alpha_1 = 0$. The low $x$ contribution to the spin structure function integral becomes

$$\int_{0}^{x_0} g_1^n(x)dx = x_0 g_1^n(x_0). \qquad (63)$$

Here we assumed a Regge behavior up to $x_0 = 0.1$ and used a weighted fit of $g_1^n$ to the lowest three $x$ bins to reduce the statistical uncertainties. The low $x$ extrapolation yields the result at Q$^2$=2 (GeV/$c$)$^2$,

$$\int_{0}^{0.03} g_1^n(x)dx = -0.0053 \pm 0.0053 \qquad (64)$$

in which we assign a 100% uncertainty to the extrapolation. If one uses an alternative low $x$ behavior of $g_1^n$ [94], namely $g_1^n \approx a \log(1/x)$, and perform the extrapolation using only the lowest $x$ bin data point, one obtains $\int_0^{0.03} g_1^n(x)dx = -0.012$. The assigned error encompasses this result, statistical errors on the low $x$ points and the result obtained using the other boundary of the Regge intercept $\alpha = -0.5$. Our fit of $g_1^n$ is also consistent with the low $x$ results from SMC [95, 96, 97] over the measured range within the statistical uncertainties.

The total neutron integral at $Q^2$=2 (GeV/$c$)$^2$ becomes the sum of the three integrals Eqs. (61, 62, 64),

$$\Gamma_1^n = \int_0^1 g_1^n(x)dx = -0.031 \pm 0.006\ (stat) \pm 0.009\ (sys). \quad (65)$$

A comparison of the E142 data with those of E143 [89] and SMC [98] shows no significant disagreement, though there is some interesting behavior. Fig. 26 presents a comparison of x$g_1^n$ versus $x$ for E142 and E143. The data from both experiments are in reasonable agreement over the measured range. The integrals over $x$ of $g_1^n$ are $\Gamma_1^n = -0.031 \pm 0.006(stat) \pm 0.009(sys)$ for E142 at $Q^2 = 2$ (GeV/$c$)$^2$ and $\Gamma_1^n = -0.037 \pm 0.008(stat) \pm 0.011(sys)$ for E143 $Q^2 = 3$ (GeV/$c$)$^2$. Fig. 27 compares the spin structure function $xg_1^n$ extracted from SMC at $Q^2 = 10$ (GeV/$c$)$^2$ and from E142. When the SMC and E142 neutron results are combined, the shape of the structure function is interesting, with small negative results for $g_1^n$ over the range in $x$ covered by E142 followed by relatively large negative values at low $x$ measured by SMC just below the kinematic range accessible to E142. The behavior is a strong motivator for future measurements at low $x$. The integrals of $g_1^n$ over the mid $x$ range common to E142 and SMC differ, however, by approximately two standard deviations. We extract $\int g_1^n(x)dx = -0.027 \pm 0.004\ (stat) \pm 0.006\ (sys)$ from the E142 data over the $x$ range from 0.04 to 0.3, where the statistical error bars are relatively small. We compare this result to the SMC result over the same range, $\int g_1^n(x)dx = +0.007 \pm 0.015\ (stat)$. Systematic errors are neglected from the SMC data, since they are expected to be small compared to the statistical uncertainty. We do not assign any special significance to the difference but point out that it should not be ignored and needs to be monitored in future measurements where the $Q^2$ of the measured data is investigated.

In order to test the Bjorken sum rule, the proton and the neutron first moments $\Gamma_1^p$ and $\Gamma_1^n$ are evaluated at the same $Q^2$. The available experimental proton data span a different range of $Q^2$, making it necessary to evolve the proton or the neutron data to a common value of $Q^2$.

Since our neutron results and the E143 proton results [88] are at similar $Q^2$, we combine the two to test the Bjorken sum rule. The E143 proton results reads $\Gamma_1^p = 0.127 \pm 0.011$ at $Q^2 = 3$ (GeV/$c$)$^2$, while evolving the E142 neutron result to $Q^2 = 3$ (GeV/$c$)$^2$, we find $\Gamma_1^n = -0.033 \pm 0.011$. These results lead to the Bjorken integral $\Gamma^{Bj}_{exp} = \Gamma_1^p - \Gamma_1^n = 0.160 \pm 0.015$ where correlations between the two experiments, primarily from the beam polarization determination, have been taken into account.

There is agreement with the Bjorken sum rule prediction $\Gamma^{Bj}_{th} = 0.176 \pm 0.008$ using Eq. 11 from Section 2.1. assuming three flavors and choosing $\alpha_s(Q^2 = 3(\mathrm{GeV}/c)^2) = 0.32 \pm 0.05$ [99]. Fig. 28 shows tests of the Bjorken sum rule from different experiments. The present determination is the most accurate test of the Bjorken sum rule to date.

We can rewrite the Bjorken sum rule including the higher twist contributions and extract a value for $\alpha_s$.

$$\Gamma^{Bj} = \frac{1}{6} g_A [1 - \frac{\alpha_s(Q^2)}{\pi} - 3.58(\frac{\alpha_s(Q^2)}{\pi})^2 - 20.2(\frac{\alpha_s(Q^2)}{\pi})^3] + \frac{1}{6}\frac{C_{HT}}{Q^2} \quad (66)$$

where $C_{HT}$ is the higher twist contribution to the Bjorken sum rule. Recent estimates of $C_{HT}$ show that it is very model dependent. For example, using QCD sum rules methods several authors have evaluated $C_{HT}$ and found it to be $C_{HT} = -0.09 \pm 0.06$[22] in one case or $C_{HT} = -0.015 \pm 0.02$[100] in another. The sensitivity of the result can be described by the change in sign in $C_{HT}$ found when the estimate is made using a bag model [101]. Therefore, an additional theoretical uncertainty equal in magnitude to the present size of the higher twist correction should be included in the theoretical estimate of $\alpha_s$.

We use the Bjorken sum rule with perturbative QCD corrections up to third order in $\alpha_s$ without higher twist term corrections to extract a value of $\alpha_s$ at $Q^2 = 3$ (GeV/$c$)$^2$ for polarized deep inelastic scattering,

$$\alpha_s(Q^2 = 3(\mathrm{GeV}/c)^2) = 0.408^{+0.070}_{-0.085}. \quad (67)$$

If we consider the higher twist corrections to $\Gamma^{Bj}$ and choose an average value and error of $C_{HT}$ from the QCD sum rules, that is $C_{HT} = -0.10 \pm 0.05$, we find

$$\alpha_s(Q^2 = 3(\mathrm{GeV}/c)^2) = 0.312^{+0.098}_{-0.130}. \tag{68}$$

Both results (with or without a higher twist correction) for $\alpha_s$ are in agreement with the world average[99].

For the Ellis-Jaffe sum rule test, we compare at the average $Q^2$ of the experiment, namely $Q_0^2 = 2$ $(\mathrm{GeV}/c)^2$. From Eq. 12, and using $\alpha_s = 0.35 \pm 0.05$ and $3F - D = 0.58 \pm 0.12$, we obtain the theoretical value of $\Gamma_1^n = -0.016 \pm 0.016$ where the error on the result is dominated by the error on the quantity $3F - D$. We see that the Ellis-Jaffe sum rule is one standard deviation away from the experimental result, and the experimental result is consistent with the results of E143 and SMC.

Higher order perturbative QCD corrections to this sum rule have had a significant impact on the interpretation of the experimental results. For the $Q^2$ at which the SLAC experiment E142 is performed, these corrections are quite large. At this time, these corrections have been given up to third order in the expansion of $\alpha_s(Q^2)$. For example, at the average $Q^2$ of the E142 experiment (2 $(\mathrm{GeV}/c)^2$), the corrections change the Ellis-Jaffe sum rule prediction for the neutron from $-0.020$ (without corrections) to $-0.011$ (with corrections), assuming a value of $\alpha_s = 0.35$.

Using $\Gamma_1^n$ experimental results and the same values for $g_A$ and $3F - D$ as earlier, one can extract a value for the total quark spin contribution to the nucleon using the E142 results, $\Delta\Sigma(2\ (\mathrm{GeV}/c)^2)$=0.43 $\pm$ 0.12 and similarly, we can extract the fraction of polarized strange sea contribution $\Delta s$ (2 $\mathrm{GeV}/c)^2$) = $-0.05 \pm 0.06$. We also find $\Delta\Sigma_{inv}$=0.39 $\pm$ 0.11 with the corresponding fraction of polarized strange sea contribution $\Delta s_{inv}$ = $-0.06 \pm 0.06$. Table 15 gives the total quark flavors contributions to the nucleon's spin using Eq. 12 in one case and Eq. 15 in the other assuming a conservative uncertainty on $F/D$. The uncertainty on the determination of $\Delta s$ is still large even when we take a more optimistic uncertainty on the value of $F/D$. For $F/D = 0.576 \pm 0.059$ as quoted by Close and Roberts [32], and using Eq. 15 we find $\Delta s = -0.06 \pm 0.04$ while the uncertainty on the other quark flavors contributions remains the same (see

Table 16).

Including the higher order perturbative QCD corrections, this result is in agreement with the extraction of $\Delta\Sigma$ from the E143 ($\Delta\Sigma_{inv} = 0.30 \pm 0.06$) and SMC ($\Delta\Sigma_{inv} = 0.20 \pm 0.11$) experiments [89, 98]. Care must be taken when comparing these numbers since different authors make different assumption in extracting $\Delta\Sigma$ from their data. In addition, the good agreement does depend on the validity of the perturbative QCD corrections in the low $Q^2$ region and the estimated small size of the higher twist corrections.

# 8 Conclusions

We report final results on the first determination of the neutron spin structure function using a polarized $^3$He nucleus. Over the kinematic range accessible to the experiment, we find small negative asymmetries similar to the predictions from the quark parton model [16, 17]. Within the statistics of the experiment, we are not able to distinguish any clear shape as a function of $x$ to the neutron spin asymmetries, although significant deviations are expected, particularly at low and high $x$. For example, as $x$ approaches unity, the neutron asymmetry $A_1^n$ is predicted to approach unity. As $x$ approaches zero, $A_1^n$ should approach zero. The results are in agreement with an extraction of the neutron spin structure function from the deuteron as performed by SLAC experiment E143. We see no dependence on $Q^2$ within the limited precision of the data sample. In addition, we present results on $A_2^n(x)$ which are compatible with zero and significantly better than the unitarity limit given by $\sqrt{R}$ in the range from $x$ of 0.03 to $x$ of 0.3. The $A_2^n(x)$ results, however, are less precise than what one could extract from the E143 proton and deuteron data.

From our measurement we proceed to extract the first moment of $g_1^n$, namely $\int g_1^n(x)dx$. Since our asymmetries over the measured region are small, $\int g_1^n(x)dx$ is small. We proceed to use the results for the neutron integral to extract the quark flavor distributions, $\Delta u$, $\Delta d$, $\Delta s$ and $\Delta\Sigma$ with some caveats. In our extraction of the $g_1^n$ integral, we assume that one can do a Regge theory extrapolation of the contribution to the integral between $x$ of 0 and the lowest values of $x$ measured in the experiment. The implications of this fit are that the integral contribution at low $x$ is itself small. On the other hand, recent data from the SMC collaboration appears to indicate that there may be a large negative contribution to the neutron integral in the $x$ range below where we measure. If this is true, then the assumption of Regge behavior up to $x$ of approximately 0.1 underestimates the neutron contribution at low $x$. A major motivator for future measurements of spin structure functions at either higher energies or with higher precision comes from studying the spin structure functions at lower $x$. With the Regge theory assumption, we extract a value of $\Delta\Sigma$, the total quark contribution to the nucleon's spin of approximately 40%. We note that this result has a sensitive dependence on higher order perturbative and non-perturbative QCD corrections. In addition, the result depends on the scale at which $\Delta\Sigma$ is evaluated and the number of quark flavors used in the evaluation, typically three or four. We can tune for different values of $\Delta\Sigma$ ranging from $\Delta\Sigma$ of 0.36 to $\Delta\Sigma$ of 0.43 with different theoretical assumptions.

We combine the proton results from experiment E143 with the neutron results from this experiment to test the Bjorken sum rule. Ignoring the unlikely possibility for large non-singlet contributions to the proton and neutron integrals at low $x$, this comparison still stands as the most precise test of the Bjorken sum rule to date. We find that the sum rule is satisfied at the 10% level.

Future measurements of the proton and neutron spin structure functions will increase the kinematic coverage, particularly at low $x$. SLAC experiments E154[102] and E155[103] will extract the proton and neutron spin structure functions using a higher energy 50 GeV polarized electron beam. These two experiments aim to measure the spin structure functions at a higher average $Q^2$ and will extract data at lower values of $x$ with higher statistical precision. Additional measurements from the CERN SMC program will continue to increase the statistical precision, needed to draw decisive conclusions at low $x$. A collider experiment like HERA with a polarized electron and a polarized proton beam would, in principle, be ideal for reaching very low $x$ ($\approx 10^{-4}$) to extract the proton spin structure function. In addition, precision measurements of the proton and neutron spin structure functions at high $x$ ($x > 0.5$) are useful for testing Quark Parton Model predictions. The rising behavior of the neutron asymmetry $A_1^n$ at high $x$ still needs to be confirmed. Experiments at HERMES [104] and TJNAF [105] are likely to be the best grounds for these tests along with future precision measurements of $g_2$.

We conclude by noting that this first measurement of the neutron spin structure function does not complete the study, but instead has helped pave the way for future measurements with higher precision and investigations with an increasing attention to detail.

## Acknowledgements

We are indebted to G. S. Bicknell, G. Bradford, R. Boyce, B. Brau, J. Davis, S. Dyer, R. Eisele, C. Fertig, J. Hrica, C. Hudspeth, M. Jimenez, G. Jones, J. Mark, J. McDonald, W. Nichols, M. Racine, B. Smith, M. Sousa, R. Vogelsang, and J. White for their exceptional efforts in the prepartion of this experiment, to the SLAC Accelerator Operations Group for delivering the polarized beam, and to the technical staff of CE-Saclay and LPC-Clermont. We also would like to thank N.M. Shumeiko for providing us with the latest version of POLRAD to perform the electromagnetic internal radiative corrections. This work was supported by Department of Energy contracts: DE-AC03-76SF00098 (LBL), W-4705-Eng-48 (LLNL), DE-FG02-90ER40557 (Princeton), DE-AC03-76SF00515 (SLAC), DE-FG03-88ER40439 (Stanford), DE-FG02-84ER40146 (Syracuse), DE-FG02-94ER40844 (Temple), DE-AC02-76ER00881 (Wisconsin), National Science Foundation grants: 9014406, 9114958 (American), 8914353, 9200621 (Michigan); and the Bundesministerium fuer Forschung und Technologie (W. Meyer), the Centre National de la Recherche Scientifique (LPC-Clermont-Ferrand) and the Commissariat à l'Energie Atomique (Saclay).

## References

[1] V. N. Gribov and L.N. Lipatov, Yad. Fiz **15** (1972) 781 [Sov. J. Nucl. Phys. **15**, 438 (1972)].

[2] G. Altarelli and G. Parisi, Nucl. Phys. **B126**, 298 (1977).

[3] A. J. Buras, Rev. Mod. Phys. **52**, 200 (1980).

[4] W.C. Leung *et. al.*, Phys. Lett. **B317**, 655 (1993).

[5] D. Gross and C.H. Llewellyn Smith, Nucl. Phys. **B14**, 337 (1969).

[6] M.J. Alguard *et al.,* Phys. Rev. Lett. **37**, 1258 (1976); **37**, 1261 (1976).

[7] G. Baum *et al.*, Phys. Rev. Lett. **51**, 1135 (1983).

[8] V. H. Hughes and J. Kuti, Ann. Rev. Nucl. Sci **33**, 611 (1983).

[9] J.D. Bjorken, Phys. Rev. **148**, 1467 (1966); Phys. Rev. **D1**, 1376 (1970).

[10] EMC, J. Ashman *et al.*, Phys. Lett. **B206**, 364 (1988); Nucl. Phys. **B328**, 1 (1989).

[11] J. Ellis and R.L. Jaffe, Phys. Rev. **D9**, 1444 (1974); **D10**, 1669 (1974) .

[12] M. Anselmino, A. Efremov, E. Leader, Phys. Rep. **261**, 1 (1995).

[13] G.R. Farrar and D. R. Jackson, Phys. Rev. Lett. **35**, 1416 (1975).

[14] G.R. Farrar, Phys. Lett. **B70**, 346 (1977).

[15] J. Kaur, Nucl. Phys. **B128**, 219 (1977).

[16] R.D. Carlitz and J. Kaur, Phys. Rev. Lett. **37**, 673 (1977).

[17] A. Schafer, Phys. Lett. **B208**, 249 (1988).

[18] J. Kodeira *et. al.*, Phys. Rev. **D20**, 627 (1979).

[19] J. Kodeira, S. Matsuda, K Sasaki and T. Uematsu, Nucl. Phys. **B159**, 99 (1979).

[20] J. Kodeira, Nucl. Phys. **B165**, 129 (1980).

[21] E.V. Shuryak and A.I. Vainshtein, Nucl. Phys. **201**, 141 (1982).

[22] I. I. Balitsky, V.M. Braun and A. V. Kolesnichenko, Phys. Lett. **B242**, 245 (1990); Phys. Lett. **B318**, 648 (1993) Erratum.

[23] S.A. Larin and J.A.M. Vermaseren, Phys. Lett. **B259**, 345 (1991).

[24] M. Gourdin, Nucl Phys **B38**, 418 (1972).

[25] S. A. Larin, Phys. Lett. **B334**, 192 (1994).

[26] S.L. Adler, Phys. Rev. **177**, 2426 (1969).

[27] J.S. Bell and R. Jackiw, Nuovo Cimento **47**, 61 (1969).

[28] D. Kaplan and A. Manohar, Nucl. Phys. **B310**, 527 (1988).

[29] R.L. Jaffe, Phys. Lett. **B193**, 101 (1987).

[30] R.L. Jaffe and A.V. Manohar, Nucl. Phys. **B337**, 509 (1990).

[31] P.G. Ratcliffe, Phys. Lett. **B242**, 271 (1990).

[32] F. E. Close and R. G. Roberts, Phys. Lett. **B316**, 165 (1993).

[33] SLAC E142, P.L. Anthony *et. al.*, Phys. Rev. Lett. **71**, 959 (1993).

[34] R. Alley *et. al.*, Nucl. Instrum. Meth. **A365**, 1 (1995).

[35] M. Woods *et. al.*, SLAC-PUB-5965, December 1992.

[36] A.D. Yeremian *et. al.*, SLAC-PUB-6074, March 1993.

[37] C. Møller, Ann. Phys. (Leipzig) **14**, 532 (1932); J. Arrington *et. al.*, Nucl. Instrum. Meth. **A311**, 39 (1992).

[38] L.G. Levchuk, Nucl. Instrum. Meth. **A345**, 496 (1994).

[39] M. Swartz *et. al.*, SLAC-PUB-6467 (1995), to be published in Nucl. Instrum. Meth..

[40] H. R. Band, University of Wisconsin Report WISC-EX-94-341, (1994).

[41] Vacuumschmelze GMBH, Hanau, Germany.

[42] M.A. Bouchiat, T.R. Carver and C.M. Varnum, Phys. Rev. Lett. **5**, 373 (1960).

[43] N.D. Bhaskar, W. Happer, and T. McClelland, Phys. Rev. Lett. **49**, 25 (1982).

[44] W. Happer, E. Miron, S. Schaefer, D. Schreiber, W.A. van Wijngaarden, and X. Zeng, Phys. Rev. **A29**, 3092 (1984).

[45] W. Happer, Rev. Mod. Phys. **44**, 169 (1972).

[46] T.E. Chupp, M.E. Wagshul, K.P. Coulter, A.B. McDonald, and W. Happer, Phys. Rev. **C36**, 2244 (1987).

[47] N.R. Newbury, A.S. Barton, P. Bogorad, G. D. Cates, M. Gatzke, H. Mabuchi, and B. Saam, Phys. Rev. **A48**, 558 (1993).

[48] M. E. Wagshul, T. E. Chupp, Phys. Rev **A49**, 3854 (1994).

[49] N. R. Newbury, A. S. Barton, G. D. Cates, W. Happer, and H. Middleton, Phys. Rev. **A48**, 4411 (1993).

[50] R. L. Gamblin and T. R. Carver, Phys. Rev. **A138**, 946 (1965); L. D. Schearer and G. K. Walters, Phys. Rev. **A139**, 1398 (1965); G.D. Cates, S.R. Schaefer and W. Happer, Phys. Rev. **A37**, 2877 (1988).

[51] K.D. Bonin, T.G. Walker, and W. Happer, Phys. Rev. **A37**, 3270 (1988).

[52] K.P. Coulter, A.B. McDonald, G.D. Cates, W. Happer, T.E. Chupp, Nucl. Instrum. Meth. in Phys. Res. **A276**, 29 (1989).

[53] T.E. Chupp, R.A. Loveman, A.K. Thompson, A.M. Bernstein, and D.R. Tieger, Phys. Rev. **C45**, 915 (1992).

[54] T. Killian, Phys. Rev. **27**, 578 (1926).

[55] A. Abragam, Principles of Nuclear Magnetism (Oxford University Press, New York, 1961).

[56] W. A. Fitzsimmons, L. L. Tankersley, and G. Walters, Phys. Rev. **179**, 156 (1969).

[57] R. S. Timsit, J. M. Daniels, and A. D. May, Can. J. Phys. **49**, 560 (1971).

[58] G.G. Petratos *et al.*, SLAC-PUB-5698 (1991).

[59] D. Kawall, Phd Thesis, Stanford University 1995 (unpublished).

[60] P. Baillon *et. al.*, Nucl. Instrum. Meth. **A126**, 13 (1975).

[61] E.L. Garwin, Y. Tomkiewicz and D. Trines, Nucl. Instrum. Meth. **107**, 365 (1973).

[62] Y. Tomkiewicz and E. L. Garwin, SLAC-PUB 1356, 1973.

[63] J. Xu, Phd Thesis, Syracuse University, 1994 (unpublished).

[64] K. Brown, SLAC Report **75**, 1982.

[65] G. T. Bartha *et. al.*, Nucl. Instrum. Meth. **A275**, 59 (1989).

[66] H. Borel *et. al.*, IEEE TRANS on Nucl. Sci. Vol. 42, No. 4, 529 (1995) and Erratum, IEEE TRANS on Nucl. Sci. Vol.42, No 6, 2347 (1995).

[67] M. Spengos, Ph.D. thesis, American University, 1994 (unpublished).

[68] J. Dunne, Ph.D. thesis, American University, 1995 (unpublished).

[69] Y. Roblin, thèse de doctorat, Université de Clermond-Ferrand, 1995 (unpublished).

[70] V. Breton *et. al.*, Nucl. Instrum. Meth. **A362**, 478 (1995).

[71] L. Whitlow *et. al.*, Phys. Lett. **B282**, 475 (1992).

[72] J. Gomez *et. al.*, Phys. Rev. **D 49**, 4348 (1994).

[73] A.C. Venuti *et. al.*, Phys. Lett. **B223**, 485 (1989); Phys. Lett. **B237**, 592 (1990).

[74] A. Bodek *et. al.*, Phys. Rev. **D21**, 1471 (1979).

[75] P. Amaudruz *et. al.*, Phys. Lett. **B295**, 159 (1992).

[76] L. W. Mo and Y. S. Tsai, Rev. Mod. Phys. **41**, 205 (1969).

[77] Y. S. Tsai, Report No. SLAC-PUB-848 (1971).

[78] T.V. Kuchto and N. M. Shumeiko, Nucl. Phys. **B219**, 412 (1983).

[79] I. V. Akushevich and N. M. Shumeiko, J. Phys. **G**: Nucl. Part. Phys. **20**, 513 (1994).

[80] T. De Forest and D.J. Walecka, Adv. in Phys. **15** (1966).

[81] F. W. Brasse, W. Flauger, J. Gauler, S. P. Goel, R. Haiden, M. Merkwitz and H. Wriedt, Nucl. Phys. **B110**, 413 (1976).

[82] V. Burkert and Zh. Li, Phys. Rev. **D47**, 46 (1993).

[83] SLAC E143, K. Abe *et. al.* Phys. Rev. Lett. **76**, 587 (1996).

[84] J. L. Friar *et. al.*, Phys. Rev. **C42**, 2310 (1990).

[85] C. Ciofi degli Atti *et. al.*, Phys. Rev. **C48**, 968 (1993).

[86] B. Blankleider and R.M. Woloshyn, Phys. Rev. **C29**, 538 (1984).

[87] R.-W. Shulze and P.U. Sauer, Phys. Rev. **C48**, 38 (1993).

[88] SLAC E143, K. Abe *et. al.,* Phys. Rev. Lett. **74**, 346 (1995)

[89] SLAC E143, K. Abe *et. al.*, Phys. Rev. Lett. **75**, 25 (1995).

[90] SLAC E143, K. Abe *et. al.*, Phys. Lett. **B364**, 61 (1995).

[91] S. J. Brodsky, M. Burkhardt, and I Schmidt, Nucl. Phys. **B441**, 197 (1995).

[92] R. L. Heinmann, Nucl. Phys. **B64**, 429 (1973)

[93] J. Ellis and M. Karliner, Phys. Lett. **B213**, 73 (1988).

[94] F. E. Close and R. G. Roberts, Phys. Lett. **B336**, 257 (1994).

[95] SMC, D. Adeva *et. al.*, Phys. Lett. **B302**, 533 (1993).

[96] SMC, D. Adams *et. al.*, Phys. Lett. **B329**, 399 (1994).

[97] SMC, D. Adams *et. al.*, Phys. Lett. **B336**, 125 (1994).

[98] SMC, D. Adams *et. al.*, Phys. Lett. **B357**, 248 (1995).

[99] Particle Data Group, M. Aguilar-Benitez *et. al.*, Phy. Rev. D**50**, 1173 (1994).

[100] G. G. Ross and R. G. Roberts, Phys. Lett. **B322**, 425 (1994).

[101] X. Ji and P. Unrau, Phys. Lett. **B333**, 228 (1994). (1993).

[102] SLAC proposal E154, E. W. Hughes, spokesperson, October 1993.

[103] SLAC proposal E155, R.G. Arnold and J. McCarthy, spokespersons, October 1993.

[104] Hermes proposal, DESY No. PRC-90-01, R. Milner and K. Rith, spokespersons, 1990.

[105] CEBAF proposal 94-101, Z.-E. Meziani and P. Souder, spokespersons, December 1994.

Table 1: Table of parameters for the E142 experiment.

| Description | |
|---|---|
| Average beam polarization using a AlGaAs source | 36% |
| Average current | 1.5 $\mu$A |
| High density polarized $^3$He target | Pressure=8.6 atm, Vol=90 cc, Pol=33% |
| Thin windows to minimize the dilution factor | 110$\mu$m/window |
| High counting rate due to low spectrometer angle | 4.5° and 7° |
| Large statistics with polarized beam on polarized target | 300 million events |
| Target windows cooling in vacuum | No explosions due to glass radiation damage |

Table 2: Systematic uncertainties contributing to the beam polarization measurement.

| | Value(%) |
|---|---|
| Foil Magnetization | ±1.7 |
| Kinematic Acceptance | ±0.1 |
| Model Dependence | ±1.0 |
| Gain and Nonlinearity Correction | ±2.2 |
| Fit Range | ±1.0 |
| TOTAL | ±3.1 |

Table 3: Target polarimetry systematicuncertainties.

| | |
|---|---|
| Proton signal magnitude | ±5.6% |
| Proton polarization ($\xi$) | ±1.5% |
| Electronic gain | ±1.5% |
| Cell geometry | ±2.6% |
| Lock-in time constant correction | ±1.0% |
| $^3$He density | ±2.5% |
| Total | ±7.1% |

Table 4: Polarized Target parameters used in calculating the dilution factor. A run corresponds typically to one or two hours of data taking.

| | Target | | |
|---|---|---|---|
| | 1 | 2 | 3 |
| Runs used | 1000-1117,1320-1771 | 1118-1181 | 1182-1319 |
| Length (mm) | 295 ± 2 | 297 ± 2 | 303 ± 2 |
| Front window ($\mu$m) | 110 ± 7.5 % | 110 ± 7.5% | 110 ± 7.5 % |
| Rear window ($\mu$m) | 124 ± 7.5 % | 107 ± 7.5 % | 110 ± 7.5 % |
| Glass density (g/cm$^2$) | 2.52±1% | 2.52±1% | 2.52±1% |
| $^3$He density (amagats) | 8.63±2.5% | 8.90±2.0% | 8.74±2.0% |
| $N_2$ density (amagats) | 0.070±1.7% | 0.069± 1.8% | 0.082± 1.8% |
| $^3$He density (cm$^{-3}$) | $2.32 \times 10^{20}$±2.5% | $2.39 \times 10^{20}$±2.0% | $2.35 \times 10^{20}$± 2.0% |
| $N_2$ density (cm$^{-3}$) | $1.88 \times 10^{18}$±1.7% | $1.85 \times 10^{18}$± 1.8% | $2.20 \times 10^{18}$±1.8% |

Table 5: Radiation lengths, $t_{out}$, seen by electrons exiting the target.

| Spectrometer | Fraction of Events | Radiation Lengths ($t_{out}$) |
|---|---|---|
| 4.5° | 36.50% | 0.001 |
| | 54.55% | 0.085 |
| | 6.85% | 0.169 |
| | 2.10% | 0.291 |
| 7.0° | 23.39% | 0.001 |
| | 48.36% | 0.055 |
| | 18.75% | 0.269 |
| | 9.50% | 0.399 |

Table 6: The radiative corrections to $A_1^{^3He}(x)$ at the average $Q^2$ of each $x$ bin are tabulated here, with the corrections to the 4.5° spectrometer given first, then those of the 7.0°. The Δ's are the additive correction to be made to the asymmetry for internal (int), external (ext), and $f_i$ are the fractions of events in a particular bin coming from quasielastic and inelastic tails including internal and external radiative effects. The elastic contribution is small.

| $x$ | $\Delta_{RC}^{int}$ | $\Delta_{RC}^{ext}$ | $f_{quel}$ | $f_{inel}$ | $f_{ext}$ |
|---|---|---|---|---|---|
| 0.035 | -0.003 | -0.000 | 0.08 | 0.15 | 0.10 |
| 0.050 | -0.003 | -0.001 | 0.04 | 0.13 | 0.08 |
| 0.080 | -0.004 | -0.002 | 0.02 | 0.10 | 0.08 |
| 0.125 | -0.004 | -0.003 | 0.01 | 0.07 | 0.08 |
| 0.175 | -0.004 | -0.003 | 0.00 | 0.04 | 0.05 |
| 0.250 | -0.004 | -0.003 | 0.00 | 0.01 | 0.04 |
| 0.350 | -0.003 | -0.002 | 0.00 | 0.02 | 0.02 |
| | | | | | |
| 0.080 | -0.004 | -0.001 | 0.04 | 0.11 | 0.09 |
| 0.125 | -0.004 | -0.002 | 0.01 | 0.07 | 0.09 |
| 0.175 | -0.004 | -0.003 | 0.01 | 0.04 | 0.06 |
| 0.250 | -0.004 | -0.003 | 0.00 | 0.00 | 0.05 |
| 0.350 | -0.003 | -0.002 | 0.00 | 0.03 | 0.02 |
| 0.466 | -0.002 | -0.000 | 0.00 | 0.06 | 0.01 |

Table 7: Results for $A_1^{^3He}$ and $g_1^{^3He}$.

| $x$ range | $\langle x \rangle$ | $\langle Q^2 \rangle$ (GeV/c)$^2$ | $A_1^{^3He} \pm \delta$(stat) $\pm \delta$(sys) | $g_1^{^3He} \pm \delta$(stat) $\pm \delta$(sys) |
|---|---|---|---|---|
| 0.03-0.04 | 0.035 | 1.1 | $-0.0264 \pm 0.0168 \pm 0.0054$ | $-0.248 \pm 0.159 \pm 0.055$ |
| 0.04-0.06 | 0.050 | 1.2 | $-0.0238 \pm 0.0102 \pm 0.0039$ | $-0.168 \pm 0.072 \pm 0.027$ |
| 0.06-0.10 | 0.082 | 1.8 | $-0.0317 \pm 0.0081 \pm 0.0048$ | $-0.146 \pm 0.038 \pm 0.020$ |
| 0.10-0.15 | 0.124 | 2.5 | $-0.0447 \pm 0.0083 \pm 0.0068$ | $-0.141 \pm 0.027 \pm 0.018$ |
| 0.15-0.20 | 0.175 | 3.1 | $-0.0463 \pm 0.0098 \pm 0.0094$ | $-0.105 \pm 0.023 \pm 0.014$ |
| 0.20-0.30 | 0.246 | 3.7 | $-0.0333 \pm 0.0099 \pm 0.0121$ | $-0.051 \pm 0.016 \pm 0.007$ |
| 0.30-0.40 | 0.343 | 4.4 | $-0.0003 \pm 0.0172 \pm 0.0220$ | $0.000 \pm 0.017 \pm 0.004$ |
| 0.40-0.60 | 0.466 | 5.5 | $-0.0145 \pm 0.0301 \pm 0.0199$ | $-0.007 \pm 0.014 \pm 0.002$ |

Table 8: Results for $A_2^{^3He}$ and $g_2^{^3He}$. Note that the systematic uncertainties are small compared to the statistical uncertainties.

| x range | $\langle\ x\ \rangle$ | $\langle\ Q^2\ \rangle$ (GeV/c)$^2$ | $A_2^{^3He} \pm \delta(\text{stat}) \pm \delta(\text{sys})$ | $g_2^{^3He} \pm \delta(\text{stat}) \pm \delta(\text{sys})$ |
|---|---|---|---|---|
| 0.03-0.04 | 0.036 | 1.1 | $-0.0619 \pm 0.0738 \pm 0.0169$ | $-9.314 \pm 10.62 \pm 2.593$ |
| 0.04-0.06 | 0.050 | 1.2 | $0.0222 \pm 0.0516 \pm 0.0165$ | $2.106 \pm 4.280 \pm 1.399$ |
| 0.06-0.10 | 0.082 | 1.8 | $-0.0158 \pm 0.0425 \pm 0.0189$ | $-0.648 \pm 1.705 \pm 0.756$ |
| 0.10-0.15 | 0.124 | 2.5 | $-0.0193 \pm 0.0453 \pm 0.0216$ | $-0.256 \pm 0.934 \pm 0.428$ |
| 0.15-0.20 | 0.175 | 3.1 | $-0.0885 \pm 0.0559 \pm 0.0263$ | $-0.764 \pm 0.641 \pm 0.283$ |
| 0.20-0.30 | 0.246 | 3.7 | $-0.0349 \pm 0.0585 \pm 0.0286$ | $-0.088 \pm 0.336 \pm 0.153$ |
| 0.30-0.40 | 0.342 | 4.4 | $0.0756 \pm 0.1002 \pm 0.0347$ | $0.140 \pm 0.246 \pm 0.080$ |
| 0.40-0.60 | 0.466 | 5.5 | $-0.1875 \pm 0.1683 \pm 0.0344$ | $-0.188 \pm 0.170 \pm 0.036$ |

Table 9: Results on $A_1^n$ and $g_1^n$ at the measured average $Q^2$, along with $g_1^n$ evaluated at $Q^2 = 2$ (GeV/$c$)$^2$ assuming that $A_1^n$ does not depend on $Q^2$.

| $x$ range | $\langle\ x\ \rangle$ | $\langle\ Q^2\ \rangle$ | $A_1^n \pm \delta(\text{stat}) \pm \delta(\text{sys})$ | $g_1^n \pm \delta(\text{stat}) \pm \delta(\text{sys})$ | $g_1^n \pm \delta(\text{stat}) \pm \delta(\text{sys})$ ($Q^2 = 2(\text{GeV}/c)^2$) |
|---|---|---|---|---|---|
| 0.03-0.04 | 0.035 | 1.1 | $-0.092 \pm 0.061 \pm 0.022$ | $-0.269 \pm 0.182 \pm 0.065$ | $-0.311 \pm 0.207 \pm 0.074$ |
| 0.04-0.06 | 0.050 | 1.2 | $-0.082 \pm 0.038 \pm 0.017$ | $-0.177 \pm 0.083 \pm 0.033$ | $-0.195 \pm 0.090 \pm 0.036$ |
| 0.06-0.10 | 0.081 | 1.8 | $-0.109 \pm 0.031 \pm 0.021$ | $-0.151 \pm 0.044 \pm 0.025$ | $-0.154 \pm 0.044 \pm 0.026$ |
| 0.10-0.15 | 0.124 | 2.5 | $-0.162 \pm 0.033 \pm 0.030$ | $-0.146 \pm 0.031 \pm 0.022$ | $-0.142 \pm 0.031 \pm 0.022$ |
| 0.15-0.20 | 0.174 | 3.1 | $-0.170 \pm 0.041 \pm 0.042$ | $-0.105 \pm 0.026 \pm 0.017$ | $-0.099 \pm 0.026 \pm 0.016$ |
| 0.20-0.30 | 0.245 | 3.7 | $-0.113 \pm 0.044 \pm 0.055$ | $-0.045 \pm 0.018 \pm 0.009$ | $-0.042 \pm 0.018 \pm 0.009$ |
| 0.30-0.40 | 0.341 | 4.4 | $+0.050 \pm 0.083 \pm 0.107$ | $0.011 \pm 0.019 \pm 0.005$ | $+0.010 \pm 0.020 \pm 0.006$ |
| 0.40-0.60 | 0.466 | 5.5 | $+0.006 \pm 0.159 \pm 0.108$ | $0.000 \pm 0.016 \pm 0.003$ | $+0.000 \pm 0.020 \pm 0.003$ |

Table 10: Results on $A_2^n$ and $g_2^n$. Note that the systematic uncertainties are small compared to the statistical uncertainties.

| $x$ range | $\langle\ x\ \rangle$ | $\langle\ Q^2\ \rangle$ (GeV/c)$^2$ | $A_2^n \pm \delta(\text{stat}) \pm \delta(\text{sys})$ | $g_2^n \pm \delta(\text{stat}) \pm \delta(\text{sys})$ |
|---|---|---|---|---|
| 0.03-0.04 | 0.036 | 1.1 | $-0.225 \pm 0.270 \pm 0.068$ | $-10.68 \pm 12.22 \pm 3.17$ |
| 0.04-0.06 | 0.050 | 1.2 | $0.084 \pm 0.191 \pm 0.062$ | $2.44 \pm 4.92 \pm 1.63$ |
| 0.06-0.10 | 0.081 | 1.8 | $-0.058 \pm 0.163 \pm 0.073$ | $-0.74 \pm 1.96 \pm 0.87$ |
| 0.10-0.15 | 0.124 | 2.5 | $-0.074 \pm 0.181 \pm 0.087$ | $-0.29 \pm 1.07 \pm 0.49$ |
| 0.15-0.20 | 0.174 | 3.1 | $-0.365 \pm 0.234 \pm 0.119$ | $-0.88 \pm 0.74 \pm 0.34$ |
| 0.20-0.30 | 0.245 | 3.8 | $-0.147 \pm 0.261 \pm 0.128$ | $-0.11 \pm 0.39 \pm 0.18$ |
| 0.30-0.40 | 0.341 | 4.4 | $0.365 \pm 0.480 \pm 0.171$ | $0.16 \pm 0.28 \pm 0.09$ |
| 0.40-0.60 | 0.466 | 5.5 | $-0.975 \pm 0.885 \pm 0.218$ | $-0.22 \pm 0.20 \pm 0.05$ |

Table 11: Table of systematic uncertainties on $A_1^n$ for each $x$ point.

| Parameter | $x$=0.035 | $x$=0.05 | $x$=0.08 | $x$=.125 | $x$=.175 | $x$=.25 | $x$=.35 | $x$=.47 |
|---|---|---|---|---|---|---|---|---|
| $P_b$ | 0.0027 | 0.0024 | 0.0032 | 0.0048 | 0.0051 | 0.0036 | 0.0008 | 0.0023 |
| $P_t$ | 0.0063 | 0.0054 | 0.0072 | 0.0110 | 0.0117 | 0.0083 | 0.0018 | 0.0052 |
| f | 0.0071 | 0.0061 | 0.0081 | 0.0124 | 0.0131 | 0.0093 | 0.0020 | 0.0059 |
| $\Delta_{dt}$ | 0.0010 | 0.0011 | 0.0012 | 0.0012 | 0.0013 | 0.0015 | 0.0019 | 0.0013 |
| $\Delta_{RC}$ | 0.0026 | 0.0039 | 0.0055 | 0.0068 | 0.0079 | 0.0082 | 0.0068 | 0.0022 |
| R | 0.0038 | 0.0034 | 0.0038 | 0.0051 | 0.0055 | 0.0035 | 0.0006 | 0.0021 |
| $F_2$ | 0.0061 | 0.0054 | 0.0070 | 0.0063 | 0.0072 | 0.0051 | 0.0073 | 0.0090 |
| $\rho_n$ | 0.0021 | 0.0019 | 0.0025 | 0.0037 | 0.0039 | 0.0026 | 0.0012 | 0.0001 |
| $A_1^p$ | 0.0021 | 0.0018 | 0.0017 | 0.0019 | 0.0024 | 0.0035 | 0.0058 | 0.0116 |
| $\rho_p$ | 0.0010 | 0.0013 | 0.0019 | 0.0027 | 0.0036 | 0.0051 | 0.0076 | 0.0131 |
| $A_\perp$ | 0.0082 | 0.0093 | 0.0123 | 0.0191 | 0.0335 | 0.0518 | 0.1064 | 0.1053 |
| $A^{\pi^-}$ | 0.0079 | 0.0037 | 0.0024 | 0.0000 | 0.0000 | 0.0000 | 0.0000 | 0.0000 |
| $A^{e^+}$ | 0.0123 | 0.0044 | 0.0030 | 0.0033 | 0.0000 | 0.0000 | 0.0000 | 0.0000 |
| $^3$He corr. | 0.0046 | 0.0041 | 0.0054 | 0.0081 | 0.0085 | 0.0057 | 0.0025 | 0.0003 |
| Total | 0.0216 | 0.0165 | 0.0209 | 0.0295 | 0.0413 | 0.0551 | 0.1074 | 0.1075 |

Table 12: Table of systematic uncertainties on $g_1^n$ for each $x$ point.

| Parameter | $x$=0.035 | $x$=0.05 | $x$=0.08 | $x$=0.125 | $x$=0.175 | $x$=0.25 | $x$=0.35 | $x$=0.47 |
|---|---|---|---|---|---|---|---|---|
| $P_b$ | 0.0090 | 0.0055 | 0.0044 | 0.0042 | 0.0030 | 0.0014 | 0.0001 | 0.0002 |
| $P_t$ | 0.0207 | 0.0125 | 0.0099 | 0.0097 | 0.0069 | 0.0031 | 0.0003 | 0.0005 |
| f | 0.0233 | 0.0141 | 0.0112 | 0.0109 | 0.0077 | 0.0035 | 0.0003 | 0.0006 |
| $\Delta_{dt}$ | 0.0034 | 0.0027 | 0.0017 | 0.0011 | 0.0008 | 0.0006 | 0.0004 | 0.0002 |
| $\Delta_{RC}$ | 0.0089 | 0.0093 | 0.0078 | 0.0061 | 0.0047 | 0.0031 | 0.0014 | 0.0002 |
| R | 0.0200 | 0.0130 | 0.0102 | 0.0077 | 0.0054 | 0.0024 | 0.0014 | 0.0009 |
| $F_2$ | 0.0131 | 0.0085 | 0.0047 | 0.0039 | 0.0028 | 0.0013 | 0.0000 | 0.0002 |
| $\rho_n$ | 0.0071 | 0.0045 | 0.0035 | 0.0033 | 0.0023 | 0.0010 | 0.0002 | 0.0000 |
| $A_1^p$ | 0.0061 | 0.0039 | 0.0024 | 0.0018 | 0.0015 | 0.0014 | 0.0012 | 0.0011 |
| $\rho_p$ | 0.0028 | 0.0028 | 0.0026 | 0.0025 | 0.0022 | 0.0020 | 0.0016 | 0.0012 |
| $A_\perp$ | 0.0310 | 0.0165 | 0.0096 | 0.0075 | 0.0068 | 0.0050 | 0.0047 | 0.0023 |
| $A^{\pi^-}$ | 0.0268 | 0.0087 | 0.0035 | 0.0000 | 0.0000 | 0.0000 | 0.0000 | 0.0000 |
| $A^{e^+}$ | 0.0413 | 0.0104 | 0.0045 | 0.0029 | 0.0000 | 0.0000 | 0.0000 | 0.0000 |
| $^3$He corr. | 0.0155 | 0.0098 | 0.0076 | 0.0072 | 0.0050 | 0.0021 | 0.0005 | 0.0000 |
| Total | 0.0737 | 0.0363 | 0.0255 | 0.0218 | 0.0161 | 0.0088 | 0.0055 | 0.0031 |

Table 13: Results on $A_1^n$ vs. $Q^2$. Systematic uncertainties are small and have been neglected. See Table IX for other values of $x$.

| $x$ range | $\langle\ x\ \rangle$ | $\langle\ Q^2\ \rangle$ (GeV/c)$^2$ | $A_1^n \pm \delta$(stat) |
|---|---|---|---|
| 0.06-0.10 | 0.082 | 1.3 | $-0.13 \pm 0.08$ |
| | | 1.6 | $-0.11 \pm 0.05$ |
| | | 1.9 | $-0.18 \pm 0.07$ |
| | | 2.0 | $0.01 \pm 0.10$ |
| | | 2.5 | $-0.08 \pm 0.10$ |
| | | 2.9 | $-0.07 \pm 0.13$ |
| 0.10-0.15 | 0.124 | 1.5 | $-0.24 \pm 0.10$ |
| | | 2.0 | $-0.04 \pm 0.07$ |
| | | 2.4 | $-0.12 \pm 0.10$ |
| | | 2.5 | $-0.16 \pm 0.09$ |
| | | 3.1 | $-0.24 \pm 0.07$ |
| | | 3.6 | $-0.20 \pm 0.09$ |
| 0.15-0.20 | 0.175 | 1.7 | $0.01 \pm 0.15$ |
| | | 2.2 | $-0.05 \pm 0.09$ |
| | | 2.7 | $-0.14 \pm 0.15$ |
| | | 2.9 | $-0.22 \pm 0.10$ |
| | | 3.7 | $-0.23 \pm 0.07$ |
| | | 4.4 | $-0.24 \pm 0.10$ |
| 0.20-0.30 | 0.246 | 1.8 | $0.05 \pm 0.17$ |
| | | 2.4 | $-0.05 \pm 0.10$ |
| | | 3.0 | $-0.34 \pm 0.17$ |
| | | 3.4 | $-0.13 \pm 0.12$ |
| | | 4.3 | $-0.16 \pm 0.07$ |
| | | 5.2 | $-0.02 \pm 0.11$ |
| 0.30-0.40 | 0.343 | 2.0 | $0.30 \pm 0.34$ |
| | | 2.6 | $0.00 \pm 0.18$ |
| | | 3.3 | $-0.25 \pm 0.38$ |
| | | 3.8 | $-0.12 \pm 0.24$ |
| | | 5.0 | $-0.05 \pm 0.13$ |
| | | 6.1 | $0.47 \pm 0.20$ |

Table 14: Systematic uncertainties on $\Gamma_1^n$. The total uncertainty on $\Gamma_1^n$ coming from systematics is 0.0060.

| $P_b$ | $P_t$ | f | $\Delta_{RC}$ | R | $F_2$ | $A_1^p$ | $\rho_p$ | $A_\perp$ | $^3$He corr. |
|---|---|---|---|---|---|---|---|---|---|
| 0.0012 | 0.0021 | 0.0024 | 0.0016 | 0.0021 | 0.0010 | 0.0009 | 0.0010 | 0.0032 | 0.0015 |

Table 15: Total quark flavor contributions to the nucleon's spin using a conservative uncertainty on $F/D$ as described in the text.

| $\Delta u$ | $\Delta d$ | $\Delta s$ | $\Delta\Sigma$ | $Q^2 = 2\ (\mathrm{GeV}/c)^2$ |
|---|---|---|---|---|
| $0.87 \pm 0.04$ | $-0.39 \pm 0.04$ | $-0.05 \pm 0.06$ | $0.43 \pm 0.12$ | |
| $\Delta u$ | $\Delta d$ | $\Delta s$ | $\Delta\Sigma_{inv}$ | Invariant quantities |
| $0.86 \pm 0.04$ | $-0.40 \pm 0.04$ | $-0.06 \pm 0.06$ | $0.39 \pm 0.11$ | |

Table 16: Total quark flavor contributions to the nucleon's spin using an uncertainty as quoted by Close and Roberts (see text).

| $\Delta u$ | $\Delta d$ | $\Delta s$ | $\Delta\Sigma_{inv}$ | Invariant quantities |
|---|---|---|---|---|
| $0.86 \pm 0.04$ | $-0.40 \pm 0.04$ | $-0.06 \pm 0.04$ | $0.39 \pm 0.11$ | |

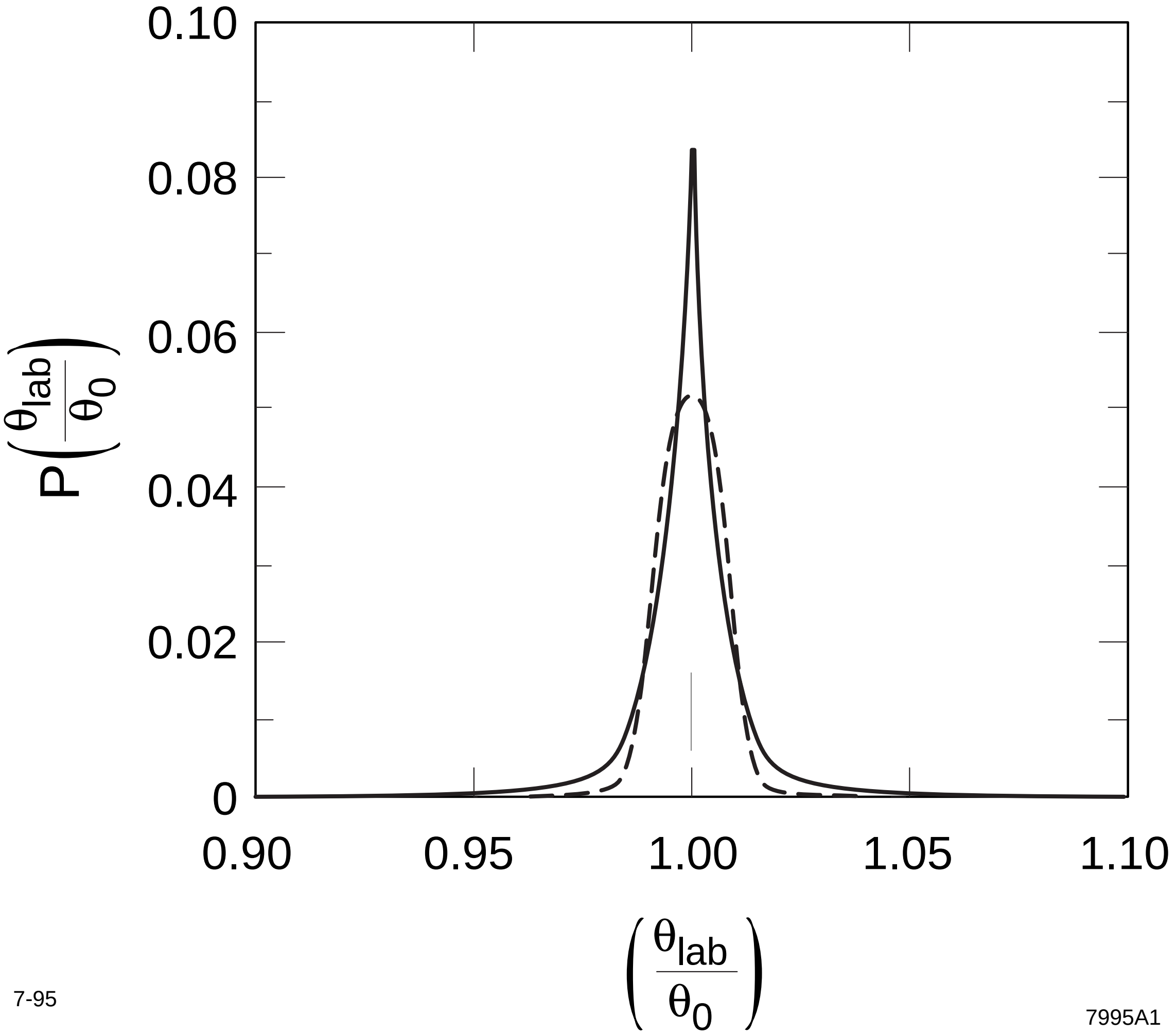

Figure 1: The calculated probability distribution $P\,(\theta_{lab}/\theta_0)$ for laboratory scattering angles due to atomic electron motion. The horizontal axis is the laboratory scattering angle $(\theta_{lab})$ in units of the central scattering angle $(\theta_0)$. The dashed curve is for M shell polarized target electrons. The solid curve is an average for all target electrons.

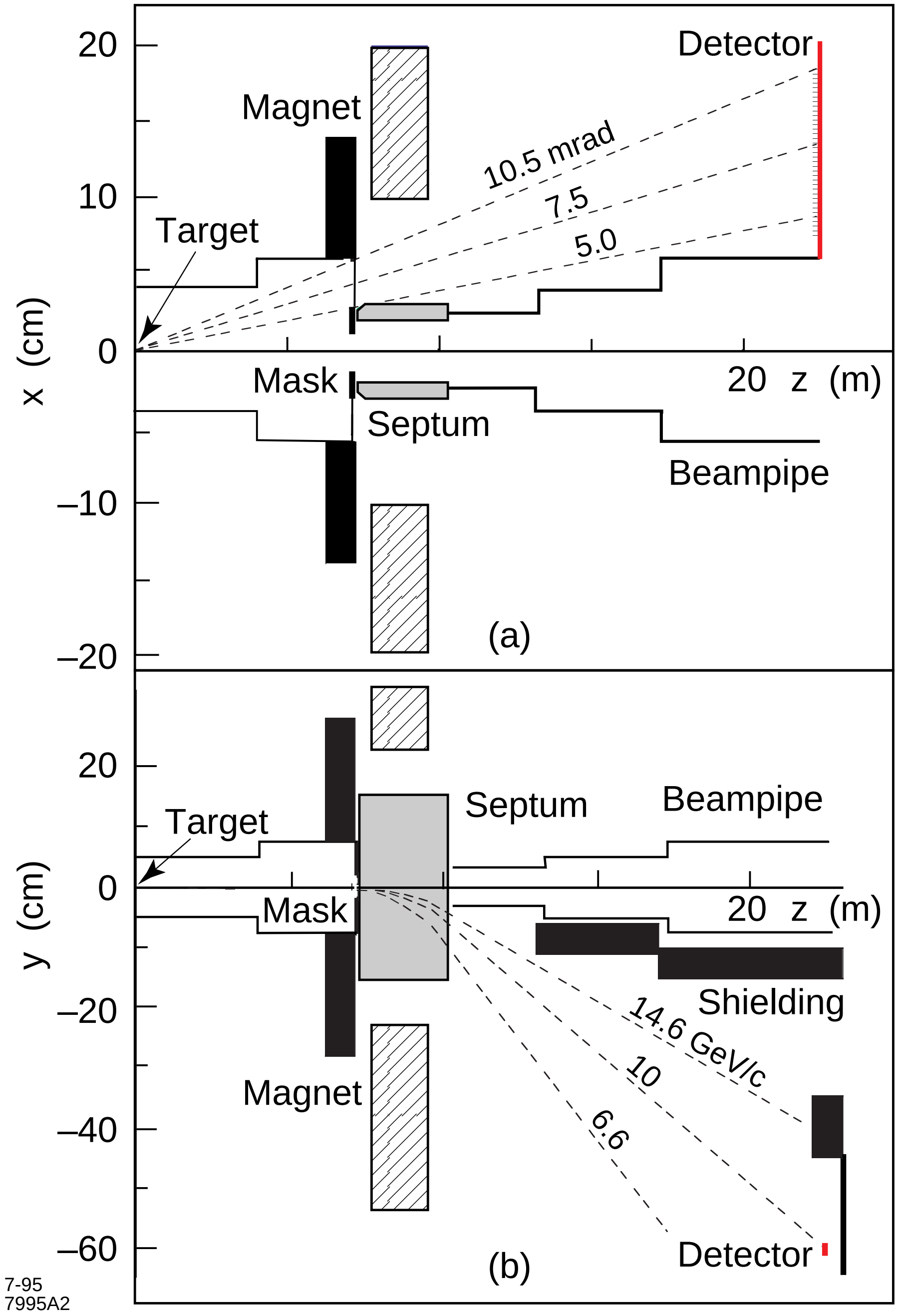

Figure 2: Top (a) and side (b) views of the E-142 Møller Polarimeter. The mask selects Møller scattered electrons near the horizontal plane which are then dispersed vertically by the magnet. The detector uses gas proportional tubes embedded in lead to sample the Møller signal over a specific momentum range and to measure the scattering angle.

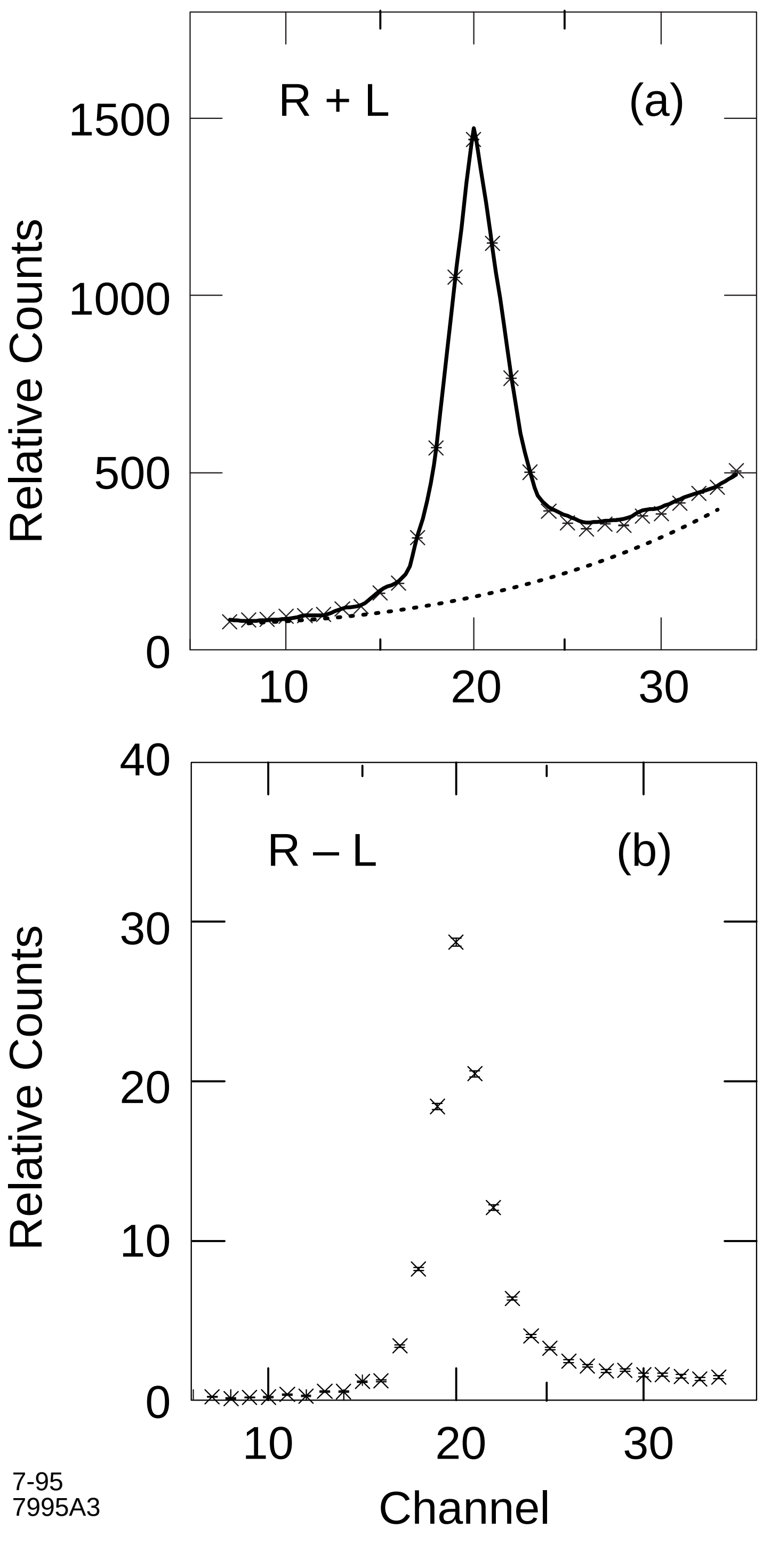

Figure 3: R+L (a) and R–L (b) scattering distributions for an average of 12 Møller runs at E = 22.66 GeV/c with a 30 $\mu$m thick target. This solid line in (a) is the fitted R+L line shape and the dotted line is the fitted background.

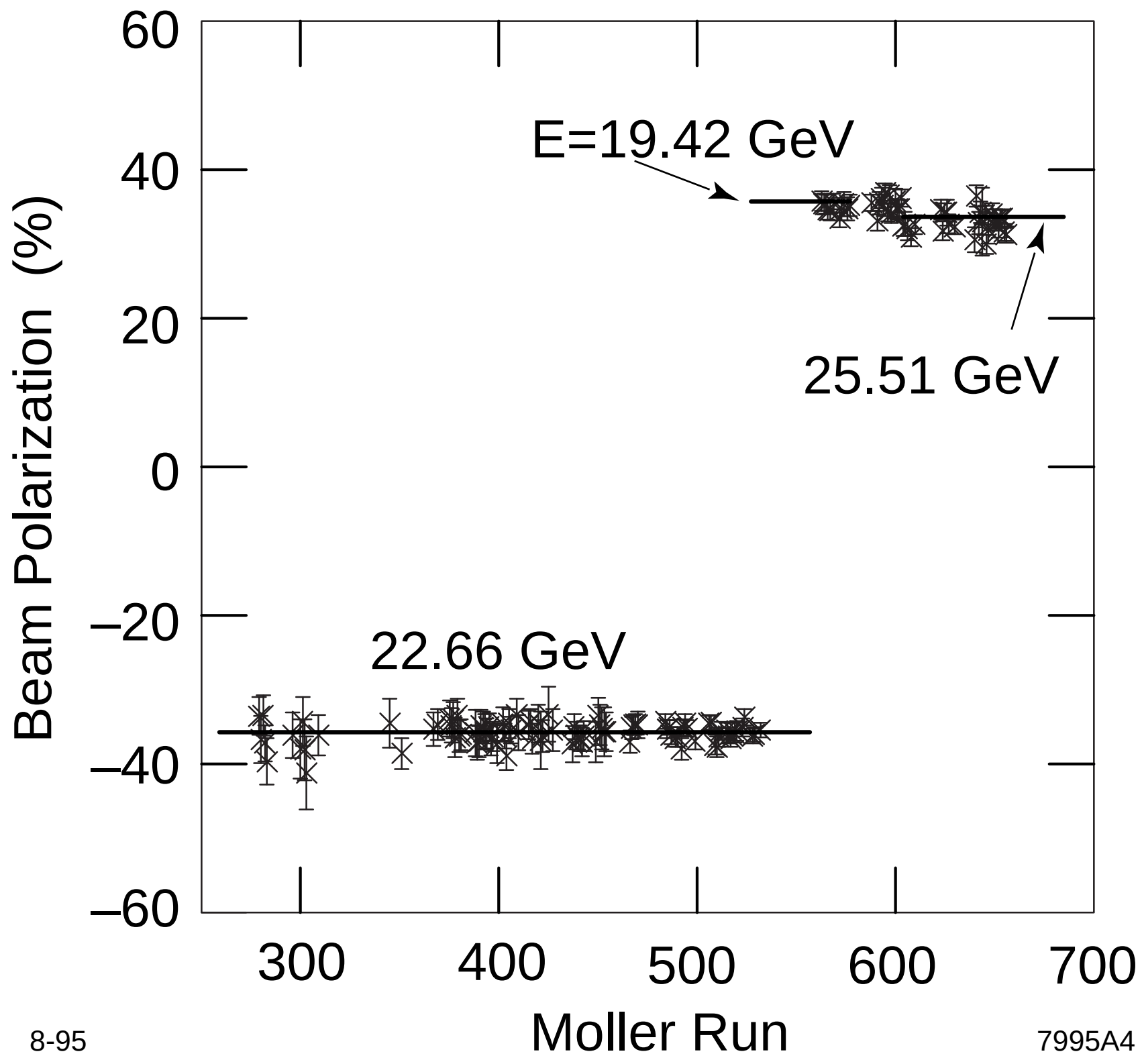

Figure 4: The measured longitudinal beam polarization $P_B$ versus Møller polarimeter run number for runs passing the cuts described in the text. The sign is relative to the polarization direction at the source.

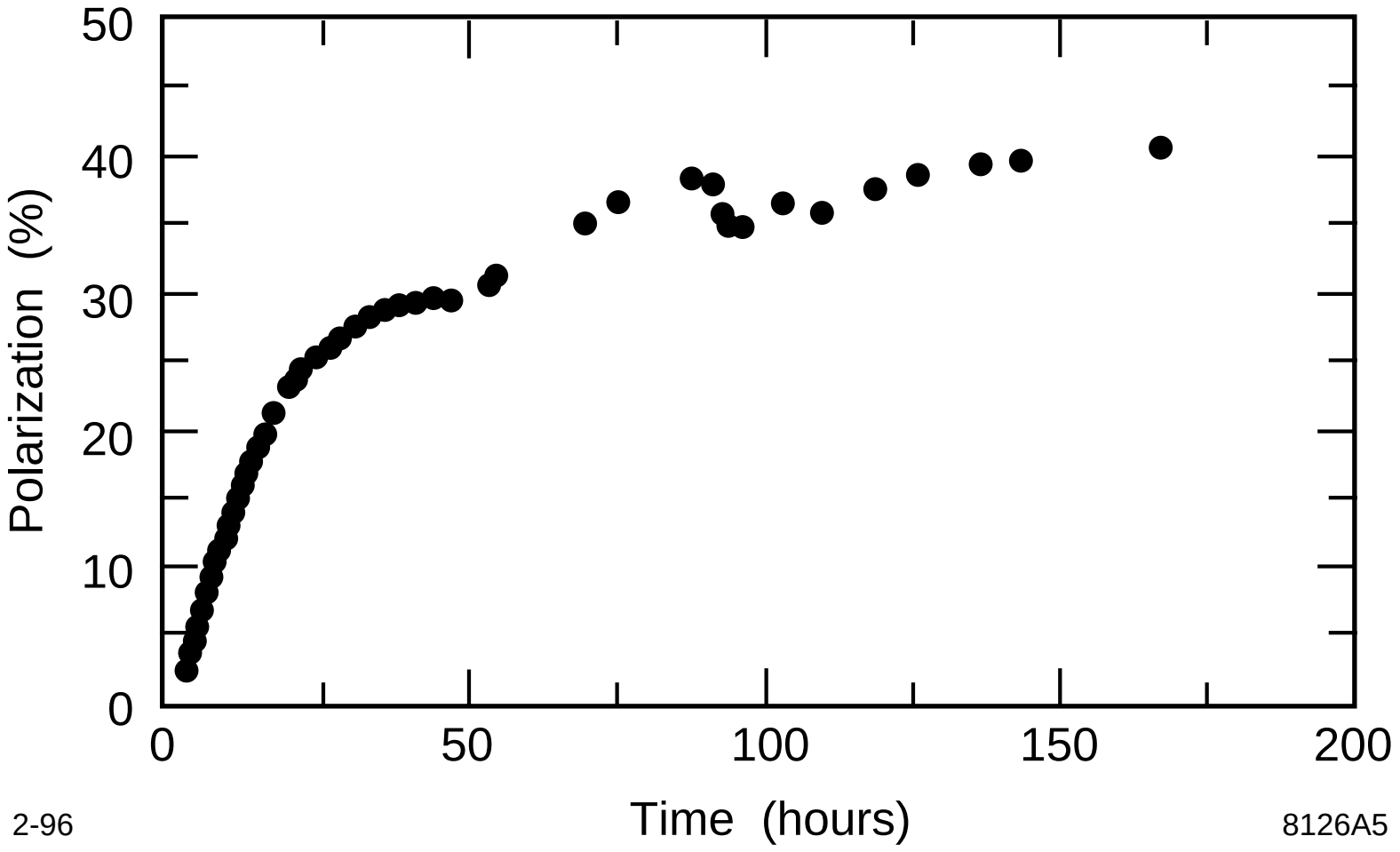

Figure 5: Typical polarization spin-up curve for the longitudinal $^3$He polarization in one of the target cells used in the experiment.

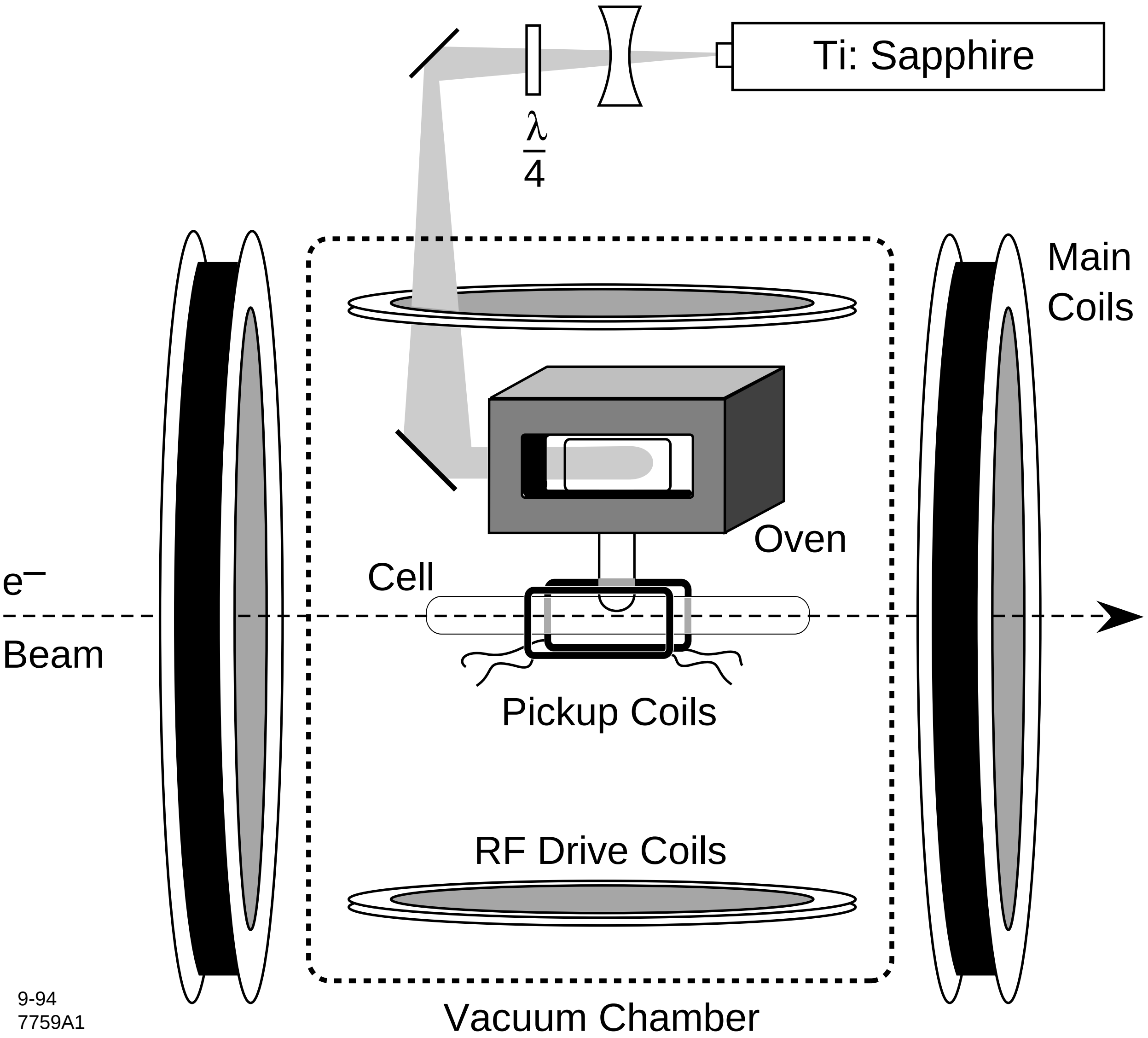

Figure 6: Schematic of the E142 $^3$He target system.

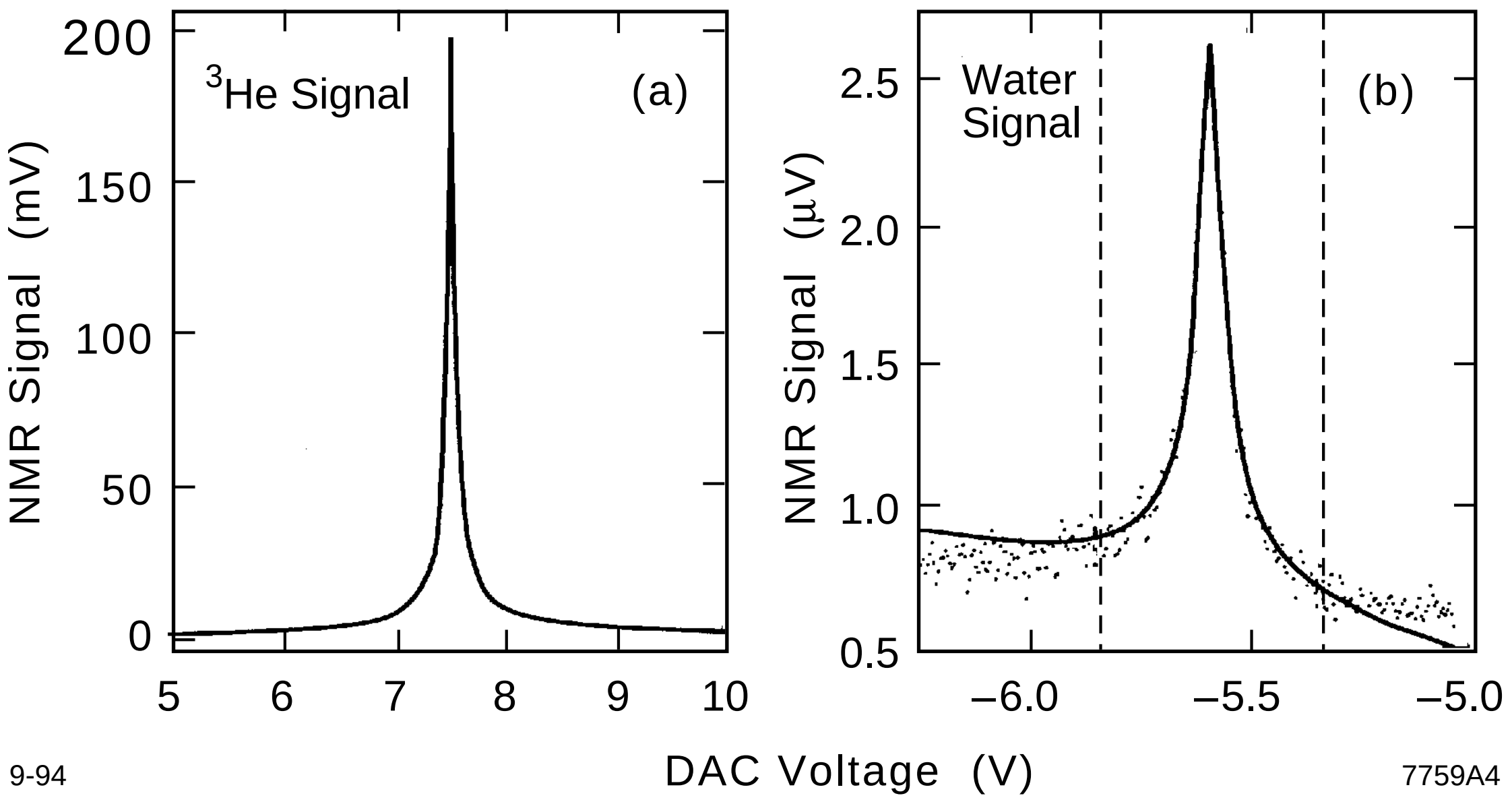

Figure 7: (a) $^3$He NMR-AFP signal obtained using one sweep of the main holding field. (b) Water NMR-AFP average signal obtained using twenty five sweeps of the main holding field. Note the difference in scale between the two signals. The curve corresponds to a Lorentzian fit to the data.

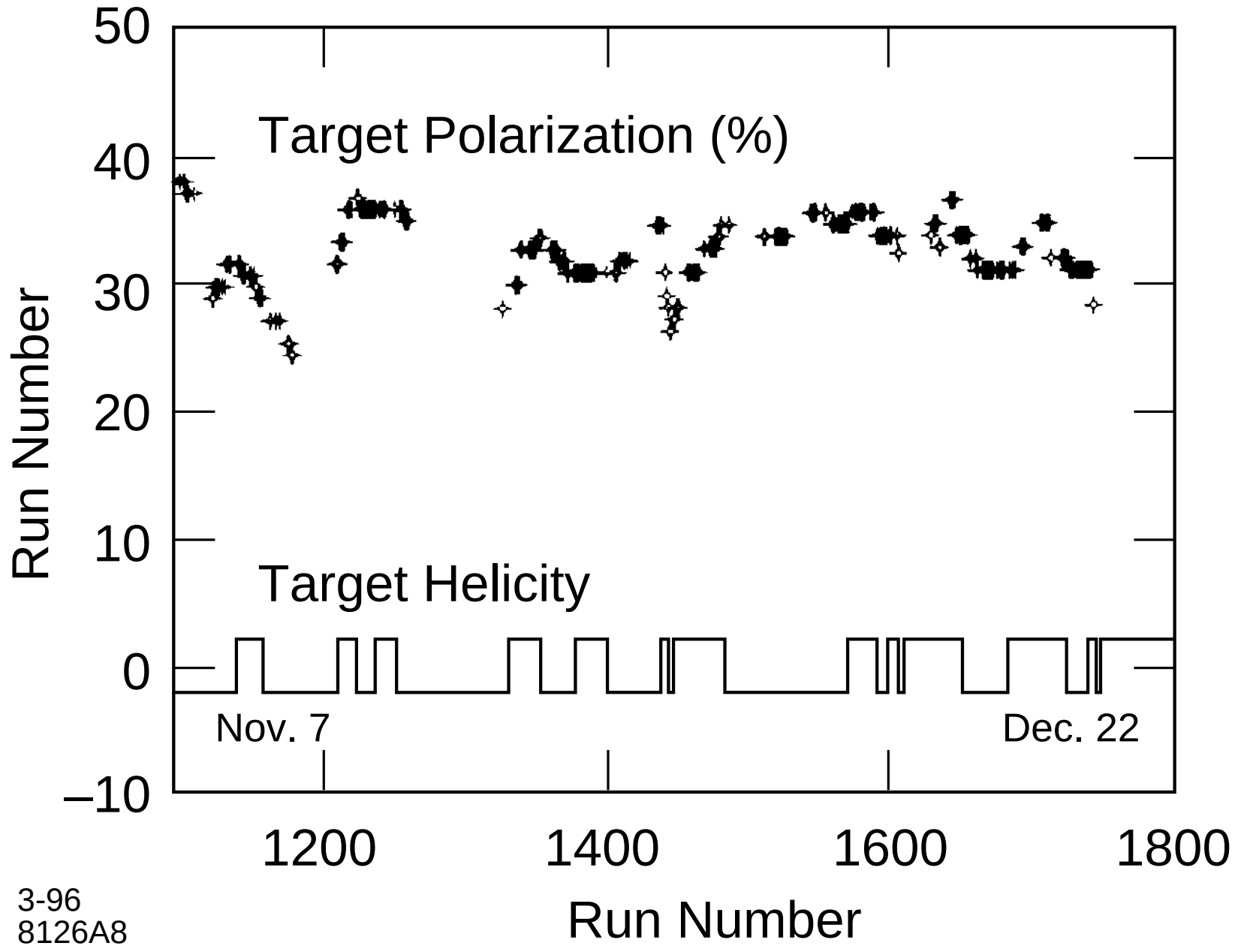

Figure 8: $^3$He target polarization versus time.

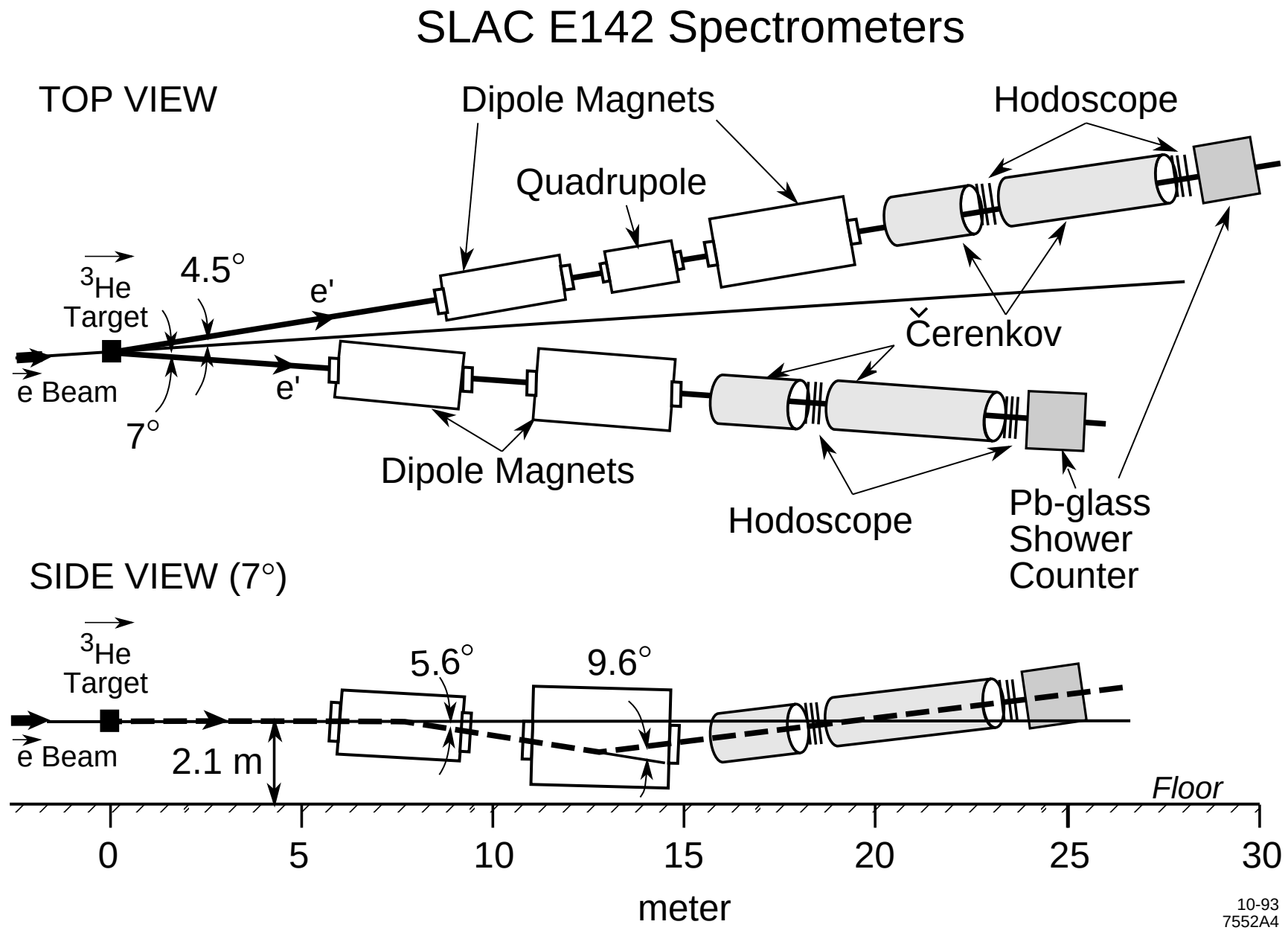

Figure 9: Layout of the magnets and detectors used in E142 experiment.

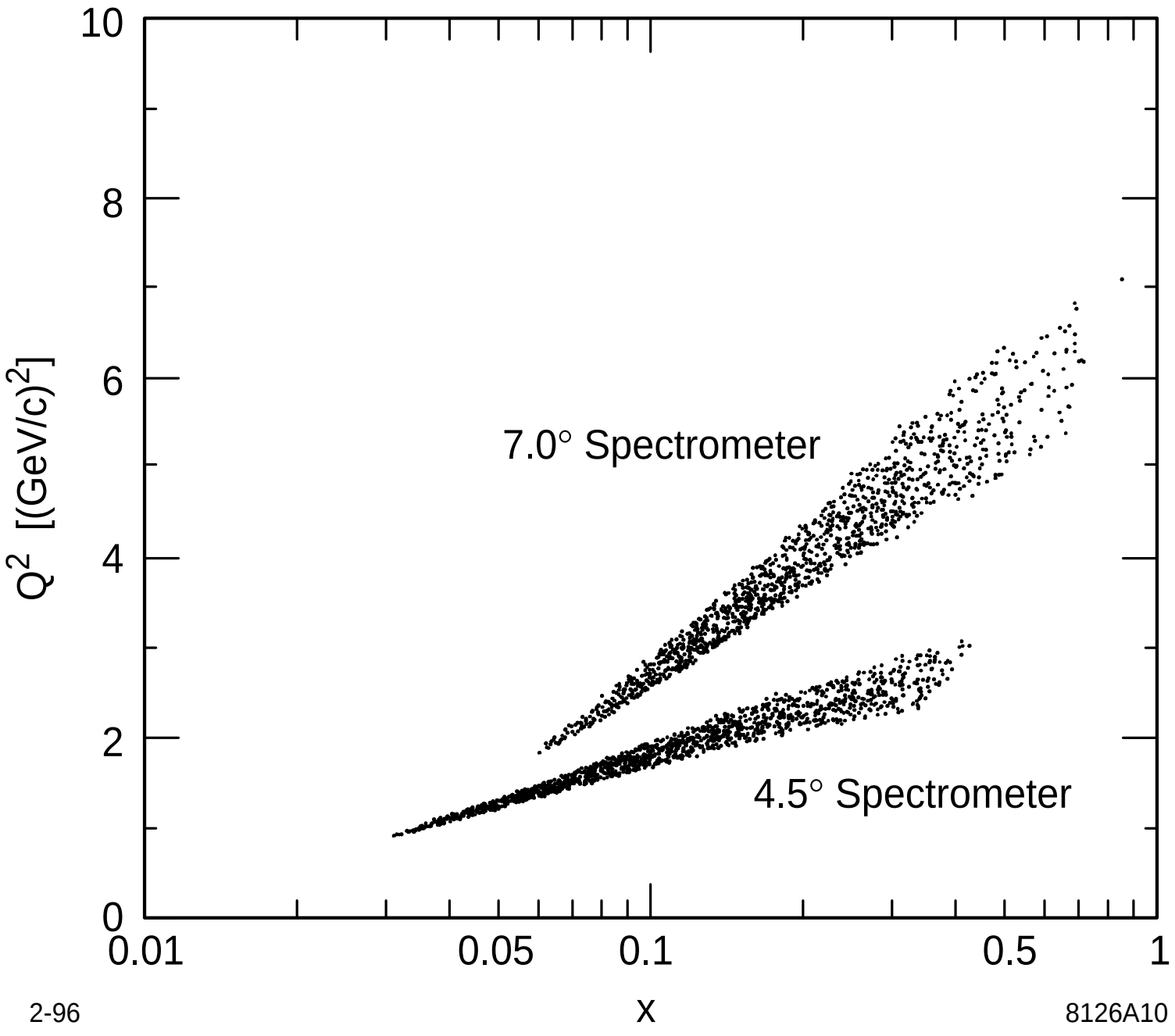

Figure 10: $Q^2$ versus $x$ range covered by the experiment for E = 22.66 GeV. This range is defined by the beam incident energy, scattering angle and spectrometers momentum and angular acceptances. The density of points corresponds to the relative electron scattering rates.

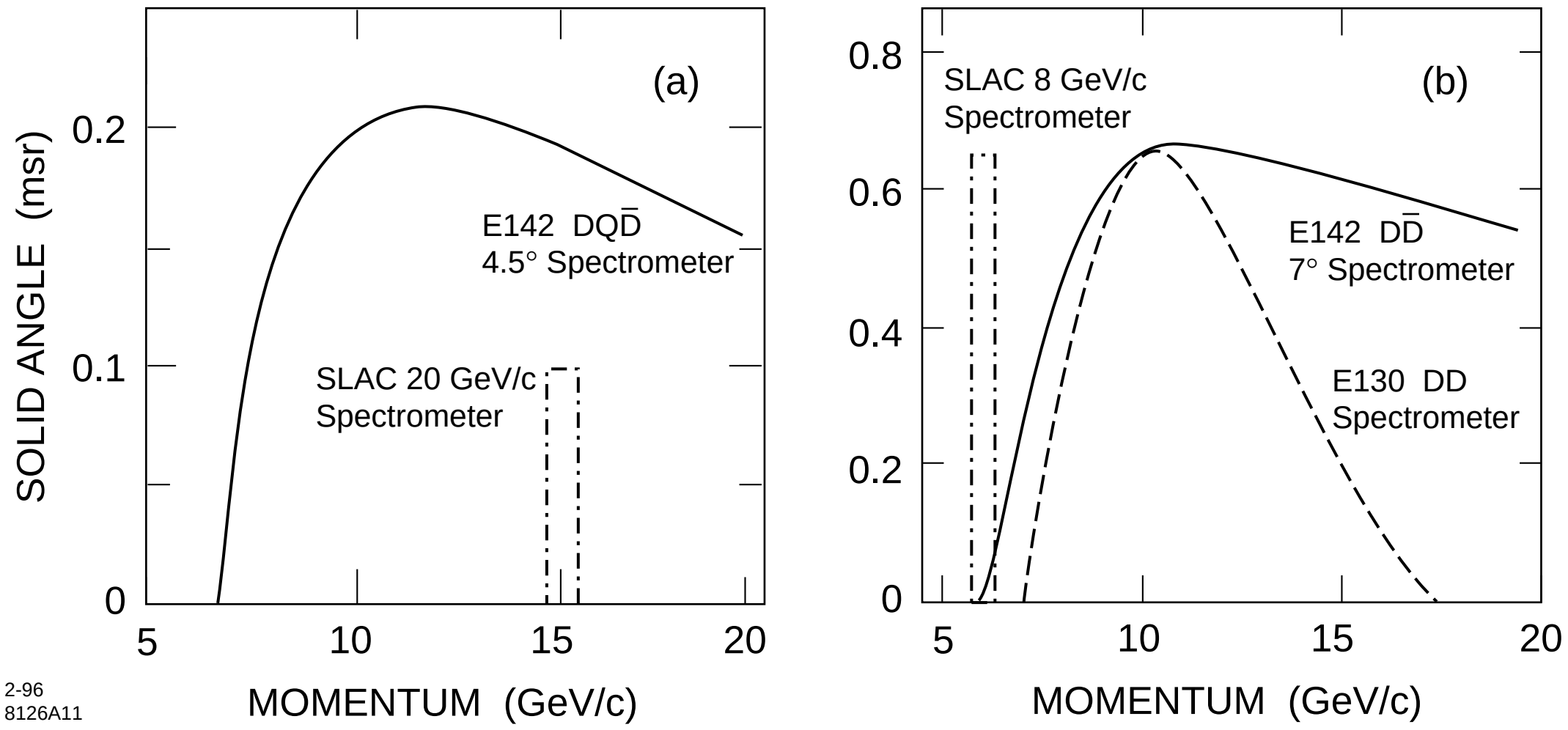

Figure 11: Momentum acceptance of (a) the 4.5° and (b) the 7° spectrometer. Acceptances for the SLAC 20 GeV (a), the 8 GeV (b) and the E130 spectrometers are presented for comparison.

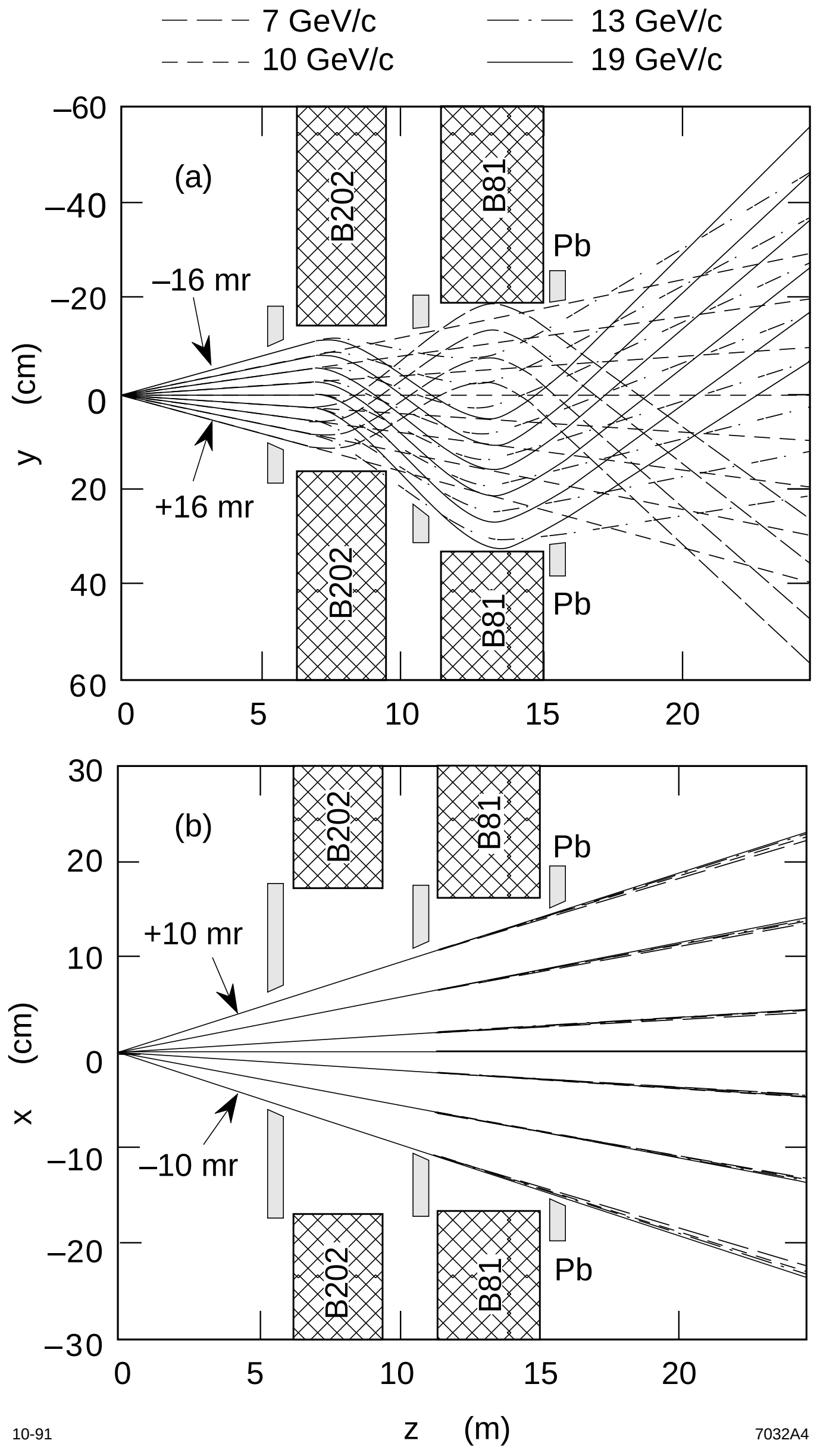

Figure 12: Bend plane (a) and non-bend plane (b) raytrace for the 7.0° spectrometer for rays of different momenta originating from the center of the polarized target. All rays are drawn with respect to the central trajectory of the system ( (a) $\phi_0 = 0$ mr, (b) $\theta_0 = 0$ mr and $p_0 = 10$ GeV/$c$ ). Also shown are the iron magnet poles and lead collimators.

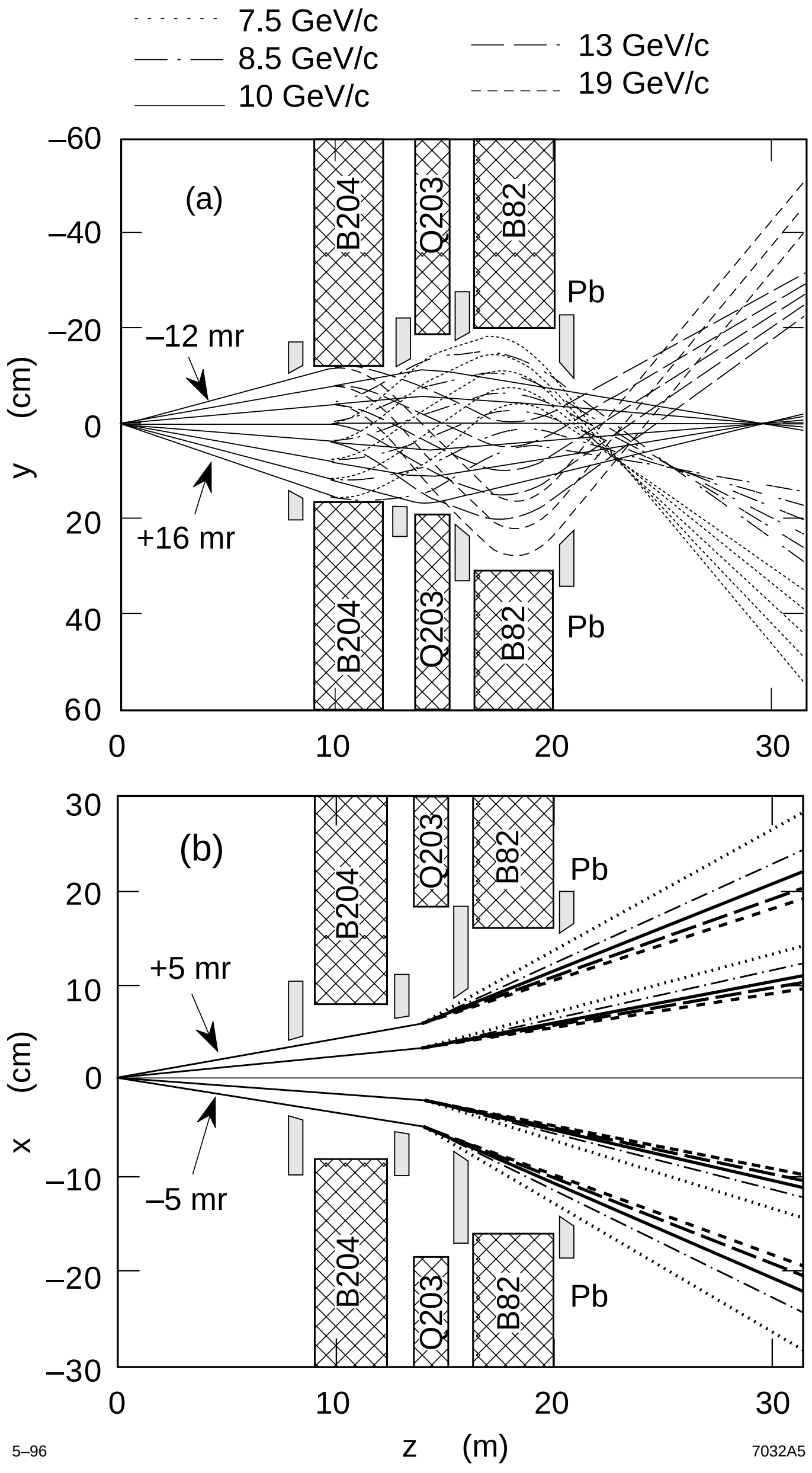

Figure 13: Bend plane (a) and non-bend plane (b) raytrace for the 4.5° spectrometer for rays of different momenta originating from the center of the polarized target. All rays are drawn with respect to the central trajectory of the system ( (a) $\phi_0 = 0$ mr, (b) $\theta_0 = 0$ mr and $p_0 = 10$ GeV/$c$ ). Also shown are the iron magnet poles and lead collimators.

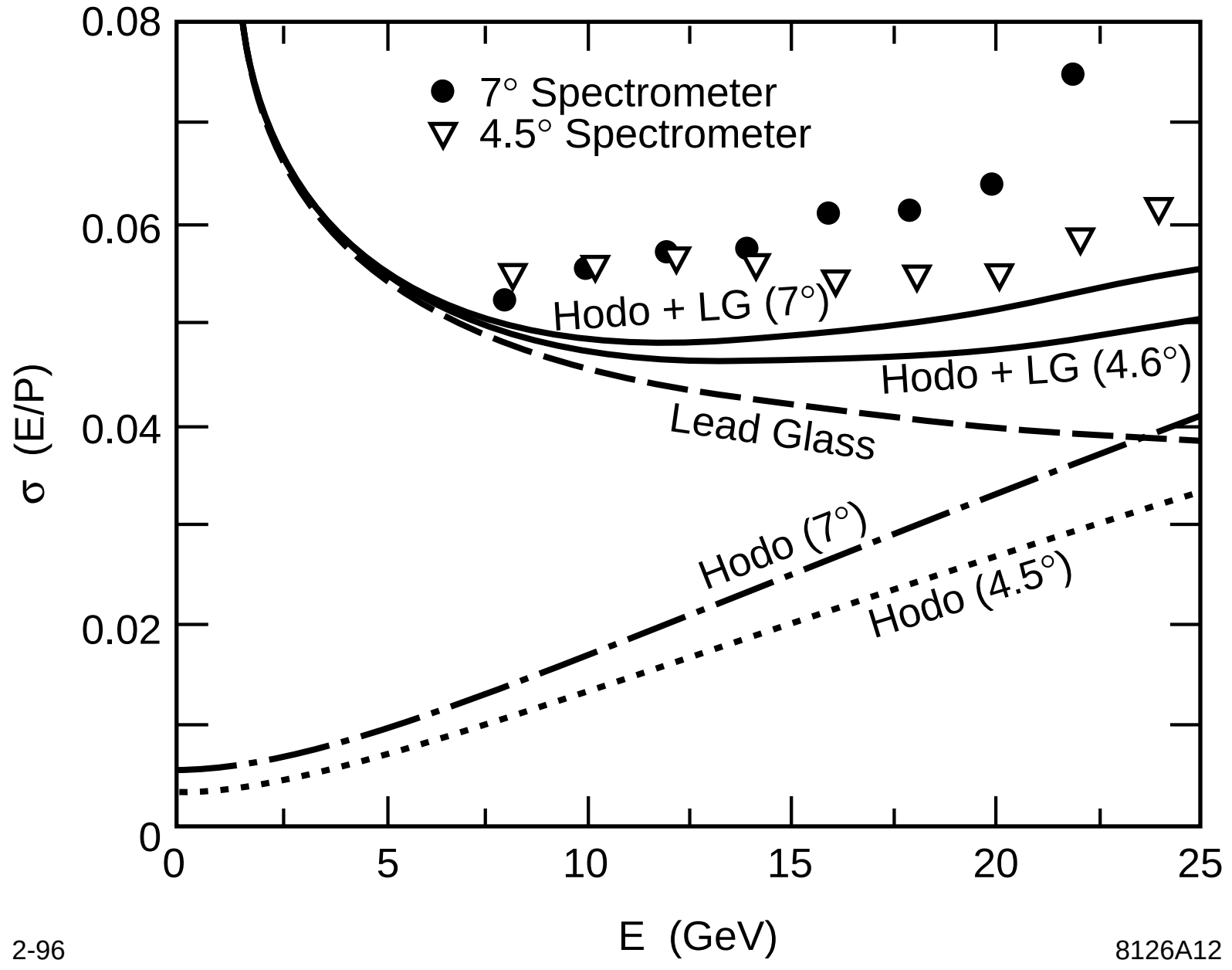

Figure 14: Data for the resolution of the ratio of energy in the shower counter divided by momentum from tracking $\sigma(E'/P)$ versus $E'$ for each spectrometer. Expected contributions from tracking ("hodoscope"), from energy deposition in the lead glass (LG) and from both combined (hodoscope + lead glass calorimeter) are also shown by the curves.

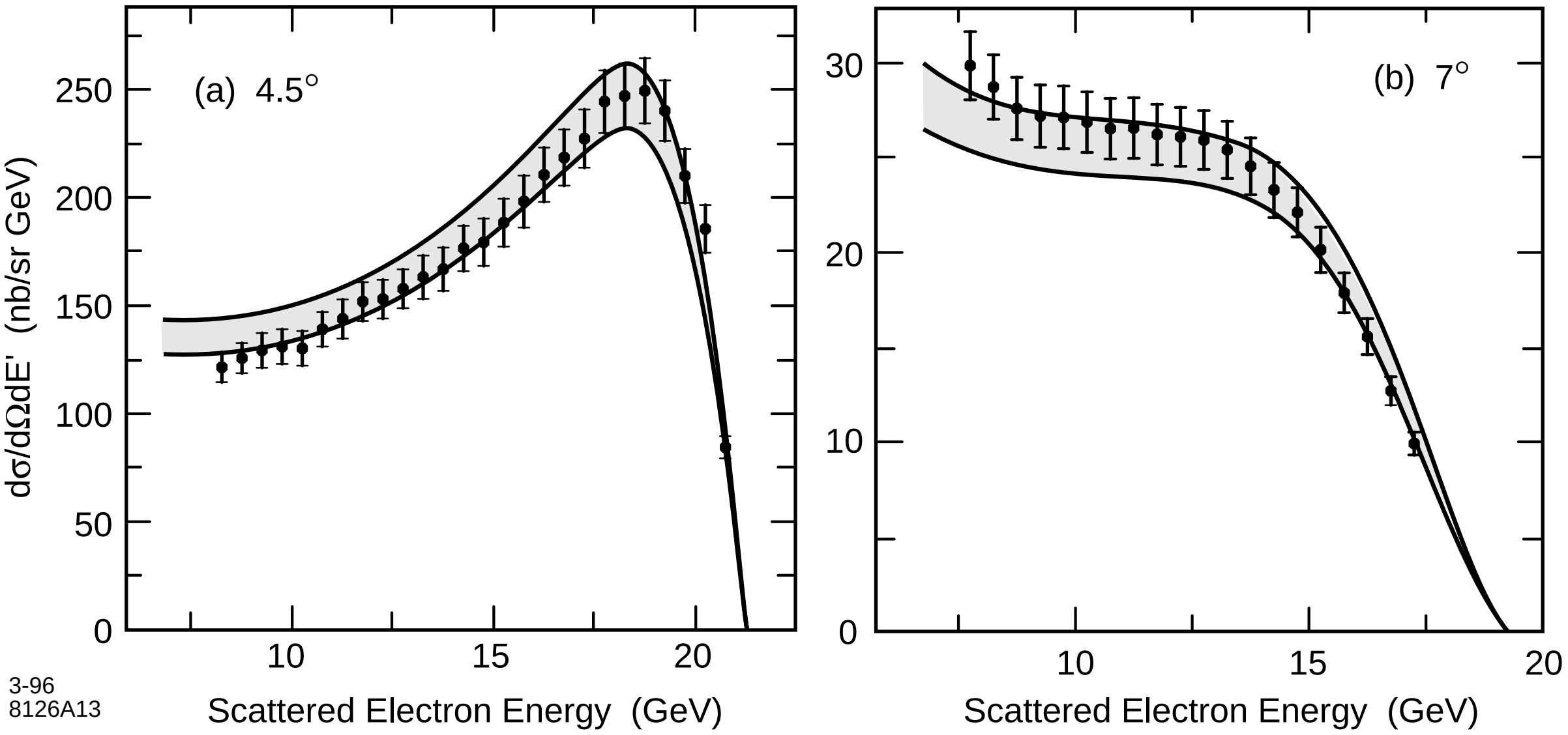

Figure 15: Comparison between measured (solid circles) and Monte Carlo simulated (highlighted band) deep inelastic cross sections for the 4.5° spectrometer (a) and for the 7° spectrometer (b). The data error bars are dominated by uncertainties in the spectrometer solid angle. The expected cross section is based on a model which relies on previous SLAC and CERN measurements. The width of the band is due to uncertainty in the target density.

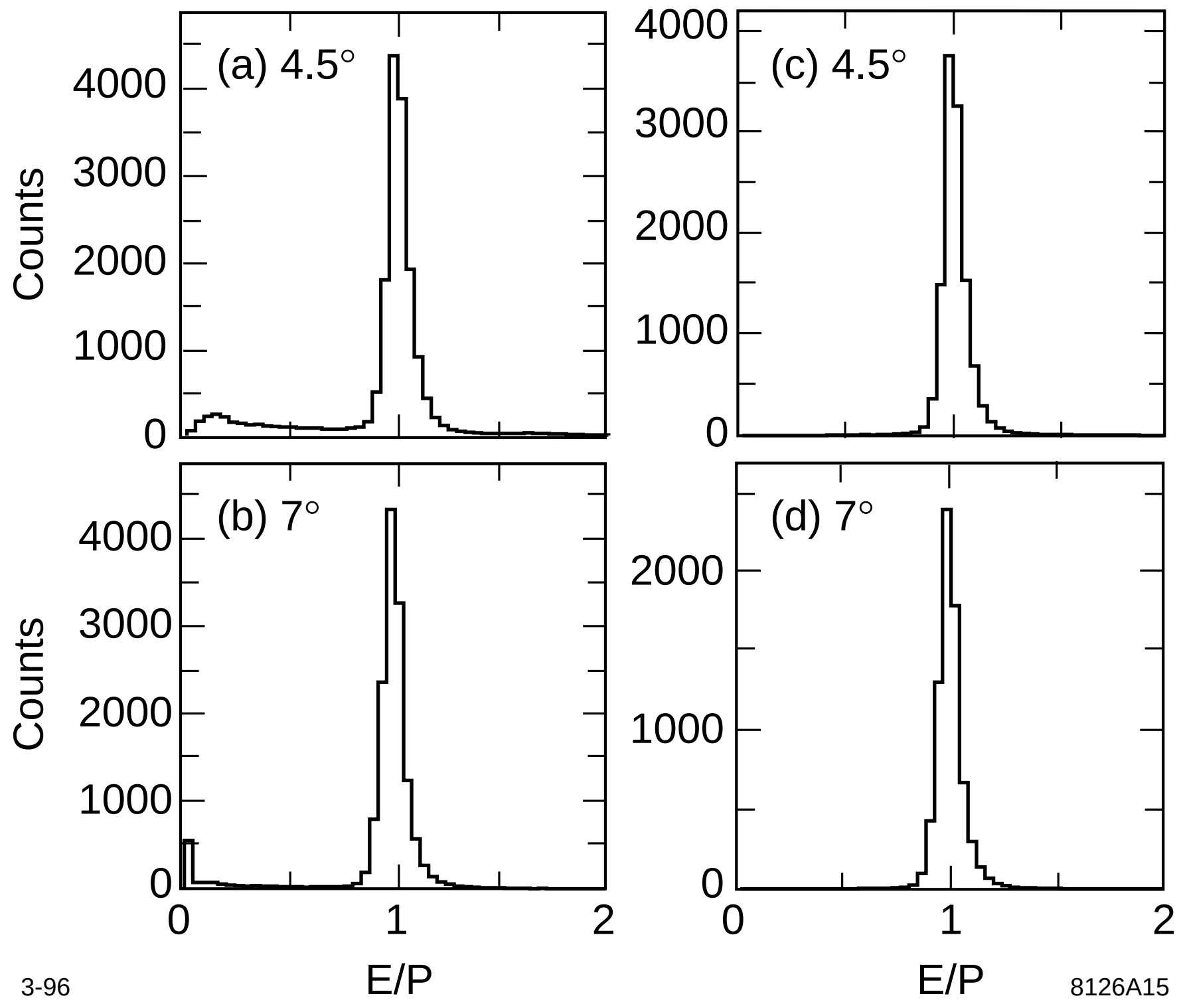

Figure 16: Plots a) and b) represent the ratio of energy (determined by the calorimeter) to the momentum (determined by the hodoscopes) of events detected in the 4.5° and 7.0° spectrometers. The electrons are identified by the peak centered around $E'/P = 1$, whereas the pions, which deposit less energy in the calorimeters, are in the region $E'/P < 0.8$. Plots c) and d) show events with the highest energy cluster for a given trigger and requiring an electron hardware trigger. These events define greater than 99% pure electrons sample and are those used in the physics analysis.

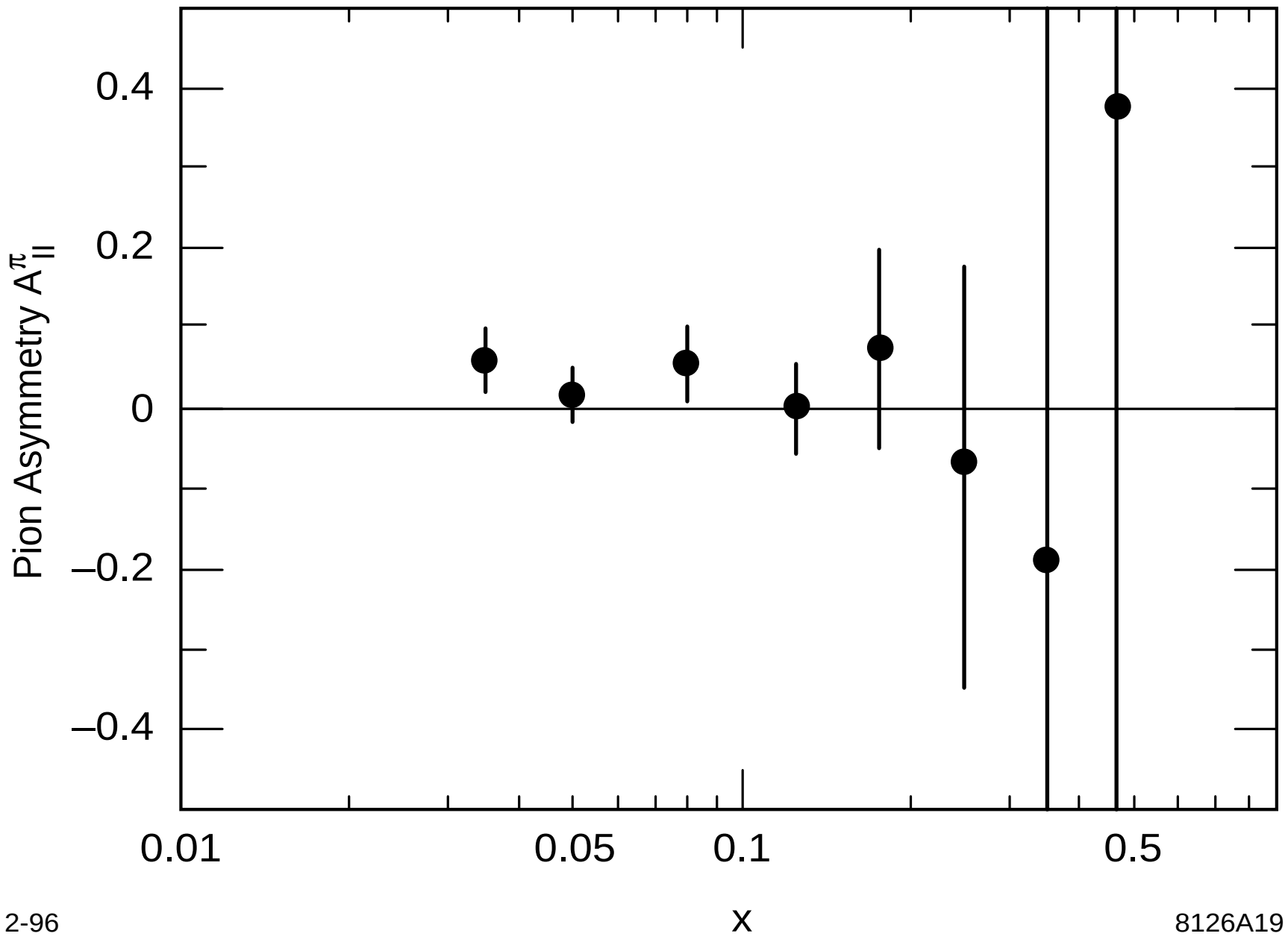

Figure 17: Inclusive pion asymmetry $A_{||}^{\pi}$ versus $x$. The results are consistent with zero ($\chi^2$ /d.f. = 0.6).

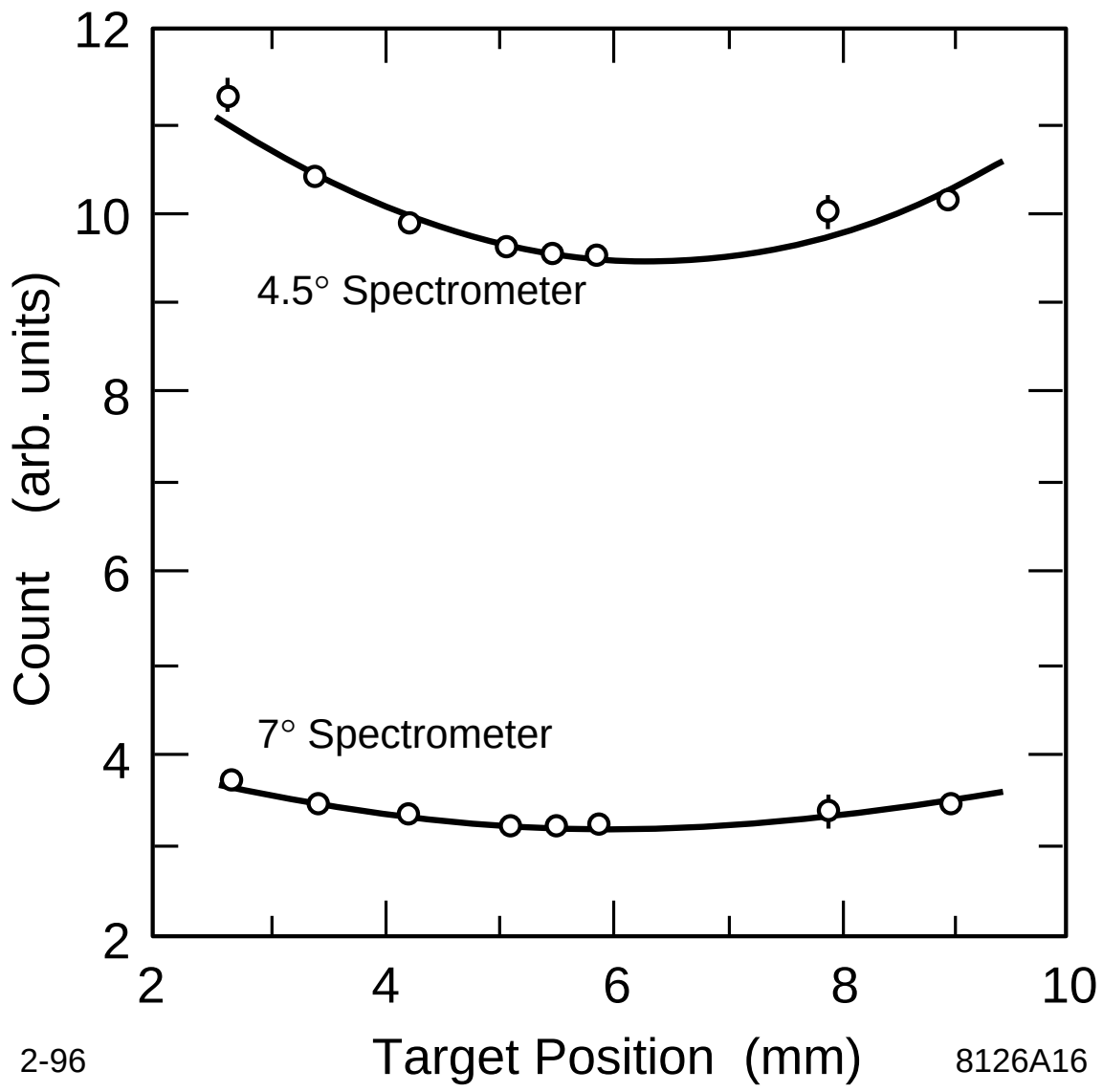

Figure 18: Spectrometer electron event rates as the target is swept vertically through the beam (6 mm is the normal position). The solid curves are quadratic fits.

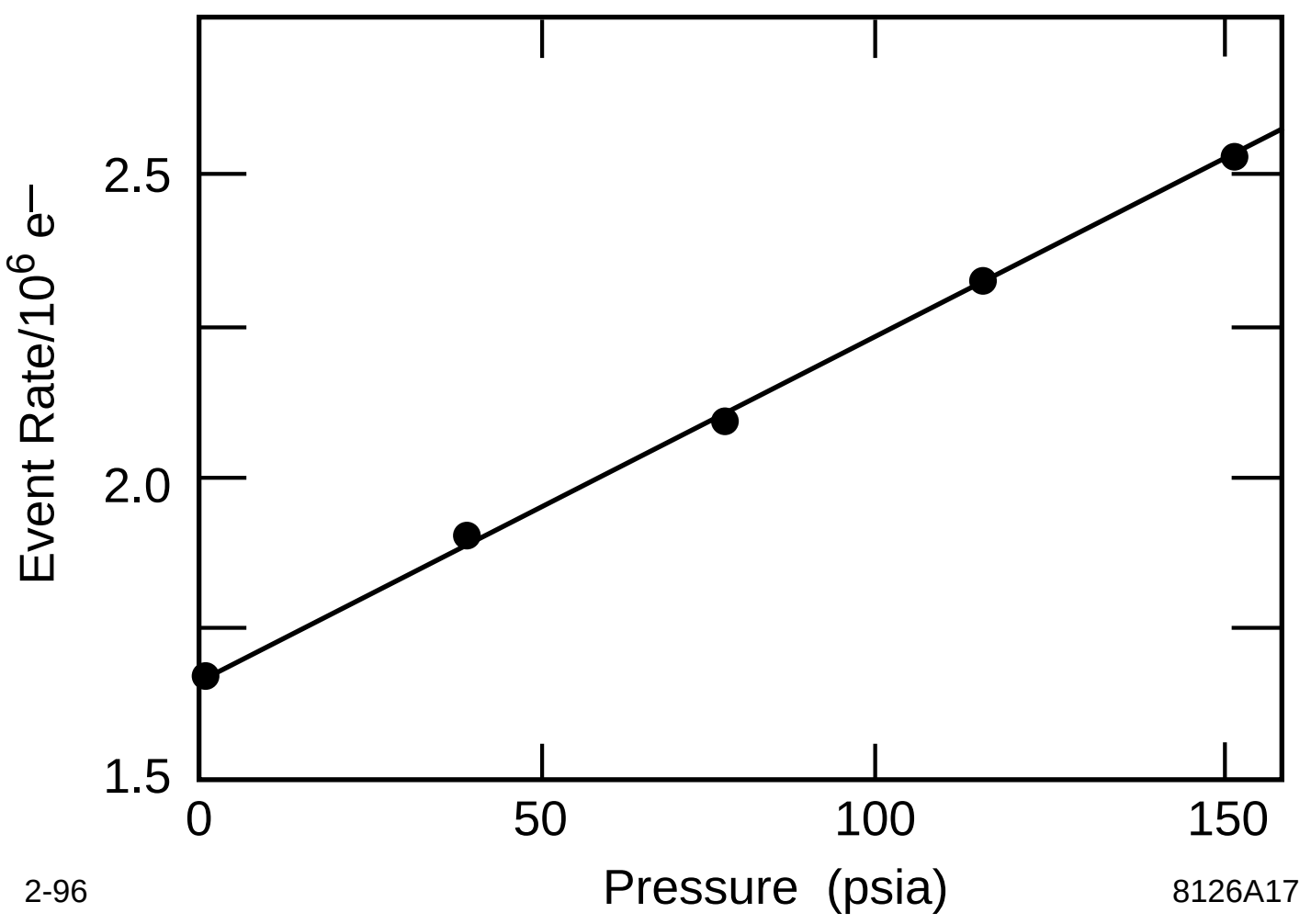

Figure 19: Event rate at $x = 0.175$ in the $4.5°$ arm versus pressure for a sequence of reference cell runs used to measure the dilution factor.

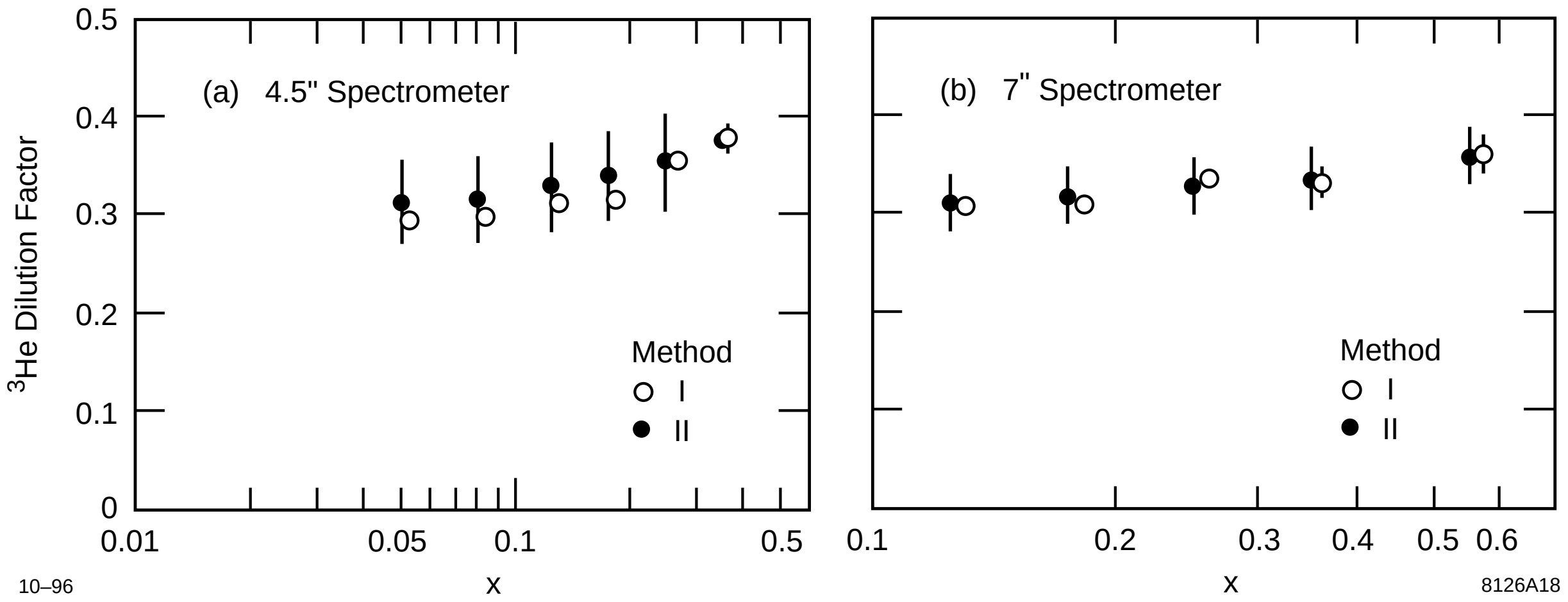

Figure 20: Material (method I) and reference cell (method II) results for the dilution factor of the reference cell as measured using the 4.5° arm (a) and the 7° arm (b) for a $^3$He gas pressure of 147 psi.

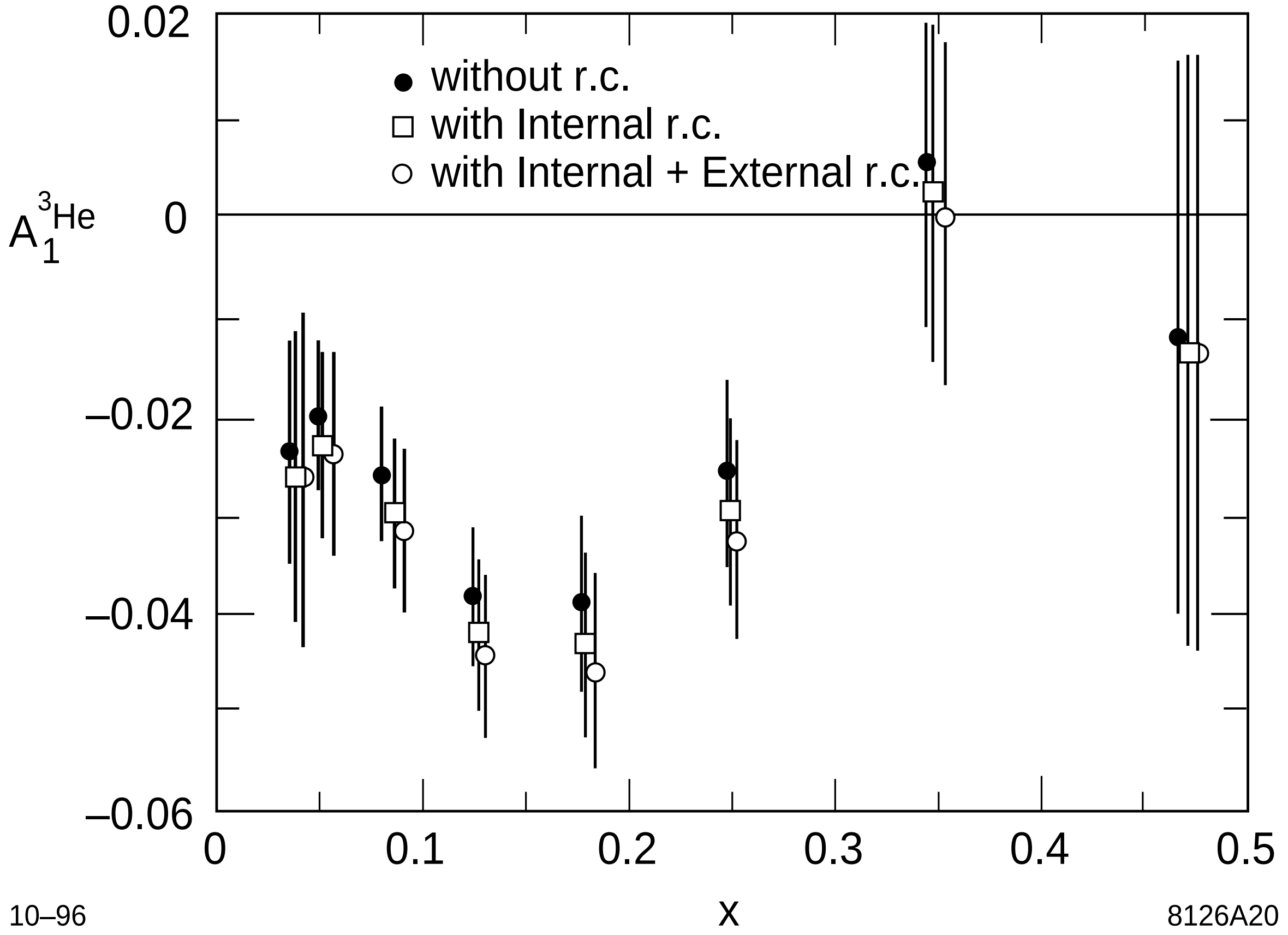

Figure 21: Change in the asymmetry $A_1^{^3He}$ (averaged over $E$ and $\theta$) as the radiative corrections (r.c.) are added. Only the statistical error on the final results are shown for comparison. The $x$ values of each data set are the same but they have been shifted for clarity.

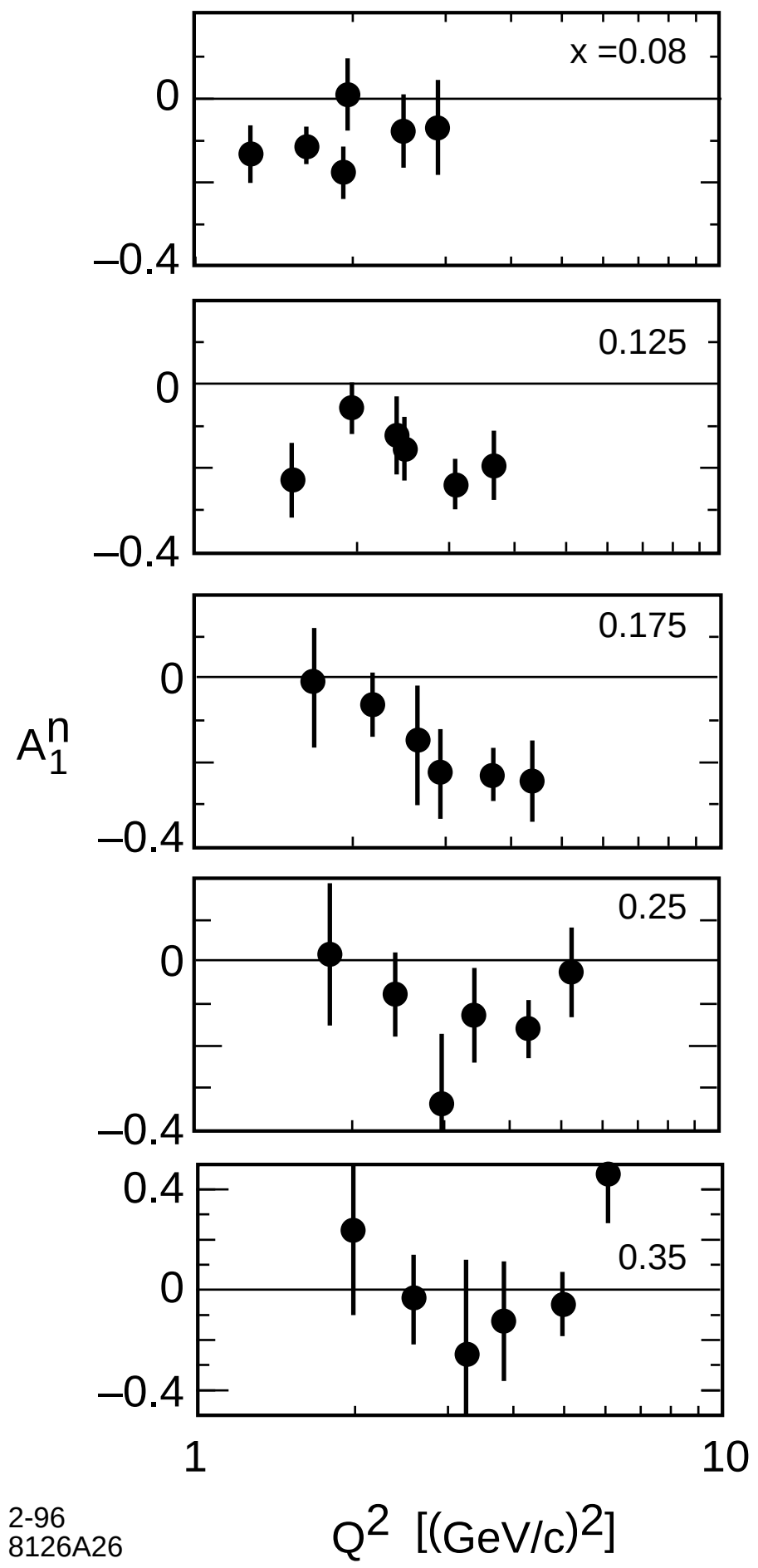

Figure 22: The asymmetry $A_1^n$ is plotted versus $Q^2$ for five different values of $x$. The results are consistent with $A_1^n$ being independent of $Q^2$. The data comes from the two spectrometer arms and three beam energies.

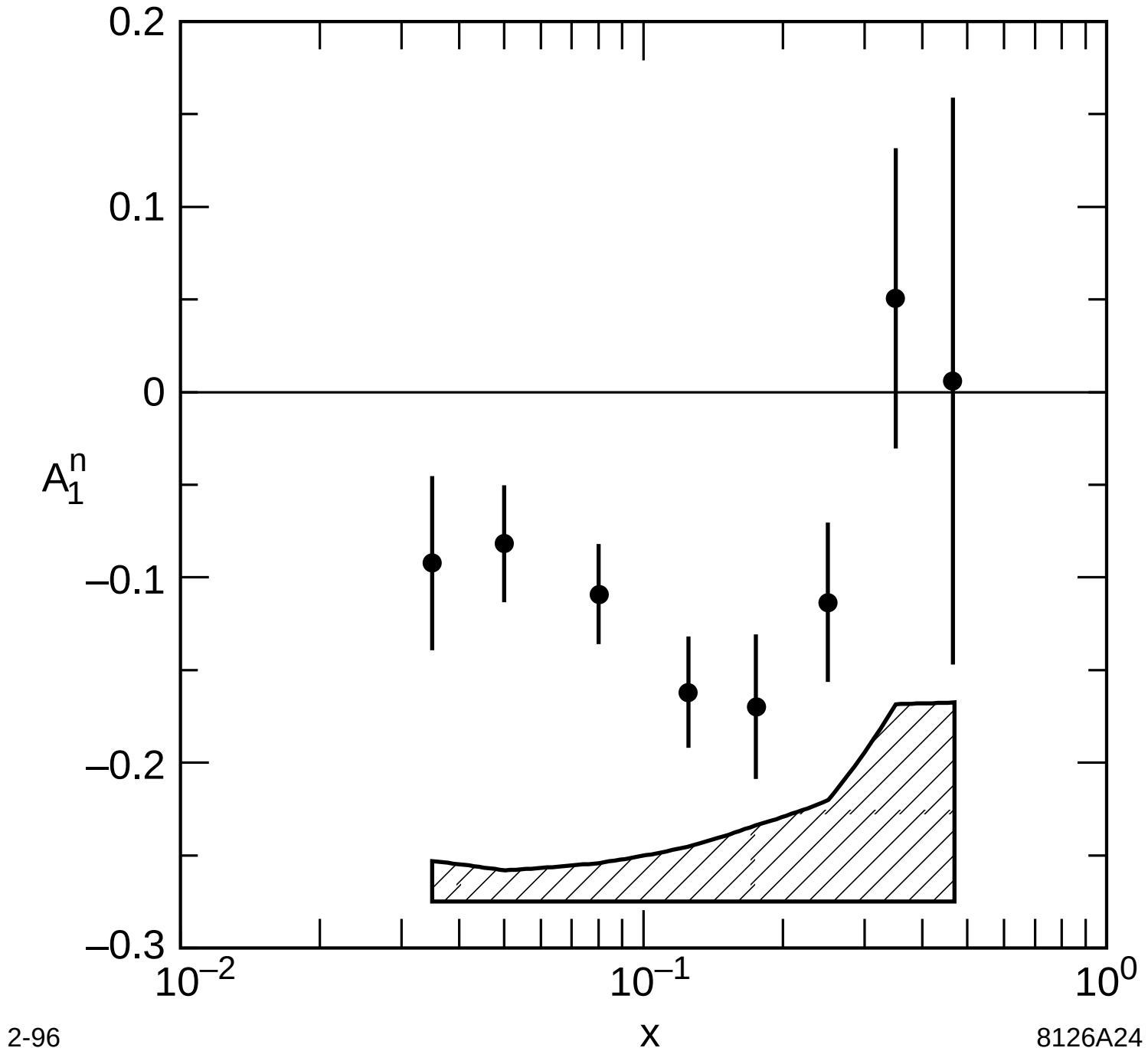

Figure 23: The neutron asymmetry $A_1^n$ versus $x$. The error bars are statistical, with the enclosed region at the bottom representing the size of the systematic errors.

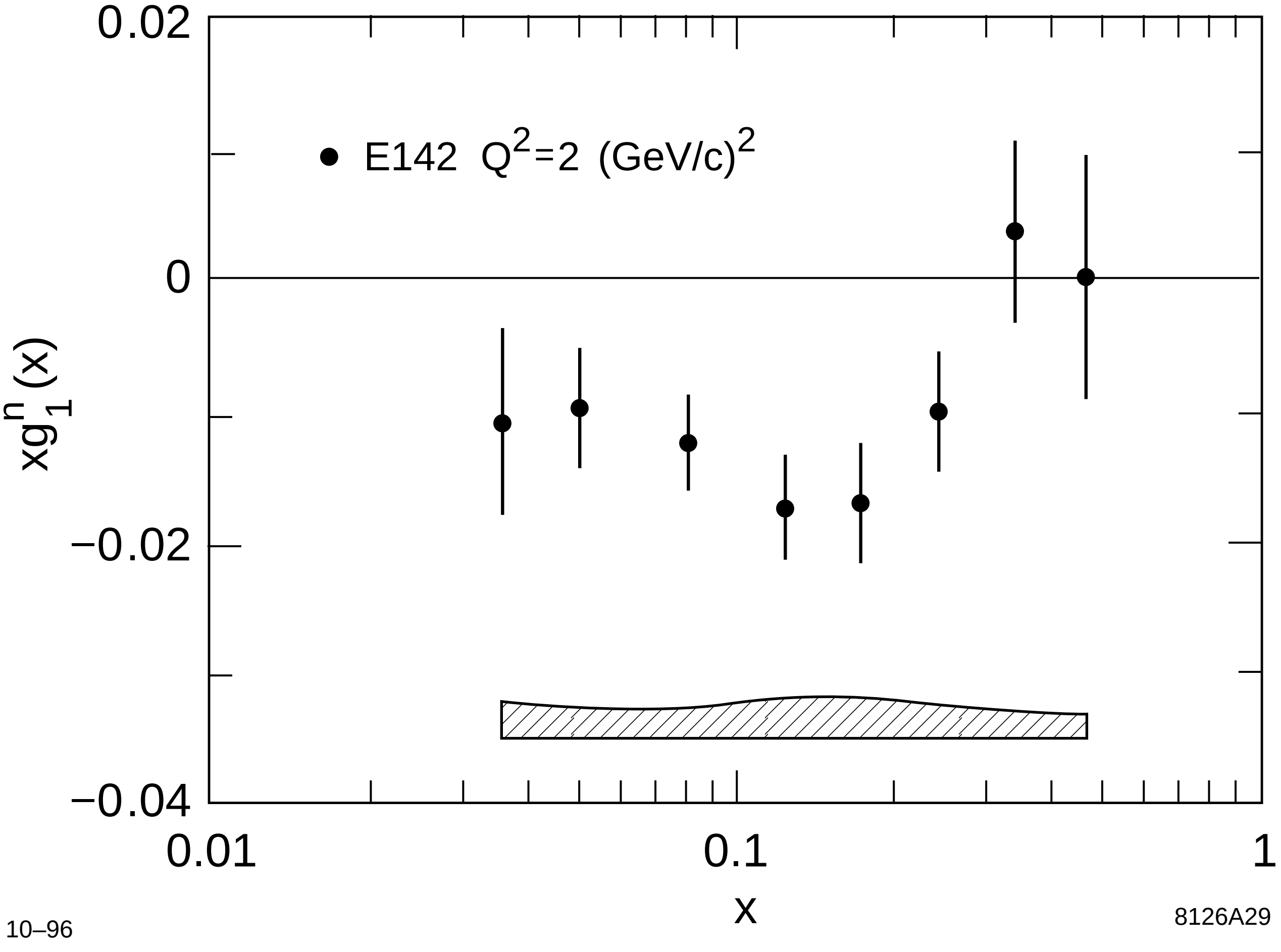

Figure 24: The spin structure function $xg_1^n$ evaluated at fixed $Q_0^2 = 2$ $(\mathrm{GeV}/c)^2$. The error bars are statistical while the band at the bottom represents the size of the systematic uncertainties $(1\sigma)$.

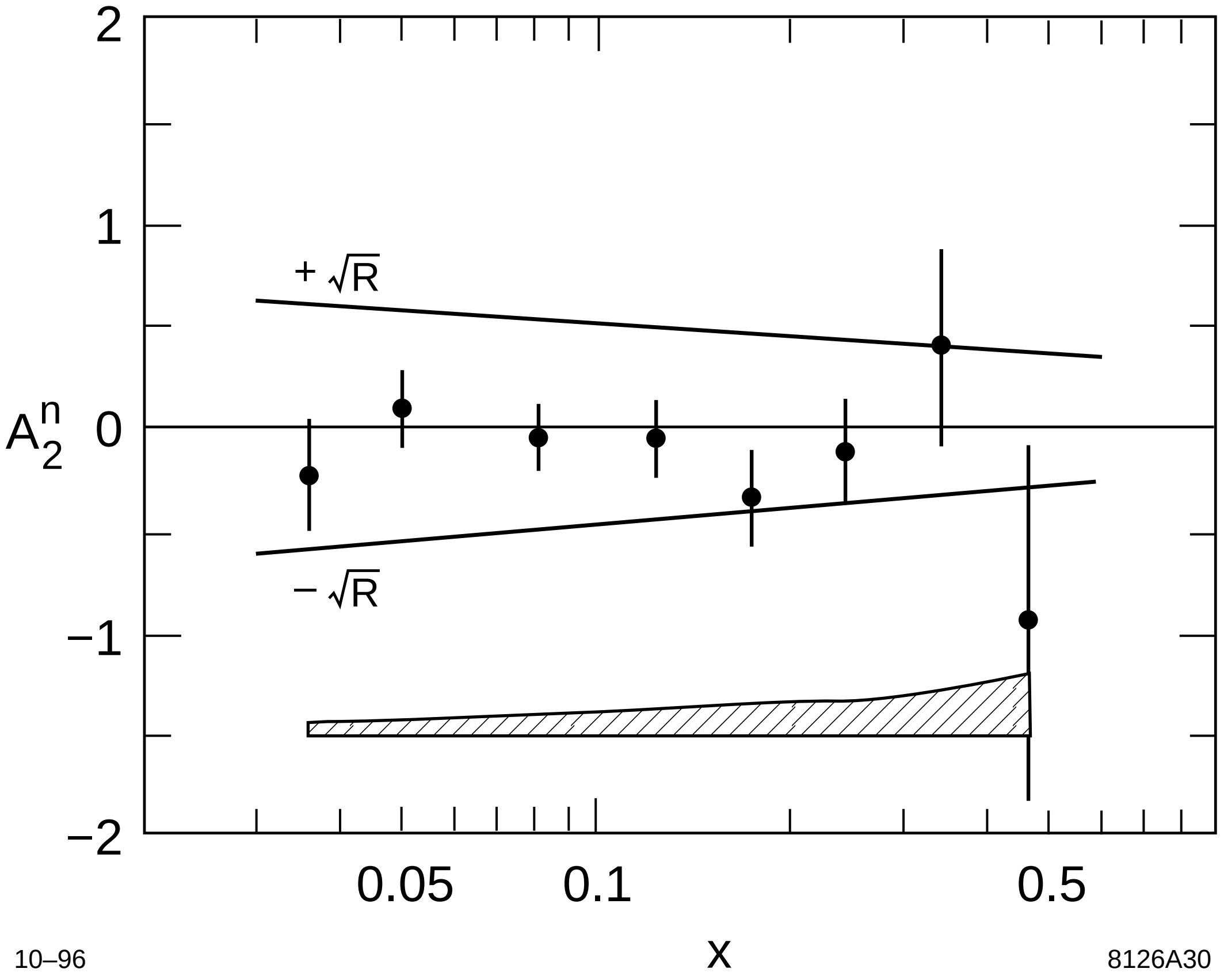

Figure 25: $A_2^n$ versus $x$ averaged over both spectrometers is shown. The error bars are statistical only (systematic errors are small in comparison).

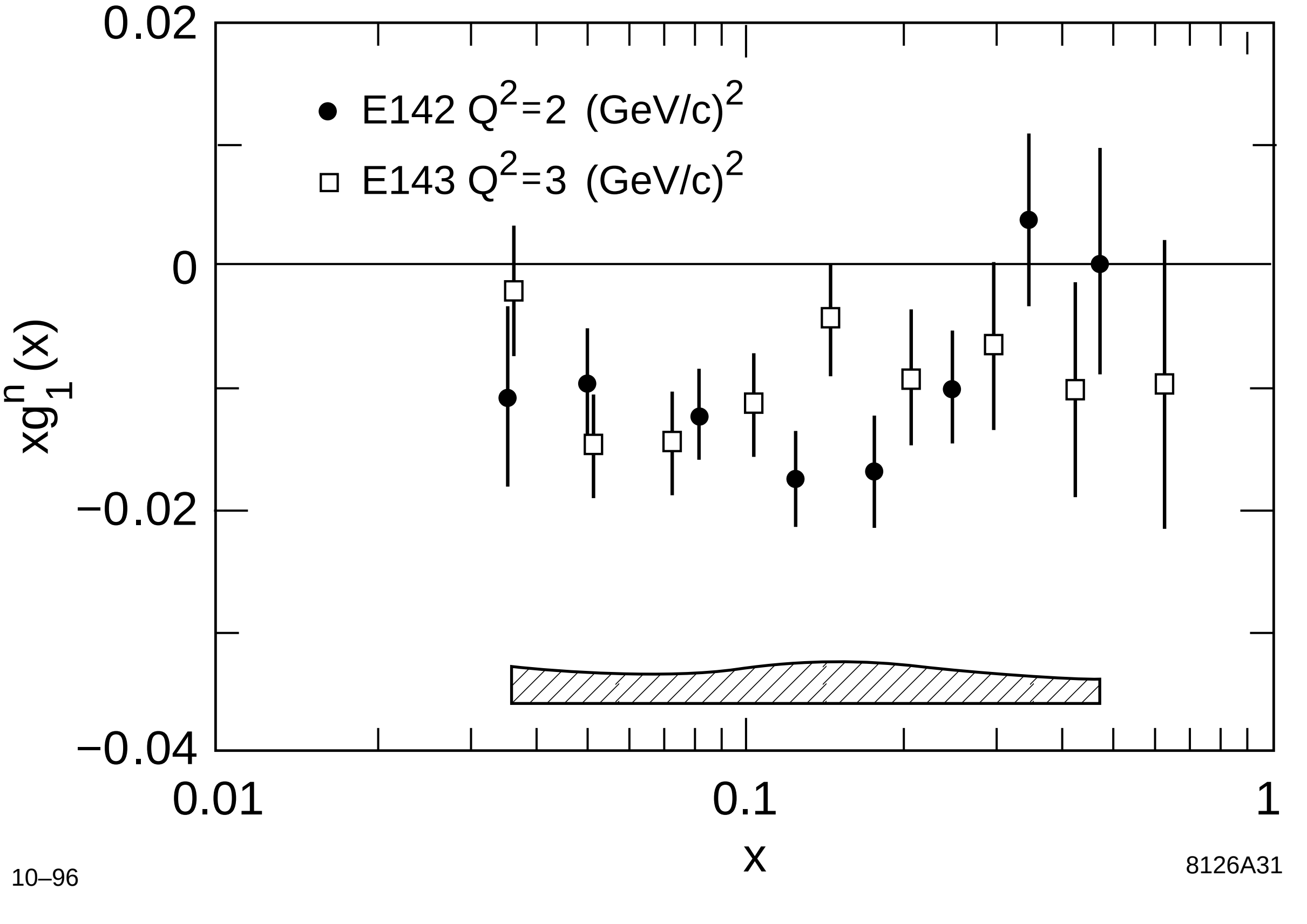

Figure 26: Comparison of $xg_1^n$ between E142 and E143.

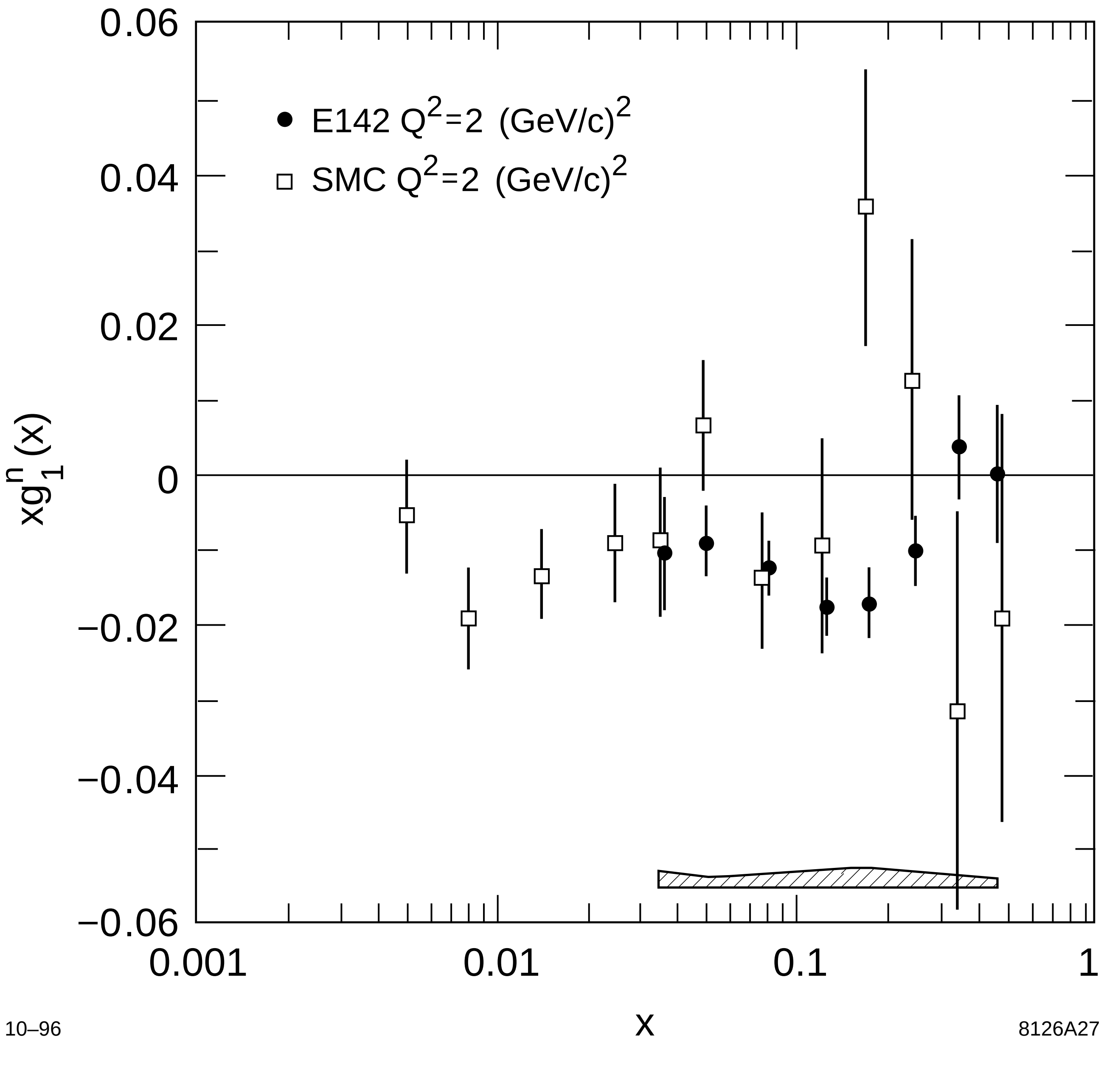

Figure 27: Comparison of $xg_1^n$ between E142 and SMC.

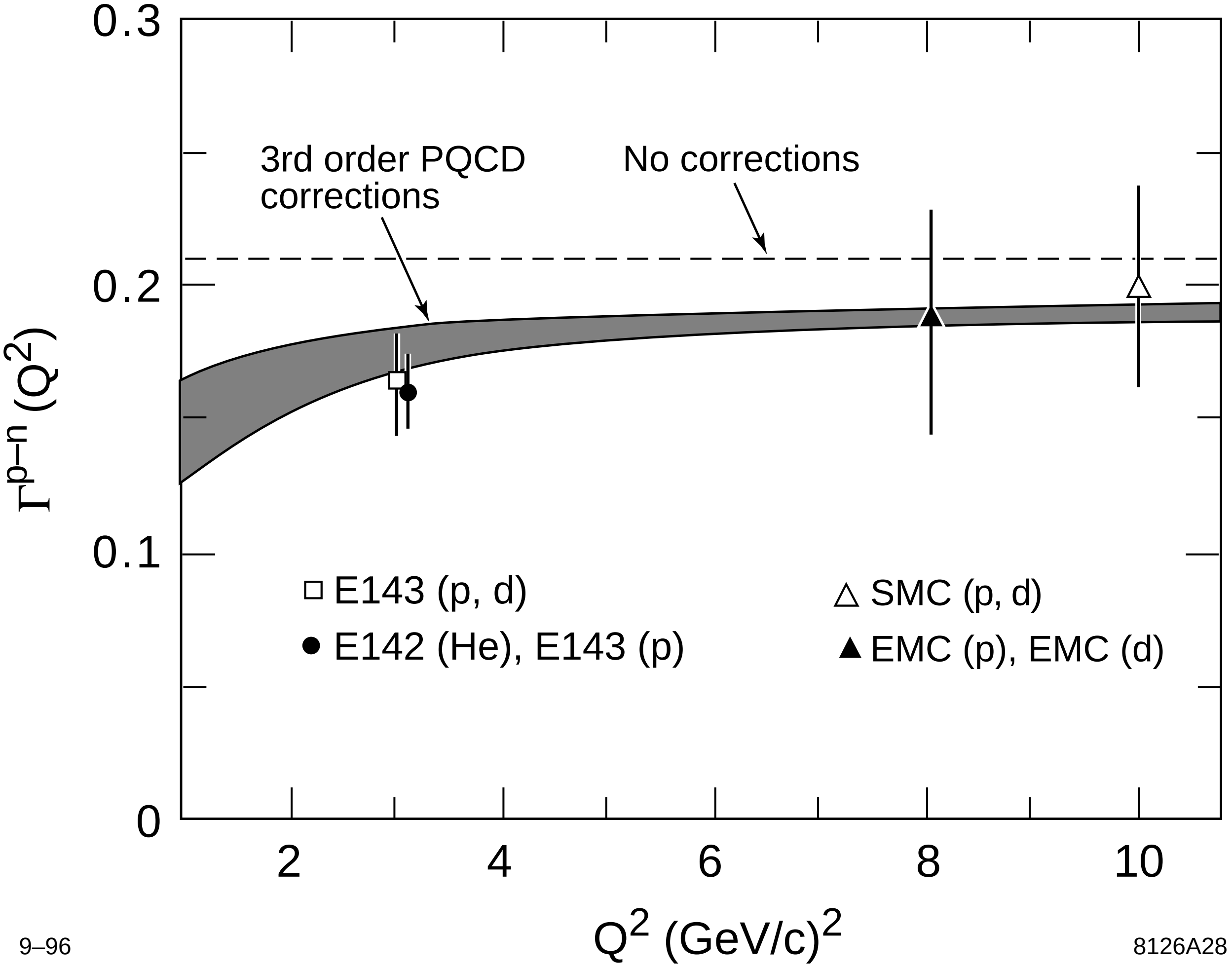

Figure 28: Comparison of data for $\Gamma_1^p - \Gamma_1^n$ using different experiments compared to the Bjorken sum rule with $3^{rd}$ order QCD corrections and no higher twist corrections. The E142 results are at $Q^2 = 3(\mathrm{GeV}/c)^2$ but have been shifted for clarity.